\documentclass{aa}  
\usepackage{graphicx}
\usepackage{placeins}
\usepackage{txfonts}
\usepackage{float}
\usepackage{rotating}
\usepackage[dvipsnames]{xcolor}
\usepackage[colorlinks=true,linkcolor=bluea,allcolors=blue]{hyperref}

\begin{document}

   \subtitle{Investigating the impact of quasar-driven outflows on galaxies at z$\sim$0.3-0.4}

   \author{K. Hervella Seoane\inst{1}
          \and C. Ramos Almeida\inst{2,1}
          \and J. A. Acosta-Pulido\inst{2,1}
          \and G. Speranza\inst{2,1}
          \and C. N. Tadhunter\inst{3}
             \and P. S. Bessiere\inst{2,1}
          }

\titlerunning{Investigating the impact of quasar-driven outflows}

   \institute{Departamento de Astrof\'{\i}sica, Universidad de La Laguna, 38206, La Laguna, Tenerife, Spain\\
              \email{kiherse@gmail.com}
         \and
             Instituto de Astrof\'{\i}sica de Canarias, Calle V\' ia Láctea, s/n, 38205, La Laguna, Tenerife, Spain
             \and
             Department of Physics \& Astronomy, University of Sheffield, S3 7RH Sheffield, UK}

   \date{}


  \abstract 
   {}
   {We present a detailed study of the kinematics of 19 type 2 quasars (QSO2s) with redshifts in the range 0.3$<$z$<$0.41 and [OIII] luminosities of $L_{\rm [OIII]} > 10^{8.5}$L$_{\rm \odot}$. We aim at advancing our understanding of the AGN feedback phenomenon by correlating outflow properties with i) the presence of young stellar populations (YSPs) with ages <100 Myr, ii) the optical morphology and the environment of the galaxies, and iii) the radio luminosity.}
   {We characterize the ionized gas kinematics using the [OIII]$\lambda$5007 \AA~emission line profiles detected in intermediate spectral resolution (R$\sim$1500-2500) optical spectra of the QSO2s. To do so we employed three different outflow detection methods: multi-component parametric, flux-weighted non-parametric, and peak-weighted non-parametric.}
   {We detect ionized outflows in 18 of the 19 QSO2s using the parametric analysis, and in all of them using the non-parametric methods. We find higher outflow masses using the parametric analysis (average log M$_{\rm OF}$(M$_{\odot}$)=6.47$\pm$0.50), and larger mass rates and kinetic powers with the flux-weighted non-parametric method (\.M$_{\rm OF}$=4.0$\pm$4.4 M$_{\odot}$~yr$^{-1}$ and \.E$_{\rm kin}$=41.9$\pm$0.6 erg~s$^{-1}$). However, it is when we use the parametric method and the maximum outflow velocities (v$_{\rm max}$) that we measure the highest outflow mass rates and kinetic energies (\.M$_{\rm OF}$=23$\pm$35 M$_{\odot}$~yr$^{-1}$ and log(\.E$_{\rm kin}$)=42.9$\pm$0.6 erg~s$^{-1}$).
   We do not find any significant correlation between the outflow properties and the previously mentioned AGN and galaxy-wide properties.}
   {Four of the five QSO2s without a YSP of age <100 Myr show highly disturbed kinematics, whereas only 5 out of the 14 QSO2s with YSPs show similarly asymmetric [OIII] profiles. Despite the small sample size, this might be indicative of negative feedback. The lack of correlation between the outflow properties and the presence of mergers in different interaction stages might be due to their different dynamical timescales (Myr in the case of the outflows versus Gyr in the case of the mergers). Lastly, the small radio luminosity range covered by our sample, log(L$_{\rm 5GHz}$)=[22.1, 24.7] W~Hz$^{-1}$, may be impeding the detection of any correlation between radio emission and outflow properties.}

   \keywords{galaxies: nuclei -- galaxies: evolution -- galaxies: interactions -- galaxies: kinematics and dynamics -- ISM: jets and outflows}

   \maketitle

\section{Introduction}\label{introduction}

All galaxies, or at least the most massive ones, are thought to experience short episodes of nuclear activity, of $\leq$100 Myr \citep{2004cbhg.symp..169M,novak2011feedback,schawinski2015active}. These nuclear activity phases are considered key drivers of galaxy evolution because they can potentially regulate black hole and galaxy growth \citep{2005Natur.433..604D,harrison2017impact}. 

Cosmological simulations require active galactic nuclei (AGN) feedback for quenching star formation and preventing galaxies from becoming over-massive \citep{2005Natur.433..604D,dubois2016horizon}, thereby correctly reproducing the observed galaxy-halo mass relations \citep{1998A&A...331L...1S,croton2006many,moster2010constraints}. Furthermore, observational studies have found plenty of evidence of this feedback on different scales, from the central tens to hundreds of parsecs \citep{2021A&A...652A..98G,2022A&A...658A.155R} to hundreds of kpc \citep{2019Natur.574..643R,martin2021anisotropic}. 

Multi-phase outflows of molecular, neutral, and ionized gas \citep{rupke2013multiphase,2018NatAs...2..176C,herrera2020agn,fluetsch2021properties} are one of the primary AGN feedback mechanisms that can quench star formation by heating up, disrupting, and ultimately removing the surrounding gas available to form stars. However, AGN-driven outflows have also been found to have the opposite effect, often referred to as ``positive feedback'', as they can promote star formation by pressurizing the gas and enhancing fragmentation \citep{klamer2004molecular,cresci2015magnum,cresci2018observing,carniani2016fast,bessiere2022spatially}. Hence, we still need to advance in our understanding of their actual impact on star formation.

The drivers of these multi-phase outflows have also been a subject of study, with radio jets or AGN-winds as the main potential candidates \citep{2018MNRAS.476...80M,2019ApJ...887..200F}. For example, jetted Seyfert galaxies \citep{whittle1992virial,garcia2014molecular,garcia2019alma,morganti2015fast} have been found to show larger outflow velocities than those without jets, suggesting an influence of these jets in launching and/or accelerating the outflows \citep{wylezalek2018questions,jarvis2019prevalence}. Furthermore, even in the case of lower radio-power Seyferts and radio-quiet quasars, recent studies have found evidence that compact and modest radio jets might induce nuclear outflows \citep{aalto2016precessing,audibert2019alma,fernandez2020co,garcia2021multiphase,speranza2022warm,2023A&A...671L..12A}. The influence of jets on the ionized gas kinematics has been also studied using low angular resolution data, as e.g., from the Sloan Digital Sky Survey (SDSS): \citet{mullaney2013narrow} and \citet{zakamska2014quasar} found that the higher the radio luminosity, the more disrupted the [OIII] kinematics. However, it has also been claimed that this correlation disappears if the influence of the host galaxy gravitational potential is taken into account (see \citealt{2023ApJ...954...27A} and references therein).  

Mergers and interactions have long been known as an AGN triggering mechanism \citep{canalizo2001quasi,hopkins2008cosmological,almeida2011optical,ramos2012luminous,bessiere2012importance,satyapal2014galaxy,goulding2018galaxy,pierce2023galaxy}. Given that mergers can simultaneously enhance nuclear star formation and nuclear activity (e.g., \citealt{satyapal2014galaxy}), it might be possible that outflow incidence and properties might show a dependence with galaxy morphology and/or environment. For example, some of the most powerful outflows in the local universe are found in ultra-luminous infrared galaxies (ULIRGs; \citealt{cicone2014massive,rose2018quantifying,2022A&A...668A..45L}), which are almost uniquely associated with major mergers. Apart from these, the merger-induced gas flows may also provide a rich ISM with plenty of cool gas for the radiation-driven winds and any jets to interact with, thus increasing the coupling with the hosts and making the outflows easier to detect. 

The properties of AGN-driven outflows have been shown to depend on AGN luminosity, being faster and more powerful as the AGN luminosity increases \citep{zakamska2014quasar,fiore2017agn}. However, recent works showed that other factors, including nuclear gas concentration and the coupling between the winds/jets and the galaxy discs might also play a key role \citep{ramos2012luminous,2023A&A...671L..12A}. Type 2 quasars (QSO2s) in the local universe are ideal laboratories to characterize outflows and study their impact on the host galaxies. This is because the outflows are easier to identify that the lower velocity outflows of Seyfert galaxies, and the high obscuration (either nuclear or galaxy-wide; \citealt{almeida2017nuclear}) blocks the emission from the broad-line region and the AGN continuum, avoiding dilution of the emission line and stellar absorption features. 

The most widely studied gas outflow phase is the warm ionized (T$\sim$10$^4$ K), since it emits strong forbidden emission lines in the infrared and optical range, being the [OIII]$\lambda$5007 \AA~one of the strongest. 
QSO2s are commonly found to present complex emission line profiles, showing large blue asymmetries and deviations from Gaussianity \citep{greene2011feedback,liu2013observations,villar2011ionized,villar2014triggering,harrison2014kiloparsec,2017MNRAS.470..964R,ramos2019near}, indicating highly-disrupted kinematics. The overall majority of AGN outflow studies characterized the ionized gas kinematics following a parametric approach \citep{holt2011impact,greene2011feedback,villar2011ionized,arribas2014ionized}, considering that each gaseous component that contributes to the kinematics can be described by a Gaussian distribution. They have the advantage of making it possible to characterize the properties of different gas components, yet, for objects with highly disrupted kinematics it becomes difficult to ascribe a physical meaning to each component \citep{bessiere2022spatially}. Hence, other studies implemented a non-parametric analysis, based on measuring emission line velocities at fixed fractions of the peak intensity or the cumulative line flux function \citep{whittle1985narrow,harrison2014kiloparsec,zakamska2014quasar,speranza2021murales,bessiere2022spatially}. Non-parametric methods are better for characterizing low signal-to-noise spectra with complex emission line profiles. Although they do not allow us to separate different gaseous contributions, they can easily identify high velocity gas in tail ends of emission line wings. 
Despite all the recent works making use of both parametric and non-parametric methods to characterize AGN-drive outflows in the literature, to the best of our knowledge, no comparison between their results has been performed for a common sample of objects.

Considering the previous, in this work we use three different outflow detection methods (parametric, flux-weighted non-parametric, and the peak-weighted non-parametric from \citealt{speranza2021murales}) for characterizing the ionized gas kinematics of 19 QSO2s with redshifts in the range 0.3$<$z$<$0.41 and [OIII] luminosities of $L_{\rm [OIII]} > 10^{8.5}$L$_{\rm \odot}$. Furthermore, with the aim of advancing in our understanding of the AGN feedback phenomenon, we study potential correlations between the QSO2s outflow physical properties and different AGN and host galaxy properties, including i) the presence of young stellar populations (YSPs) with ages <100 Myr, ii) the optical morphology and environment, and iii) the AGN and radio luminosity.

In Section \ref{data} we describe the QSO2 sample and the spectroscopic data used here, as well as the previously studied AGN and host galaxy properties. In Section \ref{analysis} we explain the three different methods we used for analyzing the ionized gas kinematic and present the results. Section \ref{outflowproperties} includes the physical outflow properties derived through each of the three methods. In Section \ref{correlations} we evaluate the possible correlations between the outflow properties and different AGN and host galaxy properties.
In Section \ref{discussion} we discuss the results, and finally in Section \ref{conclusions} we summarize the findings of our work. Throughout this paper we assumed a cosmology with H$_0$=70 km~s$^{-1}$~Mpc$^{-1}$, $\Omega_{\rm M}$=0.3 and $\Omega_{\rm \Lambda}$=0.7.


\section{Sample and data}\label{data}

\begin{sidewaystable*}[!h]
\caption{Main properties of the QSO2s}
\centering
\begin{tabular}{ccccccccrcccc} \hline
\textbf{Name} & \textbf{RA} & \textbf{Dec} & \textbf{z} & \textbf{Seeing} & \textbf{Scale} & \textbf{log(L$_{\rm [OIII]}$/L$_\odot$)} & \textbf{log(L$_{\rm BOL}$)} & \textbf{log(L$_{\rm 5GHz}$)} & \textbf{YSP} & \textbf{Merger} & \textbf{Stage} & \textbf{B$_{gq}$} \\
\multicolumn{1}{l}{} & \multicolumn{1}{l}{}  & \multicolumn{1}{l}{} & \multicolumn{1}{l}{} & (arcsec) & (kpc arcsec$^{-1}$) & \multicolumn{1}{l}{} & 
(erg s$^{-1}$) & (W $Hz^{-1}$)  &  \textbf{< 100 Myr}\\ \hline
\textbf{J0025-10} & 00:25:31.46 & -10:40:22.2 & 0.303 & 0.84 & 4.48 & 9.23(8.65) & 45.48 & 22.08 & yes & yes & 2,3 & -16 \\
\textbf{J0114+00} & 01:14:29.61 & 00:00:36.7 & 0.389 & - & 5.28 & 8.90(8.46) & 45.15 & 24.70 & no & yes & 2,3 & 254 \\
\textbf{J0123+00} & 01:23:41.47 & 00:44:35.9 & 0.399 & 1.12 & 5.36 & 9.53(9.14) & 45.78 & 23.27 & yes & yes & 1,2,3 & 676 \\
\textbf{J0142+14} & 01:42:37.49 & 14:41:17.9 & 0.389 & - & 5.28 & 9.40(8.87) & 45.65 & \textless{}22.58 & yes & no & 4 & 149 \\
\textbf{J0217-00} & 02:17:58.19 & -00:13:2.7 & 0.344 & 0.86 & 4.88 & 9.48(8.81) & 45.73 & 22.15 & yes & yes & 2 & 63 \\
\textbf{J0217-01} & 02:17:57.42 & -01:13:24.4 & 0.375 & 0.97 & 5.16 & 8.90(8.37) & 45.15 & 22.35 & yes & no & 4 & -153 \\
\textbf{J0218-00} & 02:18:34.42 & -00:46:10.3 & 0.372 & 1.11 & 5.13 & 9.07(8.62) & 45.32 & \textless{}22.13 & yes & yes & 1,2 & -49 \\
\textbf{J0227+01} & 02:27:1.23 & 01:07:12.3 & 0.363 & 0.53 & 5.05 & 9.13(8.70) & 45.38 & \textless{}22.11 & yes & yes & 2 & 159 \\
\textbf{J0234-07} & 02:34:11.77 & -7:45:38.4 & 0.31 & 0.78 & 4.55 & 9.81(8.78) & 45.06 & 22.35 & no & no & 4 & -115 \\
\textbf{J0249+00} & 02:49:46.09 & 00:10:3.22 & 0.408 & 0.73 & 5.44 & 9.25(8.75) & 45.50 & 22.14 & yes & yes & 1,2 & -177 \\
\textbf{J0320+00} & 03:20:29.78 & 00:31:53.5 & 0.384 & 0.75 & 5.23 & 8.63(8.40) & 44.88 & \textless{}22.17 & yes & yes & 2 & 1297 \\
\textbf{J0332-00} & 03:32:48.5 & -00:10:12.3 & 0.31 & 0.83 & 4.55 & 8.69(8.53) & 44.94 & 23.36 & no & yes & 2,3 & -19 \\
\textbf{J0334+00} & 03:34:35.47 & 00:37:24.9 & 0.407 & 1.13 & 5.43 & 9.26(8.75) & 45.51 & \textless{}22.23 & yes & yes & 2 & 36 \\
\textbf{J0848-01} & 08:48:56.58 & 01:36:47.8 & 0.35 & 0.72 & 4.95 & 8.90(8.46) & 45.15 & 24.00 & no & yes & 2 & 159 \\
\textbf{J0904-00} & 09:04:14.1 & -00:21:44.9 & 0.353 & 0.71 & 4.98 & 9.27(8.89) & 45.52 & 23.52 & yes & yes & 2 & 463 \\
\textbf{J0923+01} & 09:23:18.06 & 01:01:44.8 & 0.386 & 0.67 & 5.27 & 9.21(8.78) & 45.46 & 22.09 & yes & yes & 2 & 256 \\
\textbf{J0924+01} & 09:23:56.44 & 01:20:2.1 & 0.38 & 0.75 & 5.22 & 10.43(8.46) & 46.68 & 23.33 & yes & yes & 2 & -55 \\
\textbf{J0948+00} & 09:48:36.05 & 00:21:4.6 & 0.324 & 0.89 & 4.71 & 8.96(8.57) & 45.21 & 22.35 & no & no & 4 & -123 \\
\textbf{J2358-00} & 23:58:18.87 & -00:09:19.5 & 0.402 & 1.58 & 5.39 & 9.71(9.29) & 45.96 & \textless{}22.21 & yes & yes & 1,2 & 40 \\ \hline
\end{tabular}
\tablefoot{Columns 1, 2, 3, and 4 correspond to the QSO2 identifier (full SDSS name can be found in \citealt{bessiere2012importance}), RA, Dec, and spectroscopic redshift from SDSS. Columns 5 and 6 correspond to the seeing FWHM of the spectroscopic observations used here, measured from stars in the acquisition images, and to the spatial scale. 
Columns 7 and 8 list the extinction-corrected [OIII] luminosities from \citet{kong2018black}, with the uncorrected [OIII] luminosities from \citet{reyes2008space} within parenthesis, and the bolometric luminosities, derived from the extinction-corrected [OIII] luminosities using the factor of 454 from \citet{lamastra2009bolometric}.
Column 9 corresponds to the radio luminosity, calculated using the integrated flux at 1.4 GHz from either the FIRST or NVSS surveys, assuming a spectral index of $\alpha$ = -0.75 when unknown. When the object is undetected in both surveys, an upper limit is given. Columns 10 indicates whether the inclusion of a young stellar population (YSP) with age <100 Myr was necessary for modelling the stellar spectral features, following \citet{bessiere2017young}. Columns 11 and 12 indicate whether there is evidence of mergers/interactions in the optical images, and the different stages of these interactions, as reported in \citet{bessiere2012importance}: (1) the quasar host is part of a galaxy pair showing signatures of tidal interaction, (2) other signs of morphological disturbance, (3) system with multiple nuclei (within 10 kpc between them), and (4) isolated galaxy. Column 13 lists the value of the B$_{\rm gq}^{\rm av}$ parameter from \citet{ramos2013environments}, that measures the excess in the number of galaxies around the QSO2 as compared with the average number of background galaxies.}
\label{Table1}
\end{sidewaystable*}

Our QSO2 sample is based on the one studied by \citet{bessiere2012importance}, which is a subset of the narrow-line AGN catalogue of \citet{zakamska2003candidate}, selected from SDSS \citep{york2000aj} using emission line ratio diagnostic diagrams. We specifically selected this QSO2 sample because several host galaxy properties, including the stellar populations, optical morphologies, and environments were already characterized   
\citep{bessiere2012importance,bessiere2017young,ramos2013environments}, allowing us to investigate potential correlations between them and the outflow properties that we report in this work.

\citet{bessiere2012importance} selected the 20 QSO2s with right ascension (RA) in the range 23<RA<10 h, declination ($\delta$) of $\delta$<+20$^{\rm \circ}$, redshifts in the range 0.3<z<0.41, and emission line luminosities of L$_{\rm [OIII]}$>10$^{8.5}$ L$_{\odot}$. The luminosity cut is equivalent to an absolute magnitude of M$_{\rm B}$ < -23 mag, which is the classical definition of quasar. Subsequent updates to the [OIII] luminosities were reported for these QSO2s in \citet{reyes2008space}, which are included in Table \ref{Table1}. These updated luminosities implied that six objects from the original selection fall marginally below the quasar L$_{\rm [OIII]}$ cut, but \citet{bessiere2012importance} chose to retain the original sample of 20 objects. In a more recent work, \citet{bessiere2017young} reported deep optical spectroscopic observations for 19 of the 20 QSO2s. SDSS J015911+143922 (J0159+1439) was excluded from their analysis because they were unable to obtain spectroscopic data of sufficient quality for modelling the stellar populations. This sample of 19 QSO2s is then 95\% complete and it constitutes our QSO2 sample. Table \ref{Table1} includes the AGN bolometric luminosities of the QSO2s, which range from log L$_{\rm BOL}$=44.9 to 46.7 erg~s$^{-1}$, with a median value of 45.5 erg~s$^{-1}$. These luminosities were calculated from the extinction-corrected [OIII] luminosities from \citet{kong2018black} and using the correction factor of 454 from \citet{lamastra2009bolometric}. 

\subsection{Spectroscopic data}\label{spectroscopic}

The optical spectroscopic data used in this work, described in \citet{bessiere2017young}, were mainly obtained using the Gemini Multi-Object Spectrograph (GMOS-S) mounted on the Gemini South telescope at Cerro Pachón, Chile. Long-slit spectra were taken for 16 objects during semesters 2010B and 2011B using a relatively wide slit of 1.5 arcsec, in both the B600 and the R400 gratings. The observations consisted of 4$\times$675 s exposures using the B600 grating
and 3$\times$400 s exposures using the R400 grating.
The average spectral resolutions measured from the sky lines were 7.2 and 11.4 \AA~for the blue and red wavelength ranges, respectively. However, since all the observations but one (J2358-00) have a seeing full width at half maximum (FWHM) smaller than the slit width, the actual spectral resolutions range from 4.0-8.2 \AA~for the different spectra.

Different spectroscopic observations were compiled for the other three QSO2s in the sample, either because no GMOS-S observations were obtained (J0217-00 and J0114+00) or because [OIII] was not covered by the GMOS-S spectrum (J0142+00, for which only the blue range was observed).
J0217-00 was observed with the Gran Telescopio Canarias (GTC), at the Roque de los Muchachos Observatory, La Palma, Spain. The Optical System for Imaging and low-Intermediate-Resolution Integrated Spectroscopy (OSIRIS) instrument was used in long-slit spectroscopy mode, using a slit of 1.23 arcsec width. The target was observed with the R2000B and R2500R grisms for integration times of 3$\times$1800 s and 3$\times$1200 s, which lead to spectral resolutions of 4 and 5 \AA, respectively. Finally, the spectra of J0114+00 and J0142+00 were obtained from SDSS, which uses a standard 3 arcsec fiber covering an observed wavelength range of 3800–9200 \AA, with a spectral resolution (R) varying from R$\sim$1500 at 3800 \AA~to R$\sim$2500 at 9000 \AA. This spectral resolution corresponds to a FWHM=2.78 \AA~at 7000 \AA.

Here we used the reduced and flux-calibrated data from \citet{bessiere2017young}, where further details can be found. Large extraction apertures, of 1.5-2 arcsec ($\approx$8 kpc) were used there with the aim of performing stellar population modelling of the QSO2s host galaxies. Here, since our goal is studying the nuclear gas kinematics, we extracted the spectra using a diameter determined by the size of the seeing of each observation. The seeing was measured from the FWHM of the stars detected in the corresponding acquisition images of the QSO2s (0.53-1.13 arcsec $\approx$ 2.7-6.1 kpc; see Table \ref{Table1}), with the exception of J2358-00, for which we used an aperture of 8.5 kpc given that the seeing of 1.58 arcsec is larger than the slit width. For extracting these nuclear 1D spectra from the  reduced and calibrated 2D spectra, we used the task \textit{apall} within the \textsc{IRAF} package \textit{twodspec}. First, we summed the flux contribution for the wavelengths centered around the ionized [OIII]$\lambda$5007 emission line. Then, by plotting this flux against the spatial axis we located the maximum of the emission, and extracted a seeing-sized aperture centered at this position. In the case of the two QSO2s with SDSS data, the spectra correspond to physical sizes 15.8 kpc, set by the size of the fiber. 
We chose the [OIII] emission line to perform the kinematic analysis of the ionized gas because it is intense and AGN-dominated in the case of QSO2s, and it is not blended with other emission lines. In the case of the targets for which the [OIII] line was detected in both the red and the blue GMOS-S spectra, we used the blue because of its higher spectral resolution.

\subsection{Galaxy properties}\label{galaxy}

The selected sample has been previously studied by \citet{bessiere2012importance,bessiere2017young} and \citet{ramos2013environments}. From these works we can get information about multiple properties of the QSO2s and their host galaxies, including the presence of YSPs, optical morphologies, environments, and radio emission (see Table \ref{Table1}). One of our goals is to search for correlations between these AGN/galaxy properties and those of the ionized outflows that we characterize here, for the first time in these QSO2s. 

\citet{bessiere2017young} presented a detailed study of the stellar populations of the host galaxies of the 19 QSO2s in our sample plus another two additional QSO2s. That study is based on the modeling of the spectra described above and extracted in a large aperture of $\sim$8 kpc, centred on the peak continuum flux. For fitting the spectra, they used a combination of up to two stellar population models that are representative of viable star formation histories, and a power-law of varying spectral index, to account for the contribution from scattered AGN light. Based on this analysis, \citet{bessiere2017young} concluded that 71\% of the 21 QSO2 host galaxies require the inclusion of a YSP with an age<100 Myr for correctly modelling the stellar features detected in the optical range. From  the sample of 19 QSO2s studied here, 14 host galaxies (74\%) need the inclusion of this YSP (see Table \ref{Table1}). In the following, we just focus on the detection/non-detection of these YSPs in the QSO2s because their ages are comparable with the current phase of quasar activity \citep{2004cbhg.symp..169M,2014ApJ...782....9H}.

A full analysis of the optical morphologies of our QSO2 sample is presented in \citet{bessiere2012importance}. They visually inspected deep optical broad-band images, also observed with the GMOS-S instrument on the Gemini South telescope. They claimed that 15 of the 19 QSO2s (79\% of the sample) show signs of galaxy interactions in the form of tails, shells, fans, irregular features, amorphous haloes, and/or double nuclei. Based on the presence or absence of these structures, \citet{bessiere2012importance} classified the host galaxies in four groups that correspond to different stages of the interaction between two galaxies. These groups are the following. 

\noindent (1) The QSO2 host is part of a galaxy pair in tidal interaction.

\noindent (2) The galaxy presents other signs of morphological disturbance such as fans, shells, and/or tails.

\noindent (3) The system has multiple nuclei separated by $\leq$10 kpc.

\noindent (4) Isolated galaxy with no signs of morphological disturbance.

\citet{bessiere2012importance} also reported the 5 GHz radio luminosities of the QSO2s, which are listed in Table \ref{Table1}. They were calculated using the integrated flux at 1.4 GHz, obtained either from the FIRST or NVSS surveys, assuming a spectral index of $\alpha$=-0.75 when unknown. These 5 GHz luminosities range from 22.1 to 24.7 W~Hz$^{-1}$, with a median value of 22.35 W~Hz$^{-1}$, excluding the six QSO2s with upper limits (<22.6 W~Hz$^{-1}$). 

Lastly, \citet{ramos2013environments} characterized and quantified the environment of the QSO2s by means of the angular clustering amplitude, B$_{gq}$, and using the deep GMOS-S optical imaging data from \citet{bessiere2012importance}. The spatial clustering amplitude measures the excess in the number of galaxies around the target as compared with the expected number of background galaxies, assuming that the galaxy clustering is spherically symmetric around each target. Both the neighbours and the background galaxies were counted as the total number of galaxies surrounding the QSO2 (or its corresponding position in the offset fields) within a projected distance of a 170 kpc radius and having magnitudes between m-1 and m+2, with m being the magnitud of a generic galaxy at the redshift of the target. More information about the procedure can be found in \citet{ramos2013environments}. In Table \ref{Table1} we show the values of B$_{gq}^{av}$, calculated using all the offset fields (i.e., not including the QSO2) to determine the average number of background galaxies. Most of the QSO2s have low values of B$_{gq}^{av}$, corresponding to low-density environments. The average value reported by \citet{ramos2013environments} for the QSO2s is 151$\pm$76, and for comparison, values of B$_{gq}^{av}\ge$400 are typical of cluster environments. Indeed, \citet{ramos2013environments} compared the  values of B$_{gq}^{av}$ obtained for the QSO2s and for a control sample of quiescent early-type galaxies and did not find significant difference between them. 


\section{Kinematic analysis}\label{analysis}

\begin{figure*}
\centering
    {\par\includegraphics[width=1.02\columnwidth]{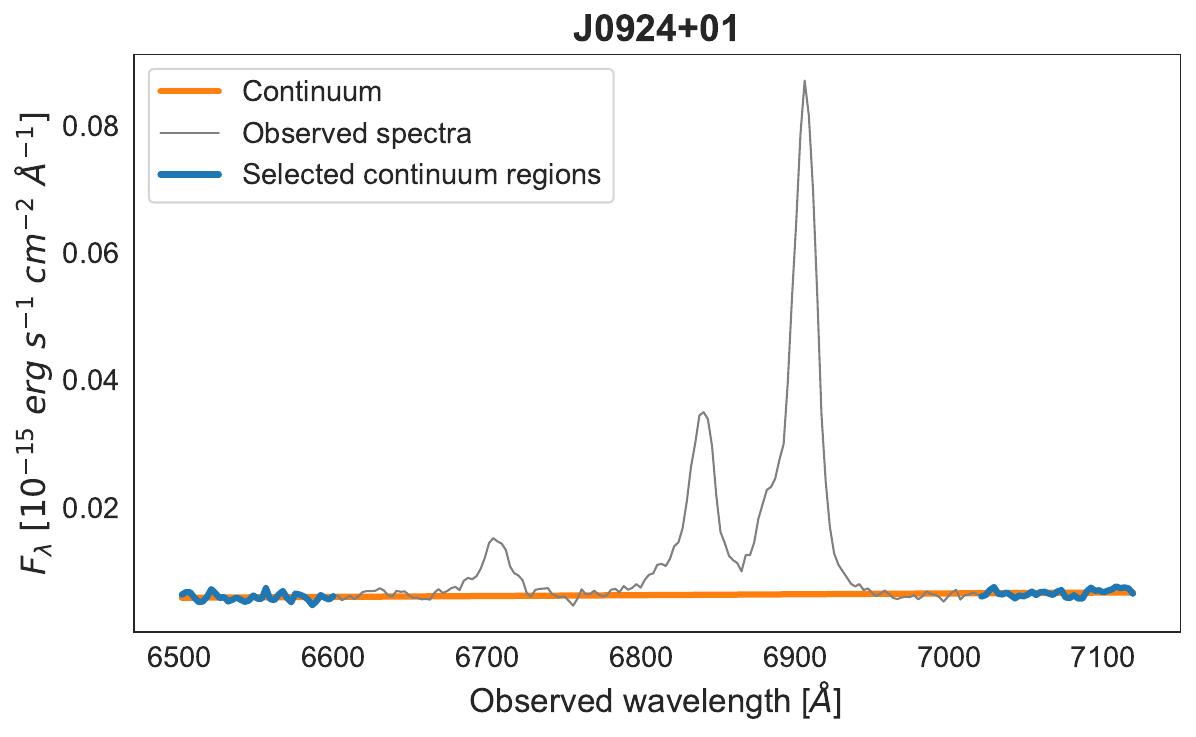}
    \includegraphics[width=0.95\columnwidth]{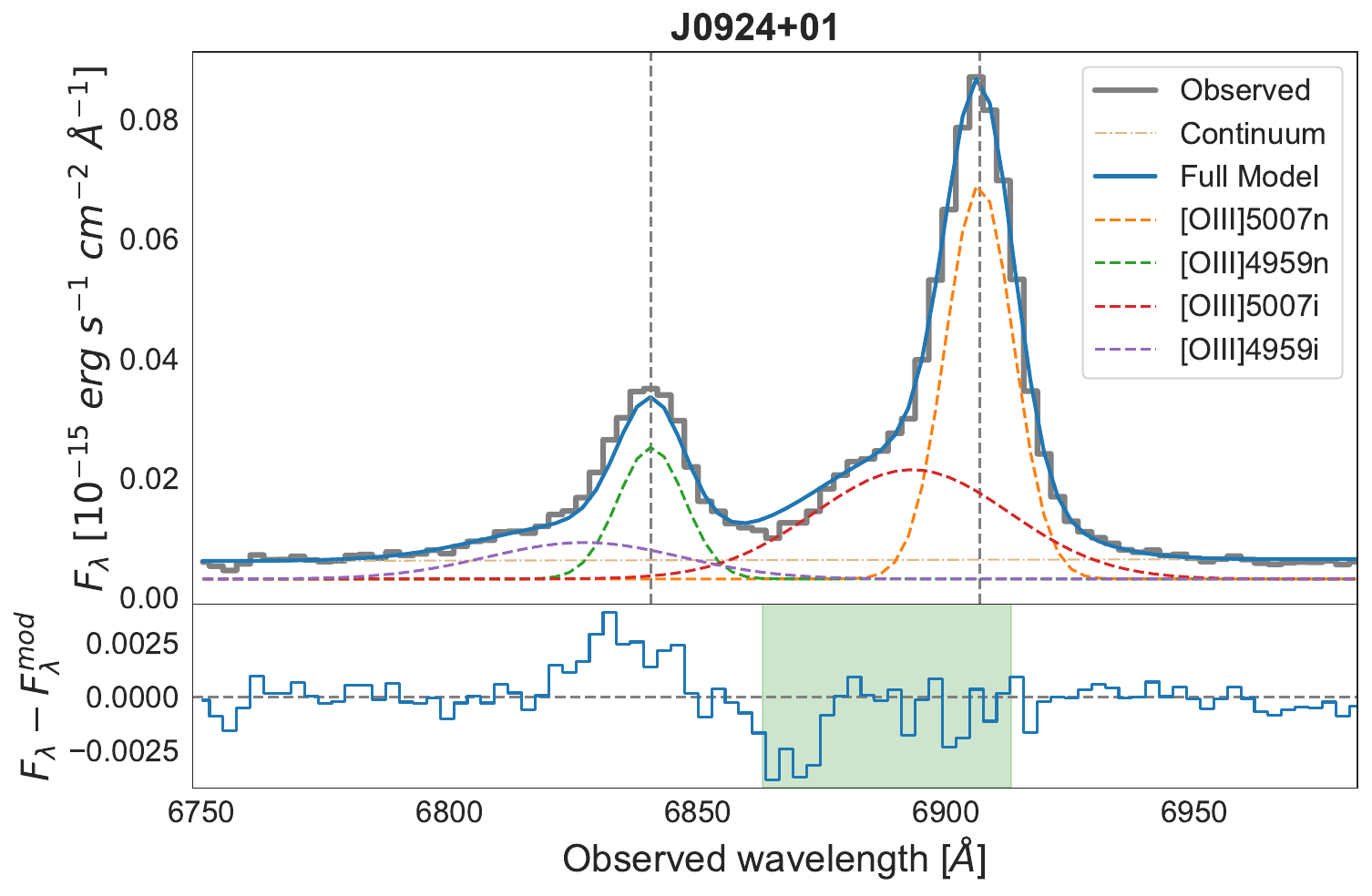}\par}
\caption{Parametric fit of the [OIII] doublet for the QSO2 J0924+01. The left panel shows the line-free regions, in blue, selected for fitting the underlying continuum, in orange. The right panel shows the two kinematic components fitted in the case of this QSO2, with the residuals included in the inset at the bottom.
The green shaded region in the residuals corresponds to the region affected by atmospheric absorption.}
\label{Fig1}
\end{figure*}

We analyzed the nuclear [OIII]$\lambda\lambda$4959,5007 \AA~profiles in order to investigate the kinematic properties of our sample, specially the possible presence of outflows. As explained in Section \ref{spectroscopic}, we extracted a nuclear spectrum of each source by choosing an aperture comparable to the seeing of each observation (see Table \ref{Table1}). These apertures correspond to physical sizes of $\sim$3-6 kpc, except for J2358-00, which is 8.5 kpc. For the two QSO2s for which only SDSS spectra are available, J0114+00 and J0142+00, the fiber size corresponds to $\sim$15.8 kpc.

For characterising the ionized gas kinematics, the main methods applied in the literature are based either on a parametric approach or a non-parametric one. At every spatial location different gaseous components with different kinematics can contribute to the flux and modify the shape of the line profiles: gas in the narrow-line region (NLR), outflowing gas, contributions from companion objects, etc. Parametric methods are based on the assumption that the different kinematic components can be described as a Gaussian distribution. They have the advantage of making it possible to characterize the properties of the different kinematic components found, including density and reddening \citep{holt2011impact,villar2014triggering,arribas2014ionized}. The challenge is that in some objects with disrupted kinematics, it is difficult to ascribe a physical meaning to all the fitted kinematic components \citep{bessiere2022spatially}. 
On the other hand, the non-parametric analysis consists of measuring emission line velocities at various fixed fractions of the peak intensity \citep{speranza2021murales} or the line cumulative flux function \citep{whittle1985narrow,harrison2014kiloparsec}. This method is most appropriate for characterizing gas properties in galaxies with multiple kinematic components, as it permits to easily identify high velocity gas in tail ends of emission line wings. However, non-parametric methods do not allow us to characterize the relative contribution of the different gaseous components at play.

Considering the previous, here we adopt both parametric and non-parametric methods to characterize the ionized gas kinematics of the QSO2s. By doing this we can investigate the dependence of the outflow properties, when present, on the method employed to measure them. First, we modelled the [OIII]$\lambda\lambda$4959,5007 \AA~emission line profiles using a parametric method (i.e., fitting Gaussians to the line profiles). Second, taking advantage of the parametric fit of the [OIII]$\lambda$5007 \AA, we used both flux-weighted and peak-weighted non-parametric methods to characterize the line profile. In Section \ref{outflowproperties} we compare the results obtained with the different methods that we describe below.


\subsection{Parametric method}\label{parametric}

We fitted the [OIII]$\lambda$$\lambda$4959,5007 profiles with multiple Gaussian components, using a \textsc{Python} program that we developed, based on the \textit{Specutils} and the \textit{Astropy} packages. 

Initially, before fitting the emission lines, the underlying continuum is subtracted by linearly interpolating the spectrum between two line-free regions close to the [OIII] doublet, redwards and bluewards (see the left panel of Figure \ref{Fig1}). To select the number of Gaussians fitted to each of the QSO2s, we added new components until the increment of the reduced $\chi^2$ ($\Delta\chi^2$) of the residuals is lower than 10\%, following \citet{bessiere2022spatially}.
We also imposed that any component must be broader than the instrumental width (see Section \ref{galaxy}), and its corresponding parameters larger than their associated uncertainties. An example of fit including two Gaussians is shown in the right panel of Figure \ref{Fig1}. 

Each kinematic component fitted to the [OIII] doublet corresponds to a set of two Gaussians simultaneously fitted (i.e., sharing the same width and velocity shift relative to systemic), with fixed wavelength separation (taking into consideration the cosmological spread of the wavelength shift between the two lines) and an amplitude ratio of [OIII]$\lambda$4959/[OIII]$\lambda$5007 = 1/3 \citep{dimitrijevic2007flux}. It is noteworthy that the GMOS-S spectra reduced and analyzed by \citet{bessiere2017young} have residual atmospheric absorption around the spectral region 6863-6913\AA~. When this residual absorption lies on top or close to one of the emission lines, as it is the case of the QSO2s J0114+00, J0142+14, J0217-01, J0218-00, J0320+00, and J0923+01, we first fit just the unaffected [OIII] line, and then use the corresponding parameters to force the fit of the affected line. In the case of J0924+01, where the atmospheric absorption lies between the two emission lines, the affected wavelength range is masked before fitting the emission line profiles (see right panel of Figure \ref{Fig1}).

All the fits and corresponding parameters derived from them (number of Gaussians, flux, FWHM, velocity shift from systemic, luminosity, and percentage luminosity of each component relative to the total luminosity of the line) are shown in Figure \ref{Fig_parametric1} and Table \ref{Tab_parametric} in Appendix \ref{Appendix par}. We note that the values of the FWHM have not been corrected from instrumental width, with the aim of keeping them comparable with W80 (see Section \ref{non-parametric}). Nevertheless, the individual instrumental FWHMs are  listed in Table \ref{Tab_parametric}. The uncertainties of the parameters were computed through a Monte Carlo simulation, creating 1000 mock spectra by varying the flux at each wavelength adding random values from a normal distribution with an amplitude given by the noise of the continuum. The errors were then computed as the 1$\sigma$ of each parameter distribution generated from the mock spectra.

Following the methodology described above, we fitted n Gaussian components to the [OIII] profiles of the QSO2s and classified them as narrow, intermediate, and broad depending on the FWHM.
\begin{itemize}
\item Narrow (n): FWHM $<$ 800 km s$^{-1}$.
\item Intermediate (i): 800 km s$^{-1}$ $\le$ FWHM $\le$ 2000 km s$^{-1}$.
\item Broad (b): FWHM $>$2000 km s$^{-1}$. 
\end{itemize}

We used these values trying to ascribe a physical meaning to the different kinematic components. Here we assume that the narrow components are associated with gas in the NLR, whose emission lines typically present FWHMs$\sim$400-600 km~s$^{-1}$ \citep{netzer1990agn,2007ASPC..373..511G}. The intermediate components are broader than the typical NLR FWHMs, but narrower than the broad components from the broad-line region (BLR), which have FWHM$>$2000 km s$^{-1}$. We note, however, that the broad components that we are measuring here for the [OIII] emission lines cannot be associated with the BLR because they are forbidden lines. Thus, the intermediate and broad components would be most likely associated with turbulent/outflowing ionized gas. With this is mind, the velocity shifts reported in Table \ref{Tab_parametric} were computed relative to the narrow component, or to the centroid of the multiple narrow components when present. We consider these velocity shifts as the outflow velocities, relative to the average kinematics of the NLR. Since these are conservative estimates, we also calculate the outflow maximum velocities as v$_{\max}$=v$_s$+2$\sigma$ \citep{rupke2013multiphase}, where $\sigma$ is FWHM/2.355.

In addition to the (n), (i), and (b) components, the fits of the QSO2s J0217-01 and J0320+00 also include a narrow and redshifted (r) component (see Table \ref{Tab_parametric} in Appendix \ref{Appendix par}) that, although do not meet all the criteria described at the beginning of this Section, were necessary to successfully reproduce the [OIII] profiles. However, these red components are not included in the outflow analysis.

We found intermediate and broad components in 18 of the 19 QSO2s, which implies an outflow incidence rate of $\sim$95\%. The only QSO2 without intermediate or broad components is J0218-00. The outflow components that we measured for the other 18 QSO2s present an average FWHM of 1800 km s$^{-1}$, with a standard deviation of 1100 km s$^{-1}$. They are mainly blueshifted, with an average value of v$_s$ = -370 km s$^{-1}$ and a standard deviation of 400 km s$^{-1}$ (see Table \ref{tab:summary}), and a maximum velocity of v$_{\rm max}$ = -1800$\pm$1400 km s$^{-1}$. 

\begin{table*}[h!]
\caption{Average and median outflow properties from the parametric and non-parametric methods.}
\centering
\setlength{\tabcolsep}{2pt}
\begin{tabular}{lcccccccccc}
\hline
\textbf{Method}  &  \multicolumn{2}{c}{\textbf{FWHM}} & \multicolumn{2}{c}{\textbf{v$_{\rm OF}$}} & \multicolumn{2}{c}{\textbf{log(M$_{\rm OF}$)}} & \multicolumn{2}{c}{\textbf{$\dot{M}_{\rm OF}$}}  & \multicolumn{2}{c}{\textbf{log($\dot{E}_{\rm kin}$)}}\\
&  \multicolumn{2}{c}{(km~s$^{-1}$)} & \multicolumn{2}{c}{(km~s$^{-1}$)} & \multicolumn{2}{c}{(M$_{\odot}$)} & \multicolumn{2}{c}{(M$_{\odot}$~yr$^{-1}$)}  & \multicolumn{2}{c}{(erg~s$^{-1}$)}\\
& Mean & Median & Mean & Median & Mean & Median & Mean & Median & Mean & Median \\
\hline
\textbf{Parametric (v$_s$)} & 1800$\pm$1100 & 1500 & -370$\pm$400 &-150 & 6.47$\pm$0.50 & 6.46 & 2.8$\pm$2.6 & 1.8 & 40.5$\pm$1.2 & 40.3\\
\textbf{Parametric (v$_{\rm max}$)} & \dots & \dots & -1800$\pm$1400 & -1500 & 6.47$\pm$0.50 & 6.46 & 23$\pm$35 & 11 & 42.9$\pm$0.6 & 42.9\\
\textbf{Flux-weighted} & 940$\pm$280 & 890 & -1120$\pm$510/680$\pm$200 & -1040/690 & 6.03$\pm$0.30 & 5.93 & 4.0$\pm$4.4 & 2.5 & 41.9$\pm$0.6 & 41.9\\
\textbf{Peak-weighted} & - & - & -840$\pm$260 & -780 & 5.75$\pm$0.32 & 5.75 & 1.9$\pm$1.8 & 1.3 & 41.4$\pm$0.5 & 41.4\\
\hline
\end{tabular}
\tablefoot{In the case of the two non-parametric methods, W$_{\rm 80}$ is listed instead of FWHM, and the average values of the outflow velocity, v$\rm _{OF}$, are given for the approaching and receding outflow sides for the flux-weighted method. In the case of the peak-weighted non-parametric method, the value of v$\rm _{OF}$ corresponds to the average value of the 18 QSO2s  with an asymmetric blue wing detected. For J0227+01, which has an assymetric red wing instead, we measure v$\rm _{OF}$=676$\pm$21 km~s$^{-1}$. The corresponding outflow physical quantities include J0227+01.}
\label{tab:summary} 
\end{table*}

We find that 14 of the QSO2s (74\% of the sample) require the inclusion of more than two Gaussian components (2 QSO2s with four and 12 with three Gaussians) to correctly model the emission line profiles. The remaining 5 QSO2s (26\% of the sample) are well fitted with just two Gaussian components. Such diversity of kinematic components detected in the [OIII] lines is common in QSO2s \citep{villar2011ionized,villar2016ionized,harrison2014kiloparsec,mcelroy2015ifu}. However, the heterogeneous results from our parametric analysis make it difficult to characterize the outflow properties of the sample. For this reason, and taking advantage of the parametric analysis here described, we performed a non-parametric analysis of the emission line profiles. This analysis, described in Section \ref{non-parametric}, provides a more homogeneous set of results, therefore allowing us to easily evaluate possible correlations between the outflow and host galaxy properties (see Section \ref{correlations}).


\subsection{Non-parametric analysis} \label{non-parametric}

Here we describe two different non-parametric methods to characterize the [OIII]$\lambda$5007 \AA~emission line: flux-weighted and peak-weighted. The two of them require of a noiseless isolated emission line profile, and hence we use the models resulting from the parametric method described in Section \ref{parametric} (i.e., the sum of the Gaussian components used to fit the [OIII]$\lambda$5007 \AA~emission line of each QSO2; see the right panel of Figure \ref{Fig1} for an example).

\subsubsection{Flux-weighted non-parametric analysis}
\label{flux-non-parametric}

\begin{figure}
\centering
\includegraphics[width=\linewidth]{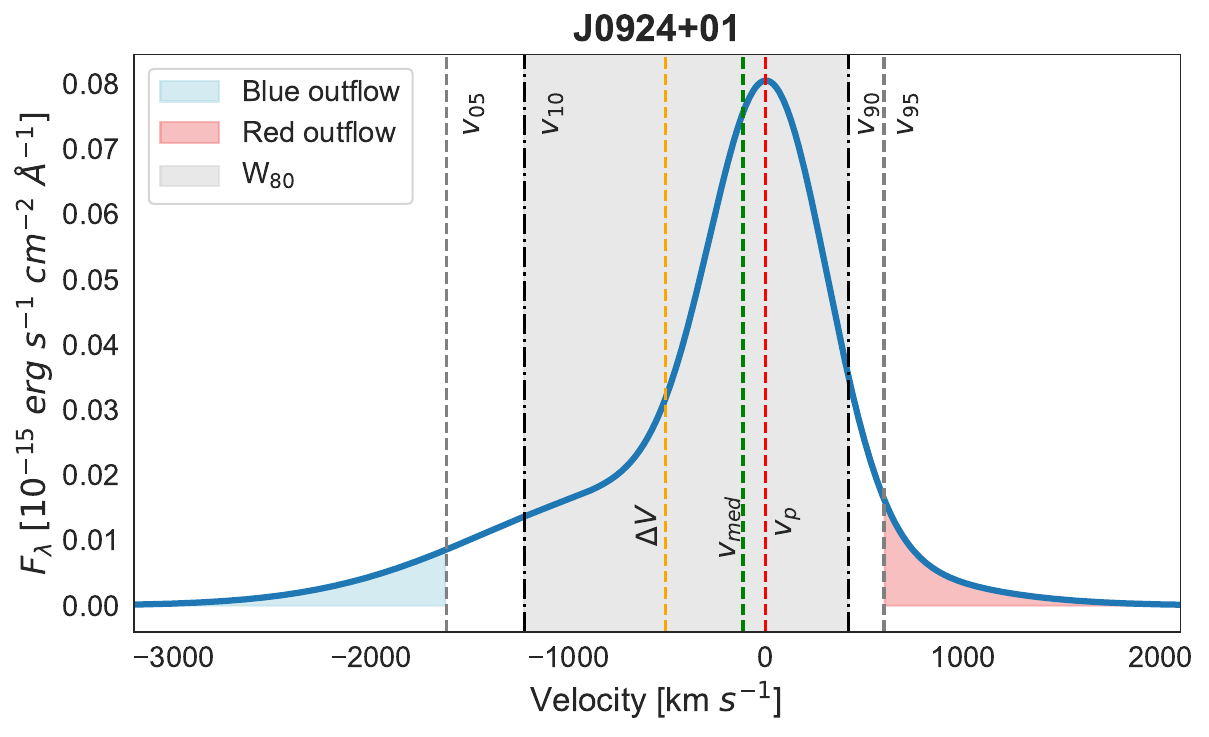}
\caption{Example of the flux-weighted non-parametric method employed in this work. The blue solid line is the parametric fit to the [OIII]$\lambda$5007 \AA~emission line shown in Figure \ref{Fig1} for J0924+01, with the continuum subtracted.  
The non-parametric velocity definitions included in this figure are the median velocity (v$_{\rm med}$), the velocities at the 5, 10, 90, and 95\% points of the normalized cumulative function of the emission line flux (v$_{05}$, v$_{10}$, v$_{90}$, and v$_{95}$), and the width of the line containing 80\% of the total flux (W$_{80}$=v$_{90}$-v$_{10}$). The grey region corresponds to the area containing 80\% of the total flux (i.e., between v$_{10}$ and v$_{90}$), and the blue and red regions correspond to high-velocity gas, which we consider outflowing.}
\label{Fig2}
\end{figure}

We first use a flux-weighted non-parametric approach \citep{heckman1981emission,harrison2014kiloparsec,zakamska2014quasar} to describe the kinematic properties of the modelled [OIII]$\lambda$5007 \AA~emission line.
The velocities v$_{05}$, v$_{10}$, v$_{90}$, and v$_{95}$ are defined as the velocities at the 5th, 10th, 90th, and 95th percentiles of the normalised cumulative function of the emission line flux. Other quantities based on this analysis that we use hereafter are the following. 

\begin{itemize}
    \item The peak velocity, v$_{\rm p}$. It corresponds to the peak flux of the emission line, and it is representative of the narrow component(s) fitted in Section \ref{parametric} (i.e., of NLR gas).

    \item The median velocity, v$_{\rm med}$, corresponding to the 50th percentile of the velocity, also representative of NLR gas.

    \item The line width, W$_{80}$, defined as the width of the line containing 80\% the total emission line flux, W$_{80}$=v$_{90}$-v$_{10}$. 

    \item The velocity offset, $\Delta$v=($v_{05}$ + $v_{95}$)/2. It quantifies the velocity offset of any blueshifted/redshifted component relative to systemic (i.e., the peak velocity, v$_p$). 

    \item The asymmetry, a=$\mid v_{90}-v_{med}\mid-\mid v_{10}-v_{med}\mid$. Negative (positive) asymmetry values indicate the presence of a blue (red) tail in the line profiles. 
    
\end{itemize}

 In the following, we consider that only the gas having velocities larger than v$_{05}$ and v$_{95}$ is outflowing, and define the corresponding outflow velocities (v$_{\rm OF}$) as the median velocity (v$_{50}$) of the blueshifted and redshifted wings (blue and red areas in Figure \ref{Fig2}). In the example shown in Figure \ref{Fig2}, the median velocity of the blue wing is -1926 $\pm$ 37 km~s$^{-1}$, and 789 $\pm$ 21 km~s$^{-1}$ for the red wing. For the whole sample, we measure v$_{\rm OF}$=-1120$\pm$510 km~s$^{-1}$ for the blue wing, and 680$\pm$200 km~s$^{-1}$ for the red.
All the parameters and figures derived using this method can be seen in Appendix \ref{Appendix nonpar}. As in the case of the parametric method, the uncertainties are computed as the 1$\sigma$ of each parameter distribution generated with a Monte Carlo simulation of 1000 mock spectra. 

We find a wide range of emission line widths, with an average W$_{80}$ = 940 $\pm$ 280 km~s$^{-1}$. 
12 of the 19 QSO2s present broad [OIII] line profiles, with W$_{80}> $800 km s$^{-1}$. Regardless of this, here we consider that all the QSO2s have outflows of ionized gas, since by definition, every emission line profile will include velocities larger than v$_{05}$ and v$_{95}$.
The average values of the asymmetry, v$_{05}$, and v$_{95}$ are -120$\pm$130 km~s$^{-1}$, -820 $\pm$ 320 km s$^{-1}$, and 530 $\pm$ 140 km s$^{-1}$. All the QSO2s show negative asymmetry values except J0227+01. These negative values are related to the presence of blue tails in the profiles (see Figure \ref{Fig_non_parametric1} and Table \ref{Tab_non_parametric} in Appendix \ref{Appendix nonpar}). Detecting blueshifted wings is more common than redshifted wings because usually we see the bulk of the outflowing gas that is coming toward us, whilst the receding side is partly hidden behind the host galaxy. In cases where the outflows subtend a small angle relative to the galaxy discs, it is possible to detect both the blueshifted and redshifted outflow components, or depending on the galaxy inclination, a dominant redshifted component. This might be the case of J0227+01.
Lastly, it should be mentioned that the quasar J0218-00, that did not need the inclusion of an intermediate or broad component in the parametric model, presents low values of the asymmetry and the velocity offset.

\subsubsection{Peak-weighted non-parametric analysis}\label{peak-non-parametric}

\begin{figure}
\centering
    \includegraphics[width=\linewidth]{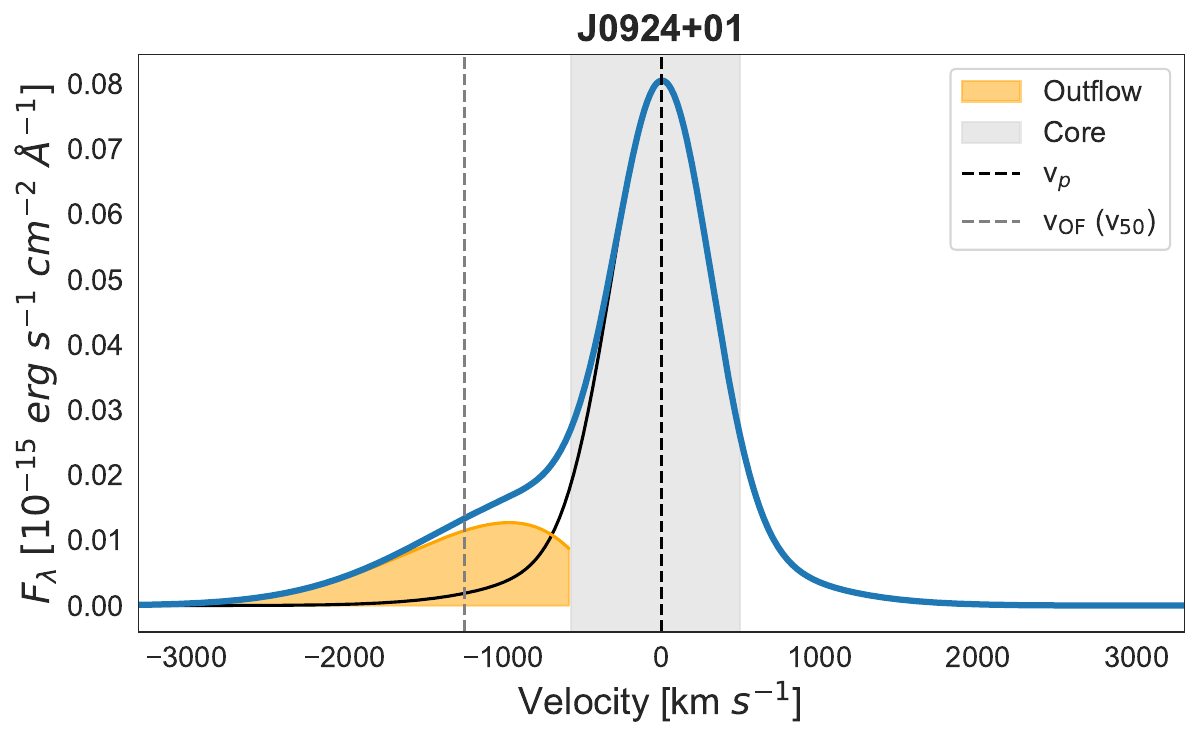}
\caption{Same as in Figure \ref{Fig2}, but using the peak-weighted non-parametric method from \citet{speranza2021murales}. The black solid line in the blue side of the emission line corresponds to the mirror image of the red side. The grey area is the core of the line, defined as the region between the peak and 1/3 of the peak flux. The orange area is the result of subtracting the black from the blue line outside the core region, which is what we consider outflowing gas using this method. The black and grey vertical dashed lines are the peak and outflow velocities (v$_p$ and v$_{\rm OF}$).}
\label{Fig3}
\end{figure}

As we mentioned in Section \ref{flux-non-parametric}, the flux-weighted non-parametric method likely overestimates the outflow mass because by definition, a certain proportion of the gas is outflowing, both blueshifted and redshifted. In addition, \citet{speranza2021murales} argued that $v_{05}$ might still be representative of rotating gas, which generally constitutes most of the emission line flux, and hence may not be fully representative of outflowing gas. Consequently, \citet{speranza2021murales} proposed a different non-parametric analysis, based on the detection of asymmetries in the emission lines (see Figure \ref{Fig3} for an example), to measure the outflow properties.

For this method we also use the model of the [OIII]$\lambda$5007 \AA~emission line profile from the parametric method. The core of the emission line is defined as the region between the peak of the line and one-third of the peak (i.e., the region that corresponds to $\sim$90\% of the total flux of the line considering a Gaussian profile).
By subtracting a mirror image of the less prominent wing from the most prominent one, the symmetric component of the emission line is removed, leaving just the asymmetric wing. Such residual is what we associate with outflowing gas (orange area in Figure \ref{Fig3}). For characterizing the outflow we use two parameters: the median velocity of the residual, v$_{\rm OF}$, (the 50th percentile of its total flux) and the flux of the residual wing F$_{\rm w}$. As in Section \ref{flux-non-parametric}, the uncertainties were estimated as 1$\sigma$ of each parameter distribution from 1000 mock spectra. All the plots and parameters derived from the peak-weighted non-parametric analysis can be seen in Figure \ref{Fig_non_parametric3} and Table \ref{Tab_speranza_energetics} of Appendix \ref{Appendix speranza}. 

Using this method we find that all the QSO2s in the sample show ionized outflows in the form of asymmetric wings. 18/19 QSO2s, including those with low values of W$_{80}$, like J0218-00, present blue wings. J0227+01 presents a red wing instead, in agreement with its positive asymmetry value. The average value of the outflow velocity computed using this method, considering the 18 QSO2s having blue wings, is v$_{\rm OF}$=-840$\pm$260 km~s$^{-1}$, while for J0227+01 we obtain a value of 676$\pm$21 km~s$^{-1}$ (see Tables \ref{tab:summary} and \ref{Tab_speranza_energetics}).


\section{Physical outflow properties}\label{outflowproperties}

In Section \ref{analysis} we described the three methods used here to characterize the ionized outflows of the QSO2s, and the corresponding results. The parameters derived from these fits are direct measurements (e.g., FWHM, W$_{80}$, a, v$_{\rm med}$), but we can use them to derive physical properties, such as the outflow mass, outflow mass rate, and kinetic power. These are key quantities for investigating the outflow impact on the surrounding environment. First, we calculated the mass of the ionized outflows using Eq. \ref{Eq1} \citep{osterbrock2006astrophysics,carniani2015ionised,fiore2017agn}.

\begin{equation}
    M_{\rm [OIII]} = 4 \times 10^7 M_{\odot} \left(\frac{C}{10^{\rm O/H}}\right)   \left(\frac{L_{\rm [OIII]}}{10^{44}}\right)   \left(\frac{10^{3}}{\langle n_e \rangle}\right)
\label{Eq1}
\end{equation}

\noindent where $n_e$ is the electronic gas density, C the condensation factor, L$_{\rm [OIII]}$ the outflowing gas luminosity and O/H = [O/H] - [O/H]$_{\odot}$ is the oxygen abundance ratio relative to solar, with [O/H]$_{\odot}$ $\sim$ 8.86 \citep{centeno2008new}. For the whole sample, we assumed that the gas clouds have the same electron density, leading to C = $<n_e^2>$ - $<n_e>^2$ = 1, and solar metallicity as in \citet{bessiere2017young}, hence O/H = 0.

For ionized outflows, an important source of uncertainty is the electron density \citep{harrison2018agn,2023MNRAS.524..886H}. The outflow mass is inversely proportional to the gas density, which can be estimated from the ratio of the [SII]$\lambda\lambda$6716,6731 doublet, from the [OIII]/H$\beta$ and [NII]/H$\alpha$ ratios, or using the fainter trans-auroral lines. The latter technique uses the flux ratios of the [SII]$\lambda\lambda$6716,6731 and [O II]$\lambda\lambda$3726,3729 doublets as well as of the trans-auroral [O II]$\lambda\lambda$7319,7331 and [SII]$\lambda\lambda$4068,4076 lines. The trans-auroral ratios have the advantage of being sensitive to higher density gas than the classical [SII] ratio \citep{holt2011impact,rose2018quantifying,ramos2019near}. Using [SII], \citet{harrison2014kiloparsec} reported densities in the range 200-1000 cm$^{-3}$ for a sample of 16 QSO2s at z$<$0.2, and \citet{singha2022close} measured outflow densities of $\sim$1900 for type-1 AGN at 0.01<z<0.06. Using the trans-auroral lines, \citet{rose2018quantifying} measured outflow densities in the range 2500 < n$_e$ (cm$^{-3}$) < 20000, for ULIRGS with 0.06 < z < 0.15. Finally, using optical line ratios, \citet{baron2019discovering} reported densities of $\sim$30000 cm$^{-3}$ for a sample of 234 nearby type-2 AGN. For this work, since our spectra do not cover either the [SII] doublet, the trans-auroral lines, or the [NII] and H$\alpha$ emission lines, we adopted a gas density of n$_e$=200 cm$^{-3}$ for all the QSO2s, as in \citet{fiore2017agn} and \citet{speranza2021murales}. Therefore, the derived outflow masses will most likely be upper limits. 

Since we are measuring only the [OIII] gas, we assume that the total ionized gas mass of the outflows is three times the [OIII] outflow mass (M$_{\rm OF}$ = 3$\times$M$_{\rm [OIII]}$; \citealt{fiore2017agn}). The outflow mass rate represents the instantaneous rate of outflowing gas at the edge of the wind. Assuming an spherical sector \citep{fiore2017agn,lutz2020molecular}, it can be defined as three times the total [OIII] mass divided by the time required to push this mass through a spherical surface of radius R$_{\rm OF}$.
\begin{equation}
    \dot{\rm M}_{\rm OF} = 3~{\rm v}_{\rm OF}~\frac{\rm M_{OF}}{\rm R_{OF}}
\label{Eq2}
\end{equation}

As there are no integral field observations that we can use to constrain the outflow radius (R$_{\rm OF}$), it might be possible to do it using averaged spatial slices of the broad line wings detected in the long-slit spectra, as in \citet{rose2018quantifying} and \citet{ramos2019near}. However, this procedure is not straightforward in the case of the [OIII] lines, because the blueshifted wing of [OIII]$\lambda$5007 is blended with [OIII]$\lambda$4959, and the blueshifted wing of the latter is much fainter. Moreover, since for the GMOS-S spectra used here we measured a median seeing of 0.83 arcsec ($\sim$3.8 kpc at the average redshift of the sample, z$\sim$0.366), the ionized outflows will most likely be unresolved, according to other studies of QSO2s at similar redshifts \citep{villar2011ionized,villar2016ionized,karouzos2016unraveling,2017MNRAS.470..964R,ramos2019near,speranza2021murales}. Early studies of quasar-driven outflows based on optical integral field observations of local QSO2s reported ionized outflows extending up to $\sim$10-15 kpc \citep{liu2013observations,harrison2014kiloparsec,mcelroy2015ifu}. However, later works claimed that these sizes were overestimated due to seeing smearing effects \citep{2016A&A...594A..44H,karouzos2016unraveling,harrison2018agn} or to selection biases \citep{villar2016ionized}. More recent studies report ionized outflows with sizes of $\sim$1-3.4 kpc for local QSO2s \citep{karouzos2016unraveling,2017MNRAS.470..964R,ramos2019near,speranza2022warm}. Hence, here we assume an outflow radius of 1 kpc for all the QSO2s in our sample. If the radii are larger than this value, our mass outflow rates will be upper limits, although more compact outflows have been reported for nearby ULIRGs and AGN \citep{tadhunter2018quantifying,tadhunter2021light,singha2022close,2023A&A...670A...3W}.

The outflow velocity (v$_{\rm OF}$) that we used to compute the outflow mass rate depends on the method: in the case of the parametric analysis we considered both v$_{\rm s}$ and v$_{\rm max}$ (see Table \ref{Tab_parametric} in Appendix \ref{Appendix par}). v$_{\rm s}$ is the velocity of the intermediate and/or broad components relative to the narrow component(s), and v$_{\rm max}$=v$_{\rm s}$+2$\sigma$ (see Section \ref{parametric}).
We compute the outflow mass rate associated with each of the intermediate and/or broad components and then added them to compute the total outflow rate (see Table \ref{Tab_parametric_energetics}). 

For the flux-weighted non-parametric analysis, we used as outflow velocities, v$_{\rm OF}$, the v$_{50}$ of the red and blue areas shown in Figures \ref{Fig2} and \ref{Fig_non_parametric1}. Then, we computed two separate outflow rates for each QSO2, red and blue, which we finally added to compute the total outflow mass rate (see Table \ref{Tab_nonparametric_energetics} in Appendix \ref{Appendix nonpar}). Finally, for the peak-weighted non-parametric analysis we used 50th velocity percentile of the residual wings (orange areas in Figures \ref{Fig3} and \ref{Fig_non_parametric3}) to calculate the outflow mass rates (see Table \ref{Tab_speranza_energetics} in Appendix \ref{Appendix speranza}).

Once we estimated the outflow mass rates using Eq. \ref{Eq2}, we can calculate the kinetic power as:
\begin{eqnarray}
    \dot{\rm E}_{\rm kin} = \frac{1}{2}~\dot{\rm M}_{\rm OF}~{\rm v}_{\rm OF}^2 \label{Eq3}
\end{eqnarray}

\begin{figure}[htb]
\centering
\includegraphics[width=0.95\linewidth]{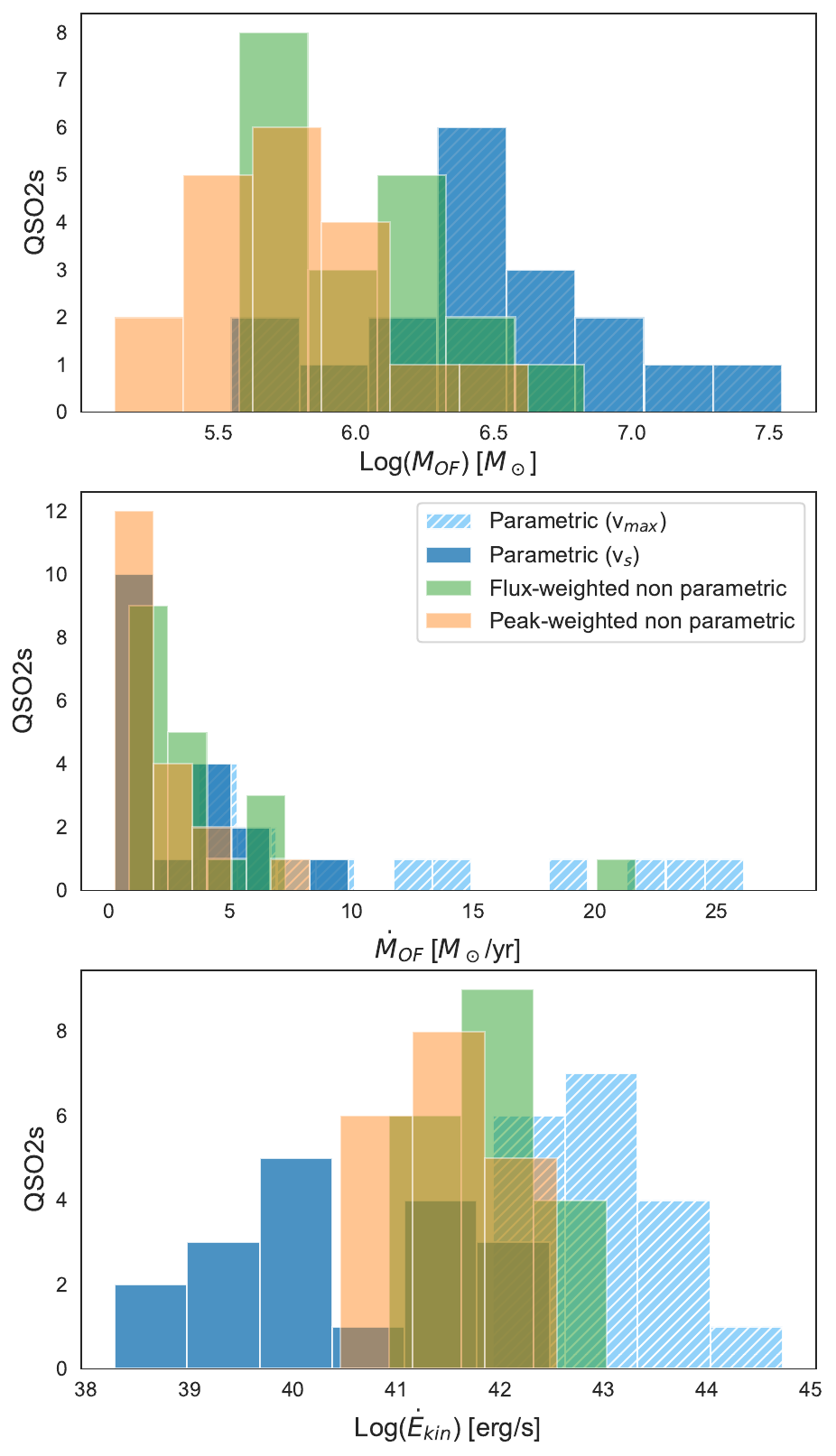}
\caption{Histograms of the outflow masses, mass rates, and kinetic powers computed through the three different methods: parametric (considering either v$_s$ or v$_{\rm max}$ as the outflow velocities), flux-weighted, and peak-weighted non-parametric. For easier comparison of the values, in the outflow mass rate histograms we omitted the largest values of 57 and 151 M$_{\odot}$~yr$^{-1}$, obtained for J0923+01 and J0142+14 using v$_{\rm max}$ in the parametric method.}
\label{Fig4}
\end{figure}

The outflow masses, outflow mass rates, and kinetic powers measured with the three different methods are shown in Tables \ref{Tab_parametric_energetics}, \ref{Tab_nonparametric_energetics}, \ref{Tab_speranza_energetics}, and Figure \ref{Fig4}. The uncertainties were computed using propagation of errors. 

The outflow masses computed through the parametric, flux-weighted, and peak-weighted non-parametric methods have average values of log(M$_{\rm OF}$)=6.47$\pm$0.50, 6.03$\pm$0.30, and 5.75$\pm$0.32 M$_{\odot}$, and the medians are 6.46, 5.93, and 5.75 M$_{\odot}$, respectively (see Table \ref{tab:summary}). From these results, and from the outflow mass histograms shown in Figure \ref{Fig4}, we conclude that using the parametric method we derive the largest outflow masses. This is  because this with this method we consider the integrated flux of the intermediate and broad Gaussian components as outflowing gas, whilst in the case of the non-parametric methods, we just use the flux included in the tails of the emission lines. 

The average outflow mass rates measured from the parametric, flux-weighted, and peak-weighted non-parametric methods are 2.8$\pm$2.6, 4.0$\pm$4.4, and 1.9$\pm$1.8 M$_{\odot}$~yr$^{-1}$, and the medians are 1.8, 2.5, and 1.3 M$_{\odot}$~yr$^{-1}$ (see Table \ref{tab:summary}).
The flux-weighted non-parametric method results in higher outflow mass rates, as a consequence of always including blueshifted and redshifted gas (see Section \ref{flux-non-parametric}) with large velocities. However, all the values measured from the three methods are between 0.2 and 9 M$_{\odot}$~yr$^{-1}$ except for J0142+14, which has an outflow rate of 20 M$_{\odot}$~yr$^{-1}$ measured from the flux-weighted non-parametric method. It is when we consider the parametric v$_{\rm max}$ values that we get the largest outflow mass rates (average value of 23$\pm$35 M$_{\odot}$~yr$^{-1}$, with a median value of 11 M$_{\odot}$~yr$^{-1}$; see middle panel of Fig. \ref{Fig4} and Table \ref{tab:summary}).

Focusing on the kinetic powers, we measure average values of log($\rm \dot{E}_{kin}$) = 40.5$\pm$1.2, 41.9$\pm$0.6, and 41.4$\pm$0.5 erg~s$^{-1}$ using the three methods. The median values are 40.3, 41.9, and 41.4 erg~s$^{-1}$ (see Table \ref{tab:summary}). In this case, the lowest values of the kinetic power are those measured with the parametric method. This is because of its dependency with v$^2_{\rm OF}$, which is higher in the case of the non-parametric methods (see Tables \ref{Tab_parametric}, \ref{Tab_nonparametric_energetics}, and \ref{Tab_speranza_energetics}). By using the velocity shift of the broad and/or intermediate components (v$_{\rm s}$) relative to the narrow component(s) as outflow velocity, we are deriving lower kinetic powers than when considering the mean velocities (v$_{50}$) of the high-velocity tails considered in the non-parametric methods. If instead of v$_s$ we use v$_{\rm max}$, as commonly done in the literature \citep{mcelroy2015ifu,fiore2017agn}, the kinetic powers are larger than the non-parametric ones (average of 42.9$\pm$0.6 erg~s$^{-1}$ and median of 42.9 erg~s$^{-1}$; see Table \ref{tab:summary}). Finally, the average values of the coupling efficiencies ($\rm \dot{E}_{kin}$/$\rm L_{BOL}$) derived from for the parametric, flux-weighted, and peak-weighted non-parametric methods are (0.014$\pm$0.032)\%, (0.080$\pm$0.16)\%, and (0.020$\pm$0.025)\%, with medians of 0.001\%, 0.02\%, and 0.01\%. These values are one order of magnitude lower than those reported by other studies of ionized outflows in AGN that have followed similar considerations \citep{fiore2017agn,baron2019discovering,davies2020ionized,speranza2021murales}. However, in the case of the parametric method using v$_{\rm max}$ (as in \citealt{fiore2017agn}), we obtain much larger values, with an average of (1.2$\pm$3.0)\%, and median of 0.22\%. The large value of the average and its dispersion come from three QSO2s with coupling efficiencies larger than one: J0142+14, J0332-00, and J0948+00 (see Table \ref{Tab_parametric_energetics}). In this case, the median value is more representative of the whole sample.

\begin{figure*}[htb]
\centering
\includegraphics[width=0.95\linewidth]{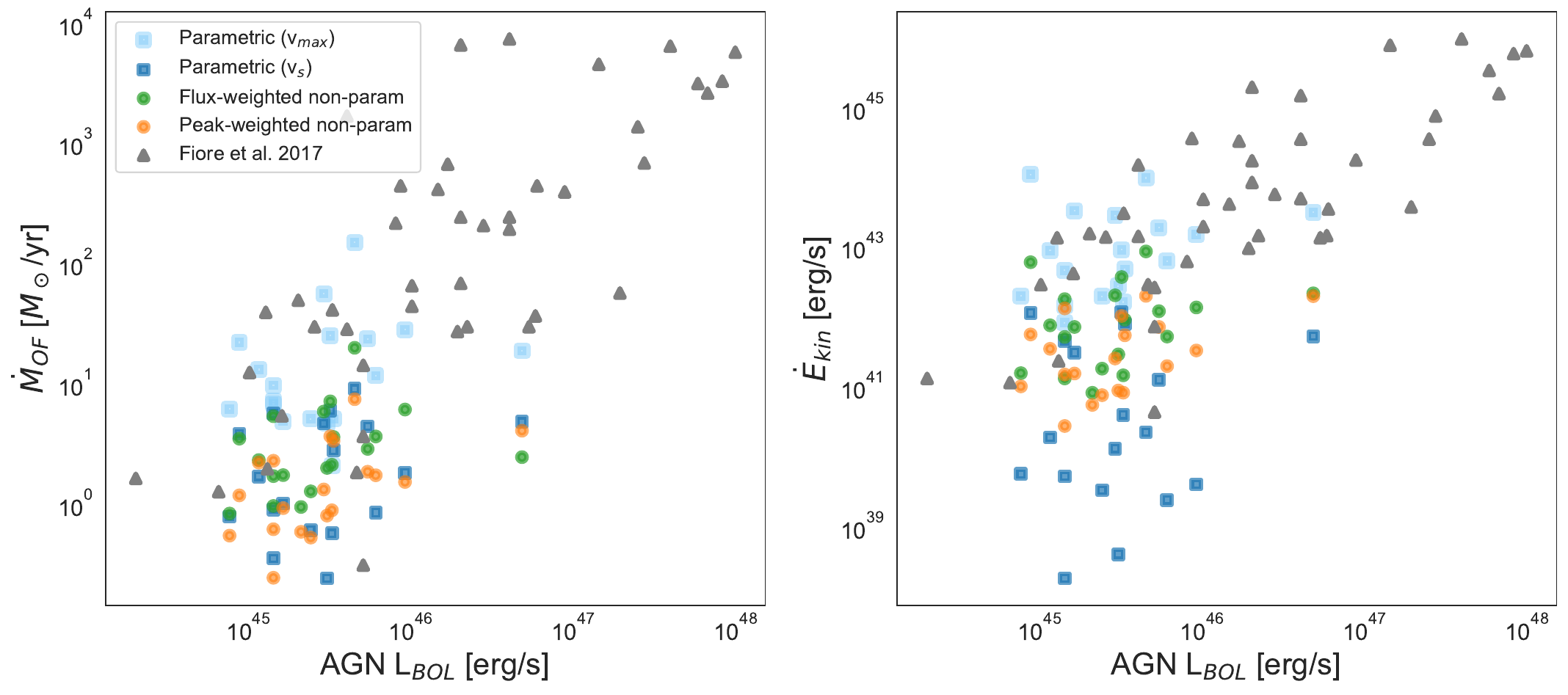}
\caption{Outflow mass rate and kinetic power as a function of the AGN bolometric luminosity. Light and dark blue squares correspond to the values derived from the parametric method using v$_s$ and v$_{\rm max}$, respectively. Green and orange circles are the values from the flux-weighted and peak-weighted non-parametric methods. The values from \citet{fiore2017agn} are shown as grey triangles for comparison.}
\label{Fig5}
\end{figure*}

From this comparison we conclude that the three adopted methods provide different physical outflow properties, as a result of the particular considerations and procedure of each one. However, these differences are consistent within the uncertainties considering that we are not accounting for by those associated with the electron density and outflow radius. 
In Figure \ref{Fig5} we plotted the outflow mass rates and kinetic powers derived from each method as a function of the bolometric luminosity.
The values from the compilation of \citet{fiore2017agn} are also shown for comparison. They also considered a fixed density of n$_e$=200 cm$^{-3}$, an outflow radius of R$_{\rm OF}$=1 kpc, and the same outflow geometry that we are assuming here. We find that the flux-weighted non-parametric results lie within the lower values of the outflow rates and kinetic powers measured by \citet{fiore2017agn} for AGN of similar luminosities, while the peak-weighted and the parametric results are smaller. However, the parametric results using v$_{\rm max}$ are the most similar to \citet{fiore2017agn}, as expected since they also used v$_{\rm max}$ to calculate the outflow physical properties.


\section{Correlations between outflow properties and galaxy properties}\label{correlations}

\begin{figure*}[htb]
\centering
\includegraphics[width=\linewidth]{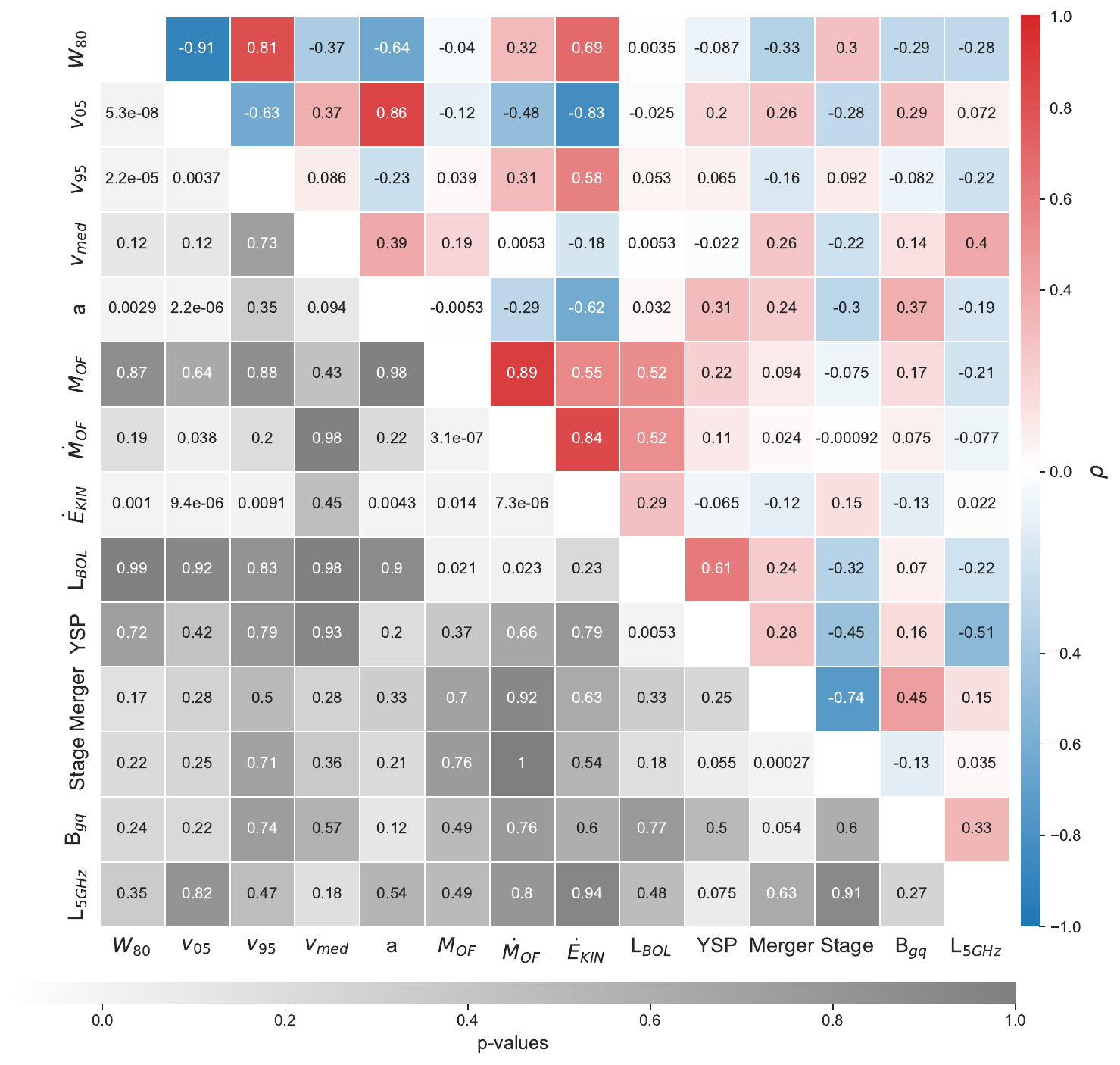}
\caption{Correlation matrix between the QSO2 outflow properties derived from the [OIII] emission line and using the flux-weighted non-parametric method, and their AGN and host galaxies properties.
The Spearman's rank correlation coefficients ($\rho$) are shown in the top right half of the matrix (blue and red colors) and the corresponding p-values in the bottom left half (grey colors).}
\label{Fig6}
\end{figure*}

As previously mentioned, apart from determining the outflow demographics and corresponding properties of our QSO2 sample, our goal is to evaluate possible correlations with different AGN and galaxy properties. These properties, available from \citet{bessiere2012importance,bessiere2017young} and \citet{ramos2013environments}, include: presence or not of stellar populations younger than 100 Myr (YSP), of mergers (including interaction stage), density of large-scale environment (B$_{\rm gq}$), and radio luminosity (L$_{\rm 5GHz}$). 

To do so we evaluated the matrix correlation, shown in Figure \ref{Fig6}, between the outflow properties derived from the flux-weighted non-parametric analysis (see Sections \ref{flux-non-parametric} and \ref{outflowproperties}) and the AGN and host galaxy properties (see Section \ref{galaxy}). From the outflow direct measurements we selected W$_{\rm 80}$, v$_{\rm 05}$, v$_{\rm 95}$, v$_{\rm med}$, and $a$; and from the physical outflow properties, M$_{\rm OF}$, $\dot{M}_{\rm OF}$, and $\dot{E}_{\rm kin}$.

In order to quantify the degree of correlation among all the properties, we computed the Spearman's rank correlation coefficient ($\rho$) for each variable pair, which is a non-parametric measurement of the strength and direction of the statistical dependence that exists between two variables. Unlike the Pearson correlation coefficient (r), the Spearman's rank correlation coeffient does not assume that the data sets are normally distributed and it assesses how well the relation between both can be described using a monotonic function. Hence, values of $\rho$ equal to -1 or 1 imply an exact monotonic relationship, with positive values indicating that as the X variable increases, Y also increases, while negative values imply that as X increases, Y decreases. In general, absolute values of $\rho$ in the range 0-0.39 are regarded as a weak or non-existent correlation, 0.40-0.59 as moderate, 0.6-0.79 as strong, and 0.8-1 as very strong correlation. The $\rho$ values found for our sample are shown in the top right half of Figure \ref{Fig6}, with darker colors indicating stronger correlations. 

We also performed a p-value test to verify the strength of the correlations, which can be seen in the lower left half of the matrix shown in Figure \ref{Fig6}. 
The p-value measures the probability of observing a correlation coefficient as extreme or more extreme than the one observed from two datasets of uncorrelated properties, i.e., it quantifies the probability of observing such correlation purely by chance. Seeking for a 90$\%$ confidence in our results, we are setting a significance level of $\alpha$ = 0.1 (100\%-confidence level). This would imply that if p-value$\leq$0.1, we can be confident within a 90$\%$ level about a genuine correlation between two properties. On the contrary, if p-value>0.1, the correlation could have arisen by chance, and hence we cannot conclude that it is significant.

As a sanity check, we first confirmed that correlations between certain outflow properties exist. For example, $W_{80}$ shows strong correlations with the outflow projected velocities, $v_{05}$ and $v_{95}$ ($\rho$=-0.91 and $\rho$=0.81, respectively). This is a consequence of broader emission line profiles having faster velocities associated with the wings.
Similarly, $v_{05}$ is also strongly correlated with the asymmetry, $a$, with $\rho$=0.86. This is not the case for $v_{95}$ 
($\rho$=-0.23), since most of the asymmetries in our sample are associated with blueshifted gas.
These outflow parameters also show strong correlations with some of the outflow physical properties, specially with the outflow mass rates and kinetics powers.

Regarding the correlations between the outflow and the AGN and host galaxy properties, we only find moderate correlations between the AGN bolometric luminosity and the 
outflow mass and outflow mass rate, both with $\rho$=0.52, respectively (p-values of 0.021 and 0.023). This implies that more luminous AGN have more massive ionized outflows, and also higher outflow mass rates \citep{cicone2014massive,hainline2014gemini,fiore2017agn,revalski2018quantifying}.

We did not find any significant correlations between the outflow properties and the host galaxy properties. Regarding the presence or not of YSPs, 4 out of the 5 quasars (80\%; J0114+00, J0234-07, J0332-00, and J0948+00) that do not require of the inclusion of a YSP < 100 Myr to reproduce their optical spectra show large values of the asymmetry: |a|>100. On the contrary, only 5 of the 14 QSO2s with such YSPs (36\%) have |a|>100. Since larger asymmetries are associated with more disrupted gas and faster outflows, this result, despite the small sample size, might be indicating that recent star formation is being supressed more efficiently in the QSO2s with the most disrupted kinematics. Nevertheless, as we noted above, there are also QSO2s with disrupted kinematics showing YSPs. 

Regarding the optical morphologies of the QSO2s, 3 out of the 4 galaxies that do not show any evidence for mergers/interactions (75\%; J0332-00, J0234-07, J0114+00, and J0948+00) show |a|>100. Considering the merging QSO2s (15/19), we do not find any trend between the outflow properties and the stage of the interaction either (i.e., pre-coalescence or post-coalescence; \citealt{2011MNRAS.410.1550R,bessiere2012importance}). This would imply that the presence of outflowing gas and/or disrupted kinematics on the spatial scales that our spectra are probing is related to the AGN and not to the mergers. Considering this, it is not surprising that we do not find any correlation between the outflow properties and the large scale environment either. In particular, we looked at the spatial clustering amplitudes (B$_{\rm gq}^{\rm av}$) from \citet{ramos2013environments}. The QSO2s with most disrupted kinematics do not show any preference for denser/sparser environments. 

Finally, we also looked at possible correlations between the outflow properties and the radio luminosity of the QSO2s. Indeed, \citet{mullaney2013narrow} and \citet{zakamska2014quasar} found a connection between the width of the [OIII] profiles and L$_{\rm 1.4GHz}$. The radio luminosity is a property that can be associated with star formation and/or nuclear activity (\citealt{2021MNRAS.503.1780J}; {\color {blue} Bessiere et al. in prep.}). For our QSO2s, we do not find any correlation between L$_{\rm 5GHz}$ and any of the outflow properties here considered. Here we are using the integrated luminosities at 5 GHz from \citet{bessiere2012importance}, measured from 5 arcsec resolution FIRST data, and we also checked that the results are the same when we use peak luminosities instead. The lack of correlation might be due to the limited range of radio luminosities that our sample is probing: 22.1 - 24.7 erg~s$^{-1}$, with an average value of 22.80$\pm$0.80 erg~s$^{-1}$ and median of 22.35 erg~s$^{-1}$ (see Table \ref{Table1} and Section \ref{discussion3}).


\section{Discussion}\label{discussion}

In this work we characterized the ionized outflow properties of a sample of 19 QSO2s at redshift 0.3<z<0.41 using three different analysis methods. We compared the results,
and looked for correlations between the outflow properties and various AGN and galaxy properties. Here we discuss these results and put them in context with others from the literature.

\subsection{Demographics and energetics of ionized outflows}
\label{discussion1}

We found signatures of ionized outflows in 18 of the 19 QSO2s using the parametric method (Section \ref{parametric}), based on the presence of at least one Gaussian component with FWHM>800 km s$^{-1}$ and generally blueshifted, as in \citet{villar2011ionized,villar2016ionized}. Using the peak-weighted non-parametric method, we find that all the QSO2s have signatures of outflowing gas: 18 in the form of blueshifted asymmetries, and one redshifted. This implies an outflow incidence rate of $\sim$95-100\%, similar to the results found in other QSO2s studies \citep{villar2011ionized,villar2016ionized,2018ApJ...856..102F,2021MNRAS.504.3890D} and higher than those generally reported for low-to-intermediate luminosity AGN (see e.g., \citealt{10.1093/mnras/stad599} and references therein). 

Some of the QSO2s show extreme [OIII] kinematics, as for example J0924+01 (see Figures \ref{Fig1}-\ref{Fig3}). From the flux-weighted non-parametric analysis of all the QSO2s, we find emission line widths ranging from 530 to 1645 km s$^{-1}$ (average W$_{80}$=940$\pm$280 km s$^{-1}$) and asymmetry values ranging from -235 to 50 km s$^{-1}$ (average a=-120$\pm$130 km s$^{-1}$). Regarding the outflow velocities, the fastest that we measure come from the flux-weighted parametric method (average v$_{\rm OF}$=-1120$\pm$510 km s$^{-1}$) and from the parametric when we compute the maximum velocity (average v$_{\rm max}$=-1800$\pm$1400 km s$^{-1}$). These velocities are typical of AGN-driven ionized outflows detected in quasars with these bolometric luminosities \citep{mullaney2013narrow,zakamska2014quasar}.

We computed outflow physical properties, such as the outflow mass, outflow mass rate, and kinetic rates, from the direct outflow measurements derived through each kinetic analysis method (see Section \ref{analysis}). This quantities are key for investigating the outflow impact on the surrounding environment of their hosts.
From the definition of these quantities (see Section \ref{outflowproperties}), it is clear that both the electron density and the outflow radius play a critical role in their determination, alongside with the outflow flux and velocity derived via each method. 
Since with our data we cannot constrain either n$_e$ or R$_{\rm OF}$, we assumed a gas density of n$_e$=200 cm$^{-3}$ and a radius of 1 kpc for all the QSO2s, as in \citet{fiore2017agn}. By assuming these values, we can just focus on the differences that are inherent to the method used to measure the outflow flux and velocity.

From our analysis of the outflow energetics using the three different methods we find that, on average, the parametric method results in the highest outflow masses (log M$_{\rm OF}$(M$_{\odot}$)=6.47$\pm$0.50), since it considers the integrated flux of the broad and intermediate components. The flux-weighted non-parametric method produces larger outflow mass rates and kinetic powers (\.M$_{\rm OF}$=4.0$\pm$4.4 M$_{\odot}$~yr$^{-1}$ and \.E$_{\rm kin}$=41.9$\pm$0.6 erg~s$^{-1}$) than the peak-weighted non-parametric method and the parametric method using v$_s$ (see Figure \ref{Fig4} and Table \ref{tab:summary}). This happens because by definition it always includes a contribution from both the blue and red emission line wings.  
This effect is more pronounced in the kinetic powers, because of its dependence on the outflow velocity, $\rm \dot{E}$$_{\rm kin}$ $\propto$ $\rm \dot{M}$$_{\rm OF}$v$_{\rm OF}^2$ $\propto$ v$_{\rm OF}^3$. Nevertheless, we find the highest values of the outflow mass rate and kinetic power when we use the parametric method and v$_{\rm max}$ instead of v$_s$ (\.M$_{\rm OF}$=23$\pm$35 M$_{\odot}$~yr$^{-1}$ and log(\.E$_{\rm kin}$)=42.9$\pm$0.6 erg~s$^{-1}$).

A comparison between the mass outflow rates and kinetic powers derived from the three methods and the values reported by \citet{fiore2017agn} for a sample of 94 AGN is shown in Figure \ref{Fig5}. The values derived from the parametric v$_{\rm max}$ method are fully consistent with \citet{fiore2017agn}, as expected since they used the same velocity definition to measure the outflow physical properties. However, we argue that these maximum velocities are not representative of outflowing gas, since they only trace the highest velocity gas instead of the bulk of the outflowing gas mass. This bias the results towards large values of the outflow physical properties.
Other works using higher electron densities for the winds and/or lower outflow velocities such as v$_s$ report outflow properties that are from one to three orders of magnitude lower \citep{holt2011impact,villar2016ionized,rose2018quantifying,spence2018quantifying,davies2020ionized,holden2023precise}. 
Besides, \citet{davies2020ionized} pointed out that the relatively low scatter find by \citet{fiore2017agn} when plotting the data shown in Figure \ref{Fig5}, despite the wide range of luminosity, is partly due to the adoption of common values of the outflow density and radius.

The median coupling efficiencies that we measure for the outflows using the three methods, are 0.001\%, 0.02\%, and 0.01\%, and when we use the parametric method and v$_{\rm max}$, it goes up to 0.22\%.
The former three values are lower than those reported in recent literature \citep{fiore2017agn,baron2019discovering,speranza2021murales}, while the parametric v$_{\rm max}$ value is similar. For example, both \citet{fiore2017agn} and \citet{speranza2021murales}  reported a coupling efficiency of $\sim$0.2\% for a sample of AGN at z<0.5 and log(L$_{\rm BOL}$)$\sim$45.5 erg s$^{-1}$, and of 3CR radio galaxies at z<0.3 with log(L$_{\rm BOL}$)=42-46 erg s$^{-1}$, respectively. Again, this is a consequence of using maximum outflow velocities and assuming a low value of the density. 

\subsection{Lack of correlations between the nuclear ionized outflow properties and galaxy properties}
\label{discussion3} 

To evaluate the outflow impact on star formation and their possible relation with other AGN and/or host galaxy properties, in Section \ref{correlations} we investigated the correlation matrix using results from the flux-weighted non-parametric method. 

We only found moderate correlations between the AGN bolometric luminosity and the outflow masses and mass rates, both with a Spearman coefficient of $\rho$ = 0.52.  
More massive and powerful outflows are usually found in luminous AGN galaxies \citep{hainline2014gemini,fiore2017agn,revalski2018quantifying}.
Indeed, since here we are considering AGN bolometric luminosities derived from the extinction-corrected [OIII] luminosities, it is not surprising to find correlations, albeit weak, with the outflow masses measured from the [OIII] emission line. Likewise, high outflow mass rates are expected for luminous AGN \citep{fiore2017agn}, although here we only find a modest correlation. This can be a consequence of the relatively small luminosity range probed by our sample, of log(L$_{\rm BOL}$)=44.9-46.7 erg~s$^{-1}$.

We have not found any significant correlations between the outflow and the host galaxies properties here considered. Since here we are assuming fixed values of the outflow density and radius for all the targets, the outflow physical quantities are not precise. Therefore, the different correlations that we evaluate here might change if individual radii and densities would be measured. This is not the case of the direct outflow measurements, such as the velocity, W80, asymmetry, etc. The outflow density is one of the parameters having the strongest impact in the derived outflow physical quantities. The outflow mass rate and kinetic power vary by one order of magnitude when we assume a density of 2000 cm$^{-3}$ instead of 200 cm$^{-3}$. However, we find the same lack of correlation between the outflow properties derived from the flux-weighted non-parametric method and different AGN and host galaxy properties when precise density measurements are used for the individual targets in the QSOFEED sample of QSO2s ({\color {blue} Bessiere et al. in prep.}).

Regarding the impact of the outflows on recent star formation, we find that 4 of the 5 QSO2s lacking a YSP with age <100 Myr \citep{bessiere2017young} present disturbed kinematics with large asymmetry values (|a|>100). On the other hand, only 5 of the 14 QSO2s with a YSP present large asymmetry values. Despite the small size of our sample, these results might be indicating that recent star formation (i.e., the one having the same dynamical timescales as the outflows) is being suppressed more efficiently in the QSO2s with most disrupted kinematics. Nevertheless, there are also QSO2s with disrupted kinematics that have YSPs. This could be due either to positive feedback \citep{klamer2004molecular,gaibler2012jet,zubovas2013agn,cresci2015magnum}, to the different spatial scales considered for the stellar population analysis ($\sim$8 kpc) and for the outflow measurements ($\sim$3.9 kpc), and/or to the time that the winds might need to supress star formation. Using integral field spectroscopic data of the QSO2 Mrk 34, \citet{bessiere2022spatially} showed that both positive and negative feedback can simultaneously occur in different parts of same galaxy, depending on the amount of energy and turbulence that the outflows inject in the ISM. This illustrates the complexity of the outflow-ISM interplay, and evidences the importance of using spatially resolved studies of AGN to evaluate the impact of feedback.

Given that interactions between galaxies disrupt the stellar and gas content of the galaxies involved, we could expect higher gas turbulence in mergers than in undisturbed galaxies.
However, we do not find any trend between the [OIII] kinematics and the optical morphologies of the QSO2s or with their environment. This is most likely due to the different timescales involved: 1-3 Gyr for the mergers \citep{2003AJ....126.1183C} and 1-100 Myr for the AGN-driven outflows \citep{2016MNRAS.462.4055Z}. A major or minor merger can efficiently transport gas towards the center the galaxy during hundreds of Myr, and this gas supply can be intermittent, leading to different phases of nuclear activity \citep{2015MNRAS.453L..46K}. This makes it challenging to look for correlations between the large-scale galaxy morphology and 
the AGN-driven outflows.

Finally, we investigated the role of jets in driving or contributing to drive outflows in our QSO2s. We do not find any significant correlation between the outflow properties and either the integrated or peak 5 GHz luminosities from FIRST. Using spectra from SDSS, \citet{mullaney2013narrow} concluded that the width of the [OIII]$\lambda$5007 line shows a maximum in AGN with radio luminosities of log(L$_{\rm 1.4GHz}$)=23-25 W~Hz$^{-1}$, i.e., moderate radio luminosities as those of our QSO2s.
The lack of correlation between these parameters is probably due to the small radio luminosity range probed by our QSO2s, of 22.1-24.7 W~Hz$^{-1}$. However, the dominant origin of this radio emission is still a matter of on-going debate \citep{zakamska2014quasar}, as it might be produced by non-thermal AGN emission \citep{jarvis2019prevalence}, star formation {\color {blue} (Bessiere et al. in prep.)}, and/or shocks induced by the quasar winds/outflows \citep{2023ApJ...953...87F}.


\section{Conclusions}\label{conclusions}

We characterized the ionized gas kinematics and energetics of a sample of 19 QSO2s at 0.3<z<0.41, using three different methods, parametric and non-parametric, to analyze the nuclear ($\sim$3.9 kpc) [OIII] emission-line profiles. The main conclusions of the work are the following.

\begin{itemize}

\item We detect ionized gas outflows in the form of asymmetric and broad emission line profiles in 95-100\% of the sample using the three methods. 18 of the 19 QSO2s show [OIII] profiles with blueshifted wings, whilst the other one shows a redshifted wing. 

\item The average physical outflow properties (e.g., outflow mass, mass rate, and kinetic energy) that we derived from the three methods are consistent within the errors. The parametric method results in the highest outflow masses, and the flux-weighted non-parametric and specially the parametric method using v$_{\rm max}$ provide the highest mass outflow rates and kinetic powers. 

\item We measure outflow mass rates ranging between 0.2 and 20 M$_{\odot}$~yr$^{-1}$ and kinetic powers between 38.3 and 42.9 erg~s$^{-1}$ for the QSO2s. For the parametric method using v$_{\rm max}$ the highest values go up to 151 M$_{\odot}$~yr$^{-1}$ and 44.0 erg~s$^{-1}$.  
These values are most likely upper limits, considering that we assumed fixed values of 200 cm$^{-3}$ and 1 kpc for the outflow density and radius, respectively. 

\item We find a modest correlation between the AGN bolometric luminosity and the outflow mass and mass rate, but we do not find any correlation with the host galaxy properties here considered. 

\item Four of the five QSO2s lacking a YSP with age <100 Myr present disturbed kinematics with large asymmetry values (|a|>100). On the other hand, only 5 of the 14 QSO2s with a YSP present large asymmetry values. Despite the small sample size, these results might be indicating that recent star formation is being suppressed more efficiently in the QSO2s with most disrupted kinematics.

\end{itemize}

Here we showed for the first time a comparison between outflow measurements using three different methods commonly used in the literature. By assuming a fix outflow density and radius, we can focus on the differences introduced by how the flux and velocity are calculated. We conclude that, although the average physical outflow properties derived from the three methods are consistent within the errors, the commonly adopted parametric measurements using maximum outflow velocities provide the highest values, not representative of the average outflow velocity. Finally, we argue that the lack of correlations between the outflow and the AGN and galaxy properties here considered is most likely due to the small luminosity ranges probed by our sample, and to the different timescales of the outflows and galaxy-wide properties. 


\begin{acknowledgements}
CRA and GS acknowledge financial support from the European Union's Horizon 2020 research and innovation programme under Marie Sk\l odowska-Curie grant agreement No 860744 (BiD4BESt). CRA, GS, JAP, and PSB acknowledge the project 
``Feeding and feedback in active galaxies'', with reference
PID2019-106027GB-C42, funded by MICINN-AEI/10.13039/501100011033. 
CRA also acknowledges the project ``Quantifying the impact of quasar feedback on galaxy evolution'', with reference EUR2020-112266, funded by MICINN-AEI/10.13039/501100011033 and the European Union NextGenerationEU/PRTR. CRA thanks the Kavli Institute for Cosmology of the University of Cambridge for their hospitality while working on this paper, and the IAC Severo Ochoa Program for the corresponding financial support. The authors thank the anonymous referee for useful and constructive suggestions.
\end{acknowledgements}


\bibliographystyle{aa}
\bibliography{references.bib}

\begin{thebibliography}{106}
\expandafter\ifx\csname natexlab\endcsname\relax\def\natexlab#1{#1}\fi

\bibitem[{Aalto {et~al.}(2016)Aalto, Costagliola, Muller, Sakamoto, Gallagher, Dasyra, Wada, Combes, Garc{\'\i}a-Burillo, Kristensen, {et~al.}}]{aalto2016precessing}
Aalto, S., Costagliola, F., Muller, S., {et~al.} 2016, Astronomy \& Astrophysics, 590, A73

\bibitem[{Arribas {et~al.}(2014)Arribas, Colina, Bellocchi, Maiolino, \& Villar-Martin}]{arribas2014ionized}
Arribas, S., Colina, L., Bellocchi, E., Maiolino, R., \& Villar-Martin, M. 2014, Astronomy \& Astrophysics, 568, A14

\bibitem[{Audibert {et~al.}(2019)Audibert, Combes, Garc{\'\i}a-Burillo, Hunt, Eckart, Aalto, Casasola, Boone, Krips, Viti, {et~al.}}]{audibert2019alma}
Audibert, A., Combes, F., Garc{\'\i}a-Burillo, S., {et~al.} 2019, Astronomy \& Astrophysics, 632, A33

\bibitem[{{Audibert} {et~al.}(2023){Audibert}, {Ramos Almeida}, {Garc{\'\i}a-Burillo}, {Combes}, {Bischetti}, {Meenakshi}, {Mukherjee}, {Bicknell}, \& {Wagner}}]{2023A&A...671L..12A}
{Audibert}, A., {Ramos Almeida}, C., {Garc{\'\i}a-Burillo}, S., {et~al.} 2023, \aap, 671, L12

\bibitem[{{Ayubinia} {et~al.}(2023){Ayubinia}, {Woo}, {Rakshit}, \& {Son}}]{2023ApJ...954...27A}
{Ayubinia}, A., {Woo}, J.-H., {Rakshit}, S., \& {Son}, D. 2023, \apj, 954, 27

\bibitem[{Baron \& Netzer(2019)}]{baron2019discovering}
Baron, D. \& Netzer, H. 2019, Monthly Notices of the Royal Astronomical Society, 486, 4290

\bibitem[{Bessiere {et~al.}(2012)Bessiere, Tadhunter, Ramos~Almeida, \& Villar~Mart{\'\i}n}]{bessiere2012importance}
Bessiere, P., Tadhunter, C.~N., Ramos~Almeida, C., \& Villar~Mart{\'\i}n, M. 2012, Monthly Notices of the Royal Astronomical Society, 426, 276

\bibitem[{Bessiere \& Ramos~Almeida(2022)}]{bessiere2022spatially}
Bessiere, P.~S. \& Ramos~Almeida, C. 2022, Monthly Notices of the Royal Astronomical Society: Letters, 512, L54

\bibitem[{Bessiere {et~al.}(2017)Bessiere, Tadhunter, Ramos~Almeida, Villar~Mart{\'\i}n, \& Cabrera-Lavers}]{bessiere2017young}
Bessiere, P.~S., Tadhunter, C.~N., Ramos~Almeida, C., Villar~Mart{\'\i}n, M., \& Cabrera-Lavers, A. 2017, Monthly Notices of the Royal Astronomical Society, 466, 3887

\bibitem[{Canalizo \& Stockton(2001)}]{canalizo2001quasi}
Canalizo, G. \& Stockton, A. 2001, The Astrophysical Journal, 555, 719

\bibitem[{Carniani {et~al.}(2015)Carniani, Marconi, Maiolino, Balmaverde, Brusa, Cano-D{\'\i}az, Cicone, Comastri, Cresci, Fiore, {et~al.}}]{carniani2015ionised}
Carniani, S., Marconi, A., Maiolino, R., {et~al.} 2015, Astronomy \& Astrophysics, 580, A102

\bibitem[{Carniani {et~al.}(2016)Carniani, Marconi, Maiolino, Balmaverde, Brusa, Cano-D{\'\i}az, Cicone, Comastri, Cresci, Fiore, {et~al.}}]{carniani2016fast}
Carniani, S., Marconi, A., Maiolino, R., {et~al.} 2016, Astronomy \& Astrophysics, 591, A28

\bibitem[{Centeno \& Socas-Navarro(2008)}]{centeno2008new}
Centeno, R. \& Socas-Navarro, H. 2008, The Astrophysical Journal, 682, L61

\bibitem[{{Cicone} {et~al.}(2018){Cicone}, {Brusa}, {Ramos Almeida}, {Cresci}, {Husemann}, \& {Mainieri}}]{2018NatAs...2..176C}
{Cicone}, C., {Brusa}, M., {Ramos Almeida}, C., {et~al.} 2018, Nature Astronomy, 2, 176

\bibitem[{Cicone {et~al.}(2014)Cicone, Maiolino, Sturm, Graci{\'a}-Carpio, Feruglio, Neri, Aalto, Davies, Fiore, Fischer, {et~al.}}]{cicone2014massive}
Cicone, C., Maiolino, R., Sturm, E., {et~al.} 2014, Astronomy \& Astrophysics, 562, A21

\bibitem[{{Conselice} {et~al.}(2003){Conselice}, {Bershady}, {Dickinson}, \& {Papovich}}]{2003AJ....126.1183C}
{Conselice}, C.~J., {Bershady}, M.~A., {Dickinson}, M., \& {Papovich}, C. 2003, \aj, 126, 1183

\bibitem[{Cresci \& Maiolino(2018)}]{cresci2018observing}
Cresci, G. \& Maiolino, R. 2018, Nature Astronomy, 2, 179

\bibitem[{Cresci {et~al.}(2015)Cresci, Marconi, Zibetti, Risaliti, Carniani, Mannucci, Gallazzi, Maiolino, Balmaverde, Brusa, {et~al.}}]{cresci2015magnum}
Cresci, G., Marconi, A., Zibetti, S., {et~al.} 2015, Astronomy \& Astrophysics, 582, A63

\bibitem[{Croton {et~al.}(2006)Croton, Springel, White, De~Lucia, Frenk, Gao, Jenkins, Kauffmann, Navarro, \& Yoshida}]{croton2006many}
Croton, D.~J., Springel, V., White, S.~D., {et~al.} 2006, Monthly Notices of the Royal Astronomical Society, 365, 11

\bibitem[{{Dall'Agnol de Oliveira} {et~al.}(2021){Dall'Agnol de Oliveira}, {Storchi-Bergmann}, {Kraemer}, {Villar Mart{\'\i}n}, {Schnorr-M{\"u}ller}, {Schmitt}, {Ruschel-Dutra}, {Crenshaw}, \& {Fischer}}]{2021MNRAS.504.3890D}
{Dall'Agnol de Oliveira}, B., {Storchi-Bergmann}, T., {Kraemer}, S.~B., {et~al.} 2021, \mnras, 504, 3890

\bibitem[{Davies {et~al.}(2020)Davies, Baron, Shimizu, Netzer, Burtscher, De~Zeeuw, Genzel, Hicks, Koss, Lin, {et~al.}}]{davies2020ionized}
Davies, R., Baron, D., Shimizu, T., {et~al.} 2020, Monthly Notices of the Royal Astronomical Society, 498, 4150

\bibitem[{{Di Matteo} {et~al.}(2005){Di Matteo}, {Springel}, \& {Hernquist}}]{2005Natur.433..604D}
{Di Matteo}, T., {Springel}, V., \& {Hernquist}, L. 2005, \nat, 433, 604

\bibitem[{Dimitrijevi{\'c} {et~al.}(2007)Dimitrijevi{\'c}, Popovi{\'c}, Kova{\v{c}}evi{\'c}, Da{\v{c}}i{\'c}, \& Ili{\'c}}]{dimitrijevic2007flux}
Dimitrijevi{\'c}, M., Popovi{\'c}, L., Kova{\v{c}}evi{\'c}, J., Da{\v{c}}i{\'c}, M., \& Ili{\'c}, D. 2007, Monthly Notices of the Royal Astronomical Society, 374, 1181

\bibitem[{Dubois {et~al.}(2016)Dubois, Peirani, Pichon, Devriendt, Gavazzi, Welker, \& Volonteri}]{dubois2016horizon}
Dubois, Y., Peirani, S., Pichon, C., {et~al.} 2016, Monthly Notices of the Royal Astronomical Society, 463, 3948

\bibitem[{Fern{\'a}ndez-Ontiveros {et~al.}(2020)Fern{\'a}ndez-Ontiveros, Dasyra, Hatziminaoglou, Malkan, Pereira-Santaella, Papachristou, Spinoglio, Combes, Aalto, Nagar, {et~al.}}]{fernandez2020co}
Fern{\'a}ndez-Ontiveros, J., Dasyra, K., Hatziminaoglou, E., {et~al.} 2020, Astronomy \& Astrophysics, 633, A127

\bibitem[{Fiore {et~al.}(2017)Fiore, Feruglio, Shankar, Bischetti, Bongiorno, Brusa, Carniani, Cicone, Duras, Lamastra, {et~al.}}]{fiore2017agn}
Fiore, F., Feruglio, C., Shankar, F., {et~al.} 2017, Astronomy \& Astrophysics, 601, A143

\bibitem[{{Fischer} {et~al.}(2019){Fischer}, {Smith}, {Kraemer}, {Schmitt}, {Crenshaw}, {Koss}, {Mushotzky}, {Larson}, {U}, \& {Rigby}}]{2019ApJ...887..200F}
{Fischer}, T., {Smith}, K.~L., {Kraemer}, S., {et~al.} 2019, \apj, 887, 200

\bibitem[{{Fischer} {et~al.}(2023){Fischer}, {Johnson}, {Secrest}, {Crenshaw}, \& {Kraemer}}]{2023ApJ...953...87F}
{Fischer}, T.~C., {Johnson}, M.~C., {Secrest}, N.~J., {Crenshaw}, D.~M., \& {Kraemer}, S.~B. 2023, \apj, 953, 87

\bibitem[{{Fischer} {et~al.}(2018){Fischer}, {Kraemer}, {Schmitt}, {Longo Micchi}, {Crenshaw}, {Revalski}, {Vestergaard}, {Elvis}, {Gaskell}, {Hamann}, {Ho}, {Hutchings}, {Mushotzky}, {Netzer}, {Storchi-Bergmann}, {Straughn}, {Turner}, \& {Ward}}]{2018ApJ...856..102F}
{Fischer}, T.~C., {Kraemer}, S.~B., {Schmitt}, H.~R., {et~al.} 2018, \apj, 856, 102

\bibitem[{Fluetsch {et~al.}(2021)Fluetsch, Maiolino, Carniani, Arribas, Belfiore, Bellocchi, Cazzoli, Cicone, Cresci, Fabian, {et~al.}}]{fluetsch2021properties}
Fluetsch, A., Maiolino, R., Carniani, S., {et~al.} 2021, Monthly Notices of the Royal Astronomical Society, 505, 5753

\bibitem[{Gaibler {et~al.}(2012)Gaibler, Khochfar, Krause, \& Silk}]{gaibler2012jet}
Gaibler, V., Khochfar, S., Krause, M., \& Silk, J. 2012, Monthly Notices of the Royal Astronomical Society, 425, 438

\bibitem[{Garc{\'\i}a-Bernete {et~al.}(2021)Garc{\'\i}a-Bernete, Alonso-Herrero, Garc{\'\i}a-Burillo, Pereira-Santaella, Garc{\'\i}a-Lorenzo, Carrera, Rigopoulou, Almeida, Mart{\'\i}n, Gonz{\'a}lez-Mart{\'\i}n, {et~al.}}]{garcia2021multiphase}
Garc{\'\i}a-Bernete, I., Alonso-Herrero, A., Garc{\'\i}a-Burillo, S., {et~al.} 2021, Astronomy \& Astrophysics, 645, A21

\bibitem[{{Garc{\'\i}a-Burillo} {et~al.}(2021){Garc{\'\i}a-Burillo}, {Alonso-Herrero}, {Ramos Almeida}, {Gonz{\'a}lez-Mart{\'\i}n}, {Combes}, {Usero}, {H{\"o}nig}, {Querejeta}, {Hicks}, {Hunt}, {Rosario}, {Davies}, {Boorman}, {Bunker}, {Burtscher}, {Colina}, {D{\'\i}az-Santos}, {Gandhi}, {Garc{\'\i}a-Bernete}, {Garc{\'\i}a-Lorenzo}, {Ichikawa}, {Imanishi}, {Izumi}, {Labiano}, {Levenson}, {L{\'o}pez-Rodr{\'\i}guez}, {Packham}, {Pereira-Santaella}, {Ricci}, {Rigopoulou}, {Rouan}, {Shimizu}, {Stalevski}, {Wada}, \& {Williamson}}]{2021A&A...652A..98G}
{Garc{\'\i}a-Burillo}, S., {Alonso-Herrero}, A., {Ramos Almeida}, C., {et~al.} 2021, \aap, 652, A98

\bibitem[{Garc{\'\i}a-Burillo {et~al.}(2019)Garc{\'\i}a-Burillo, Combes, Almeida, Usero, Alonso-Herrero, Hunt, Rouan, Aalto, Querejeta, Viti, {et~al.}}]{garcia2019alma}
Garc{\'\i}a-Burillo, S., Combes, F., Almeida, C.~R., {et~al.} 2019, Astronomy \& Astrophysics, 632, A61

\bibitem[{Garc{\'\i}a-Burillo {et~al.}(2014)Garc{\'\i}a-Burillo, Combes, Usero, Aalto, Krips, Viti, Alonso-Herrero, Hunt, Schinnerer, Baker, {et~al.}}]{garcia2014molecular}
Garc{\'\i}a-Burillo, S., Combes, F., Usero, A., {et~al.} 2014, Astronomy \& Astrophysics, 567, A125

\bibitem[{Goulding {et~al.}(2018)Goulding, Greene, Bezanson, Greco, Johnson, Leauthaud, Matsuoka, Medezinski, \& Price-Whelan}]{goulding2018galaxy}
Goulding, A.~D., Greene, J.~E., Bezanson, R., {et~al.} 2018, Publications of the Astronomical Society of Japan, 70, S37

\bibitem[{Greene {et~al.}(2011)Greene, Zakamska, Ho, \& Barth}]{greene2011feedback}
Greene, J.~E., Zakamska, N.~L., Ho, L.~C., \& Barth, A.~J. 2011, The Astrophysical Journal, 732, 9

\bibitem[{{Groves}(2007)}]{2007ASPC..373..511G}
{Groves}, B. 2007, in Astronomical Society of the Pacific Conference Series, Vol. 373, The Central Engine of Active Galactic Nuclei, ed. L.~C. {Ho} \& J.~W. {Wang}, 511

\bibitem[{Hainline {et~al.}(2014)Hainline, Hickox, Greene, Myers, Zakamska, Liu, \& Liu}]{hainline2014gemini}
Hainline, K.~N., Hickox, R.~C., Greene, J.~E., {et~al.} 2014, The Astrophysical Journal, 787, 65

\bibitem[{Harrison(2017)}]{harrison2017impact}
Harrison, C. 2017, Nature Astronomy, 1, 1

\bibitem[{Harrison {et~al.}(2014)Harrison, Alexander, Mullaney, \& Swinbank}]{harrison2014kiloparsec}
Harrison, C., Alexander, D., Mullaney, J., \& Swinbank, A. 2014, Monthly Notices of the Royal Astronomical Society, 441, 3306

\bibitem[{Harrison {et~al.}(2018)Harrison, Costa, Tadhunter, Fl{\"u}tsch, Kakkad, Perna, \& Vietri}]{harrison2018agn}
Harrison, C., Costa, T., Tadhunter, C., {et~al.} 2018, Nature Astronomy, 2, 198

\bibitem[{Heckman {et~al.}(1981)Heckman, Miley, van Breugel, \& Butcher}]{heckman1981emission}
Heckman, T.~M., Miley, G.~K., van Breugel, W.~J., \& Butcher, H.~R. 1981, Astrophysical Journal, Part 1, vol. 247, July 15, 1981, p. 403-418., 247, 403

\bibitem[{Herrera-Camus {et~al.}(2020)Herrera-Camus, Janssen, Sturm, Lutz, Veilleux, Davies, Shimizu, Gonz{\'a}lez-Alfonso, Rupke, Tacconi, {et~al.}}]{herrera2020agn}
Herrera-Camus, R., Janssen, A., Sturm, E., {et~al.} 2020, Astronomy \& Astrophysics, 635, A47

\bibitem[{{Hickox} {et~al.}(2014){Hickox}, {Mullaney}, {Alexander}, {Chen}, {Civano}, {Goulding}, \& {Hainline}}]{2014ApJ...782....9H}
{Hickox}, R.~C., {Mullaney}, J.~R., {Alexander}, D.~M., {et~al.} 2014, \apj, 782, 9

\bibitem[{{Holden} \& {Tadhunter}(2023)}]{2023MNRAS.524..886H}
{Holden}, L.~R. \& {Tadhunter}, C.~N. 2023, \mnras, 524, 886

\bibitem[{Holden {et~al.}(2023)Holden, Tadhunter, Morganti, \& Oosterloo}]{holden2023precise}
Holden, L.~R., Tadhunter, C.~N., Morganti, R., \& Oosterloo, T. 2023, Monthly Notices of the Royal Astronomical Society

\bibitem[{Holt {et~al.}(2011)Holt, Tadhunter, Morganti, \& Emonts}]{holt2011impact}
Holt, J., Tadhunter, C., Morganti, R., \& Emonts, B. 2011, Monthly Notices of the Royal Astronomical Society, 410, 1527

\bibitem[{Hopkins {et~al.}(2008)Hopkins, Hernquist, Cox, \& Kere{\v{s}}}]{hopkins2008cosmological}
Hopkins, P.~F., Hernquist, L., Cox, T.~J., \& Kere{\v{s}}, D. 2008, The Astrophysical Journal Supplement Series, 175, 356

\bibitem[{{Husemann} {et~al.}(2016){Husemann}, {Scharw{\"a}chter}, {Bennert}, {Mainieri}, {Woo}, \& {Kakkad}}]{2016A&A...594A..44H}
{Husemann}, B., {Scharw{\"a}chter}, J., {Bennert}, V.~N., {et~al.} 2016, \aap, 594, A44

\bibitem[{Jarvis {et~al.}(2019)Jarvis, Harrison, Thomson, Circosta, Mainieri, Alexander, Edge, Lansbury, Molyneux, \& Mullaney}]{jarvis2019prevalence}
Jarvis, M., Harrison, C., Thomson, A., {et~al.} 2019, Monthly Notices of the Royal Astronomical Society, 485, 2710

\bibitem[{{Jarvis} {et~al.}(2021){Jarvis}, {Harrison}, {Mainieri}, {Alexander}, {Arrigoni Battaia}, {Calistro Rivera}, {Circosta}, {Costa}, {De Breuck}, {Edge}, {Girdhar}, {Kakkad}, {Kharb}, {Lansbury}, {Molyneux}, {Mukherjee}, {Mullaney}, {Farina}, {Silpa}, {Thomson}, \& {Ward}}]{2021MNRAS.503.1780J}
{Jarvis}, M.~E., {Harrison}, C.~M., {Mainieri}, V., {et~al.} 2021, \mnras, 503, 1780

\bibitem[{Karouzos {et~al.}(2016)Karouzos, Woo, \& Bae}]{karouzos2016unraveling}
Karouzos, M., Woo, J.-H., \& Bae, H.-J. 2016, The Astrophysical Journal, 819, 148

\bibitem[{{King} \& {Nixon}(2015)}]{2015MNRAS.453L..46K}
{King}, A. \& {Nixon}, C. 2015, \mnras, 453, L46

\bibitem[{Klamer {et~al.}(2004)Klamer, Ekers, Sadler, \& Hunstead}]{klamer2004molecular}
Klamer, I.~J., Ekers, R.~D., Sadler, E.~M., \& Hunstead, R.~W. 2004, The Astrophysical Journal, 612, L97

\bibitem[{Kong \& Ho(2018)}]{kong2018black}
Kong, M. \& Ho, L.~C. 2018, The Astrophysical Journal, 859, 116

\bibitem[{Lamastra {et~al.}(2009)Lamastra, Bianchi, Matt, Perola, Barcons, \& Carrera}]{lamastra2009bolometric}
Lamastra, A., Bianchi, S., Matt, G., {et~al.} 2009, Astronomy \& Astrophysics, 504, 73

\bibitem[{{Lamperti} {et~al.}(2022){Lamperti}, {Pereira-Santaella}, {Perna}, {Colina}, {Arribas}, {Garc{\'\i}a-Burillo}, {Gonz{\'a}lez-Alfonso}, {Aalto}, {Alonso-Herrero}, {Combes}, {Labiano}, {Piqueras-L{\'o}pez}, {Rigopoulou}, \& {van der Werf}}]{2022A&A...668A..45L}
{Lamperti}, I., {Pereira-Santaella}, M., {Perna}, M., {et~al.} 2022, \aap, 668, A45

\bibitem[{Liu {et~al.}(2013)Liu, Zakamska, Greene, Nesvadba, \& Liu}]{liu2013observations}
Liu, G., Zakamska, N.~L., Greene, J.~E., Nesvadba, N.~P., \& Liu, X. 2013, Monthly Notices of the Royal Astronomical Society, 430, 2327

\bibitem[{Lutz {et~al.}(2020)Lutz, Sturm, Janssen, Veilleux, Aalto, Cicone, Contursi, Davies, Feruglio, Fischer, {et~al.}}]{lutz2020molecular}
Lutz, D., Sturm, E., Janssen, A., {et~al.} 2020, Astronomy \& Astrophysics, 633, A134

\bibitem[{Mart{\'\i}n-Navarro {et~al.}(2021)Mart{\'\i}n-Navarro, Pillepich, Nelson, Rodriguez-Gomez, Donnari, Hernquist, \& Springel}]{martin2021anisotropic}
Mart{\'\i}n-Navarro, I., Pillepich, A., Nelson, D., {et~al.} 2021, Nature, 594, 187

\bibitem[{{Martini}(2004)}]{2004cbhg.symp..169M}
{Martini}, P. 2004, in Coevolution of Black Holes and Galaxies, ed. L.~C. {Ho}, 169

\bibitem[{McElroy {et~al.}(2015)McElroy, Croom, Pracy, Sharp, Ho, \& Medling}]{mcelroy2015ifu}
McElroy, R., Croom, S.~M., Pracy, M., {et~al.} 2015, Monthly Notices of the Royal Astronomical Society, 446, 2186

\bibitem[{Morganti {et~al.}(2015)Morganti, Oosterloo, Oonk, Frieswijk, \& Tadhunter}]{morganti2015fast}
Morganti, R., Oosterloo, T., Oonk, J.~R., Frieswijk, W., \& Tadhunter, C. 2015, Astronomy \& Astrophysics, 580, A1

\bibitem[{Moster {et~al.}(2010)Moster, Somerville, Maulbetsch, Van Den~Bosch, Macci{\`o}, Naab, \& Oser}]{moster2010constraints}
Moster, B.~P., Somerville, R.~S., Maulbetsch, C., {et~al.} 2010, The Astrophysical Journal, 710, 903

\bibitem[{{Mukherjee} {et~al.}(2018){Mukherjee}, {Wagner}, {Bicknell}, {Morganti}, {Oosterloo}, {Nesvadba}, \& {Sutherland}}]{2018MNRAS.476...80M}
{Mukherjee}, D., {Wagner}, A.~Y., {Bicknell}, G.~V., {et~al.} 2018, \mnras, 476, 80

\bibitem[{Mullaney {et~al.}(2013)Mullaney, Alexander, Fine, Goulding, Harrison, \& Hickox}]{mullaney2013narrow}
Mullaney, J., Alexander, D., Fine, S., {et~al.} 2013, Monthly Notices of the Royal Astronomical Society, 433, 622

\bibitem[{Netzer(1990)}]{netzer1990agn}
Netzer, H. 1990, in Active Galactic Nuclei (Springer), 57--158

\bibitem[{Novak {et~al.}(2011)Novak, Ostriker, \& Ciotti}]{novak2011feedback}
Novak, G.~S., Ostriker, J.~P., \& Ciotti, L. 2011, The Astrophysical Journal, 737, 26

\bibitem[{Osterbrock \& Ferland(2006)}]{osterbrock2006astrophysics}
Osterbrock, D.~E. \& Ferland, G.~J. 2006, Astrophysics Of Gas Nebulae and Active Galactic Nuclei (University science books)

\bibitem[{Pierce {et~al.}(2023)Pierce, Tadhunter, Ramos~Almeida, Bessiere, Heaton, Ellison, Speranza, Gordon, O’Dea, Grimmett, {et~al.}}]{pierce2023galaxy}
Pierce, J., Tadhunter, C., Ramos~Almeida, C., {et~al.} 2023, Monthly Notices of the Royal Astronomical Society, 522, 1736

\bibitem[{Ramos~Almeida {et~al.}(2019)Ramos~Almeida, Acosta-Pulido, Tadhunter, Gonzalez-Fernandez, Cicone, \& Fernandez-Torreiro}]{ramos2019near}
Ramos~Almeida, C., Acosta-Pulido, J.~A., Tadhunter, C.~N., {et~al.} 2019, Monthly Notices of the Royal Astronomical Society: Letters, 487, L18

\bibitem[{Ramos~Almeida {et~al.}(2013)Ramos~Almeida, Bessiere, Tadhunter, Inskip, Morganti, Dicken, Gonz{\'a}lez-Serrano, \& Holt}]{ramos2013environments}
Ramos~Almeida, C., Bessiere, P., Tadhunter, C., {et~al.} 2013, Monthly Notices of the Royal Astronomical Society, 436, 997

\bibitem[{Ramos~Almeida {et~al.}(2012)Ramos~Almeida, Bessiere, Tadhunter, P{\'e}rez-Gonz{\'a}lez, Barro, Inskip, Morganti, Holt, \& Dicken}]{ramos2012luminous}
Ramos~Almeida, C., Bessiere, P., Tadhunter, C., {et~al.} 2012, Monthly Notices of the Royal Astronomical Society, 419, 687

\bibitem[{{Ramos Almeida} {et~al.}(2022){Ramos Almeida}, {Bischetti}, {Garc{\'\i}a-Burillo}, {Alonso-Herrero}, {Audibert}, {Cicone}, {Feruglio}, {Tadhunter}, {Pierce}, {Pereira-Santaella}, \& {Bessiere}}]{2022A&A...658A.155R}
{Ramos Almeida}, C., {Bischetti}, M., {Garc{\'\i}a-Burillo}, S., {et~al.} 2022, \aap, 658, A155

\bibitem[{Ramos~Almeida {et~al.}(2017)Ramos~Almeida, Piqueras~L{\'o}pez, Villar-Mart{\'\i}n, \& Bessiere}]{2017MNRAS.470..964R}
Ramos~Almeida, C., Piqueras~L{\'o}pez, J., Villar-Mart{\'\i}n, M., \& Bessiere, P.~S. 2017, \mnras, 470, 964

\bibitem[{Ramos~Almeida \& Ricci(2017)}]{almeida2017nuclear}
Ramos~Almeida, C. \& Ricci, C. 2017, Nature Astronomy, 1, 679

\bibitem[{Ramos~Almeida {et~al.}(2011)Ramos~Almeida, Tadhunter, Inskip, Morganti, Holt, \& Dicken}]{almeida2011optical}
Ramos~Almeida, C., Tadhunter, C.~N., Inskip, K., {et~al.} 2011, Monthly Notices of the Royal Astronomical Society, 410, 1550

\bibitem[{{Ramos Almeida} {et~al.}(2011){Ramos Almeida}, {Tadhunter}, {Inskip}, {Morganti}, {Holt}, \& {Dicken}}]{2011MNRAS.410.1550R}
{Ramos Almeida}, C., {Tadhunter}, C.~N., {Inskip}, K.~J., {et~al.} 2011, \mnras, 410, 1550

\bibitem[{Revalski {et~al.}(2018)Revalski, Dashtamirova, Crenshaw, Kraemer, Fischer, Schmitt, Gnilka, Schmidt, Elvis, Fabbiano, {et~al.}}]{revalski2018quantifying}
Revalski, M., Dashtamirova, D., Crenshaw, D.~M., {et~al.} 2018, The Astrophysical Journal, 867, 88

\bibitem[{Reyes {et~al.}(2008)Reyes, Zakamska, Strauss, Green, Krolik, Shen, Richards, Anderson, \& Schneider}]{reyes2008space}
Reyes, R., Zakamska, N.~L., Strauss, M.~A., {et~al.} 2008, The Astronomical Journal, 136, 2373

\bibitem[{Riffel {et~al.}(2023)Riffel, Storchi-Bergmann, Riffel, Bianchin, Zakamska, Ruschel-Dutra, Bentz, Burtscher, Crenshaw, Dahmer-Hahn, Dametto, Davies, Diniz, Fischer, Harrison, Mainieri, Revalski, Rodriguez-Ardila, Rosario, \& Schönell}]{10.1093/mnras/stad599}
Riffel, R.~A., Storchi-Bergmann, T., Riffel, R., {et~al.} 2023, Monthly Notices of the Royal Astronomical Society, 521, 1832

\bibitem[{Rose {et~al.}(2018)Rose, Tadhunter, Ramos~Almeida, Rodr{\'\i}guez~Zaur{\'\i}n, Santoro, \& Spence}]{rose2018quantifying}
Rose, M., Tadhunter, C., Ramos~Almeida, C., {et~al.} 2018, Monthly Notices of the Royal Astronomical Society, 474, 128

\bibitem[{Rupke \& Veilleux(2013)}]{rupke2013multiphase}
Rupke, D.~S. \& Veilleux, S. 2013, The Astrophysical Journal, 768, 75

\bibitem[{{Rupke} {et~al.}(2019){Rupke}, {Coil}, {Geach}, {Tremonti}, {Diamond-Stanic}, {George}, {Hickox}, {Kepley}, {Leung}, {Moustakas}, {Rudnick}, \& {Sell}}]{2019Natur.574..643R}
{Rupke}, D. S.~N., {Coil}, A., {Geach}, J.~E., {et~al.} 2019, \nat, 574, 643

\bibitem[{Satyapal {et~al.}(2014)Satyapal, Ellison, McAlpine, Hickox, Patton, \& Mendel}]{satyapal2014galaxy}
Satyapal, S., Ellison, S.~L., McAlpine, W., {et~al.} 2014, Monthly Notices of the Royal Astronomical Society, 441, 1297

\bibitem[{Schawinski {et~al.}(2015)Schawinski, Koss, Berney, \& Sartori}]{schawinski2015active}
Schawinski, K., Koss, M., Berney, S., \& Sartori, L.~F. 2015, Monthly Notices of the Royal Astronomical Society, 451, 2517

\bibitem[{{Silk} \& {Rees}(1998)}]{1998A&A...331L...1S}
{Silk}, J. \& {Rees}, M.~J. 1998, \aap, 331, L1

\bibitem[{Singha {et~al.}(2022)Singha, Husemann, Urrutia, O’Dea, Scharw{\"a}chter, Gaspari, Combes, Nevin, Terrazas, P{\'e}rez-Torres, {et~al.}}]{singha2022close}
Singha, M., Husemann, B., Urrutia, T., {et~al.} 2022, Astronomy \& Astrophysics, 659, A123

\bibitem[{Spence {et~al.}(2018)Spence, Tadhunter, Rose, \& Rodr{\'\i}guez~Zaur{\'\i}n}]{spence2018quantifying}
Spence, R., Tadhunter, C., Rose, M., \& Rodr{\'\i}guez~Zaur{\'\i}n, J. 2018, Monthly Notices of the Royal Astronomical Society, 478, 2438

\bibitem[{Speranza {et~al.}(2022)Speranza, Almeida, Acosta-Pulido, Riffel, Tadhunter, Pierce, Rodr{\'\i}guez-Ardila, Puga, Brusa, Musiimenta, {et~al.}}]{speranza2022warm}
Speranza, G., Almeida, C.~R., Acosta-Pulido, J., {et~al.} 2022, Astronomy \& Astrophysics, 665, A55

\bibitem[{Speranza {et~al.}(2021)Speranza, Balmaverde, Capetti, Massaro, Tremblay, Marconi, Venturi, Chiaberge, Baldi, Baum, {et~al.}}]{speranza2021murales}
Speranza, G., Balmaverde, B., Capetti, A., {et~al.} 2021, Astronomy \& Astrophysics, 653, A150

\bibitem[{Tadhunter {et~al.}(2021)Tadhunter, Patel, \& Mullaney}]{tadhunter2021light}
Tadhunter, C., Patel, M., \& Mullaney, J. 2021, Monthly Notices of the Royal Astronomical Society, 504, 4377

\bibitem[{Tadhunter {et~al.}(2018)Tadhunter, Rodr{\'\i}guez~Zaur{\'\i}n, Rose, Spence, Batcheldor, Berg, Ramos~Almeida, Spoon, Sparks, \& Chiaberge}]{tadhunter2018quantifying}
Tadhunter, C., Rodr{\'\i}guez~Zaur{\'\i}n, J., Rose, M., {et~al.} 2018, Monthly Notices of the Royal Astronomical Society, 478, 1558

\bibitem[{Villar-Mart\'in {et~al.}(2016)Villar-Mart\'in, Arribas, Emonts, Humphrey, Tadhunter, Bessiere, Cabrera~Lavers, \& Ramos~Almeida}]{villar2016ionized}
Villar-Mart\'in, M., Arribas, S., Emonts, B., {et~al.} 2016, Monthly Notices of the Royal Astronomical Society, 460, 130

\bibitem[{Villar-Mart\'in {et~al.}(2014)Villar-Mart\'in, Emonts, Humphrey, Cabrera~Lavers, \& Binette}]{villar2014triggering}
Villar-Mart\'in, M., Emonts, B., Humphrey, A., Cabrera~Lavers, A., \& Binette, L. 2014, Monthly Notices of the Royal Astronomical Society, 440, 3202

\bibitem[{Villar-Mart\'in {et~al.}(2011)Villar-Mart\'in, Humphrey, Delgado, Colina, \& Arribas}]{villar2011ionized}
Villar-Mart\'in, M., Humphrey, A., Delgado, R.~G., Colina, L., \& Arribas, S. 2011, Monthly Notices of the Royal Astronomical Society, 418, 2032

\bibitem[{Whittle(1985)}]{whittle1985narrow}
Whittle, M. 1985, Monthly Notices of the Royal Astronomical Society, 213, 1

\bibitem[{Whittle(1992)}]{whittle1992virial}
Whittle, M. 1992, Astrophysical Journal v. 387, p. 121, 387, 121

\bibitem[{{Winkel} {et~al.}(2023){Winkel}, {Husemann}, {Singha}, {Bennert}, {Combes}, {Davis}, {Gaspari}, {Jahnke}, {McElroy}, {O'Dea}, \& {P{\'e}rez-Torres}}]{2023A&A...670A...3W}
{Winkel}, N., {Husemann}, B., {Singha}, M., {et~al.} 2023, \aap, 670, A3

\bibitem[{Wylezalek \& Morganti(2018)}]{wylezalek2018questions}
Wylezalek, D. \& Morganti, R. 2018, Nature Astronomy, 2, 181

\bibitem[{York {et~al.}(2000)York, Adelman, Anderson~Jr, {et~al.}}]{york2000aj}
York, D., Adelman, J., Anderson~Jr, J., {et~al.} 2000, MNRAS, 414, 940

\bibitem[{Zakamska \& Greene(2014)}]{zakamska2014quasar}
Zakamska, N.~L. \& Greene, J.~E. 2014, Monthly Notices of the Royal Astronomical Society, 442, 784

\bibitem[{Zakamska {et~al.}(2003)Zakamska, Strauss, Krolik, Collinge, Hall, Hao, Heckman, Ivezi{\'c}, Richards, Schlegel, {et~al.}}]{zakamska2003candidate}
Zakamska, N.~L., Strauss, M.~A., Krolik, J.~H., {et~al.} 2003, The Astronomical Journal, 126, 2125

\bibitem[{{Zubovas} \& {King}(2016)}]{2016MNRAS.462.4055Z}
{Zubovas}, K. \& {King}, A. 2016, \mnras, 462, 4055

\bibitem[{Zubovas {et~al.}(2013)Zubovas, Nayakshin, King, \& Wilkinson}]{zubovas2013agn}
Zubovas, K., Nayakshin, S., King, A., \& Wilkinson, M. 2013, Monthly Notices of the Royal Astronomical Society, 433, 3079

\end{thebibliography}


\begin{appendix}

\section{Parametric analysis}
\label{Appendix par}

In Figure \ref{Fig_parametric1} we include all the fits performed using the parametric method (except the one shown in Figure \ref{Fig1}), which is described in full in Section \ref{parametric}. The parameters of the kinematic components fitted to the [OIII] lines are listed in Table \ref{Tab_parametric}, and the outflow physical properties in Table \ref{Tab_parametric_energetics}.

\begin{figure*}[hb]
\centering
    {\par\includegraphics[width=0.45\textwidth]{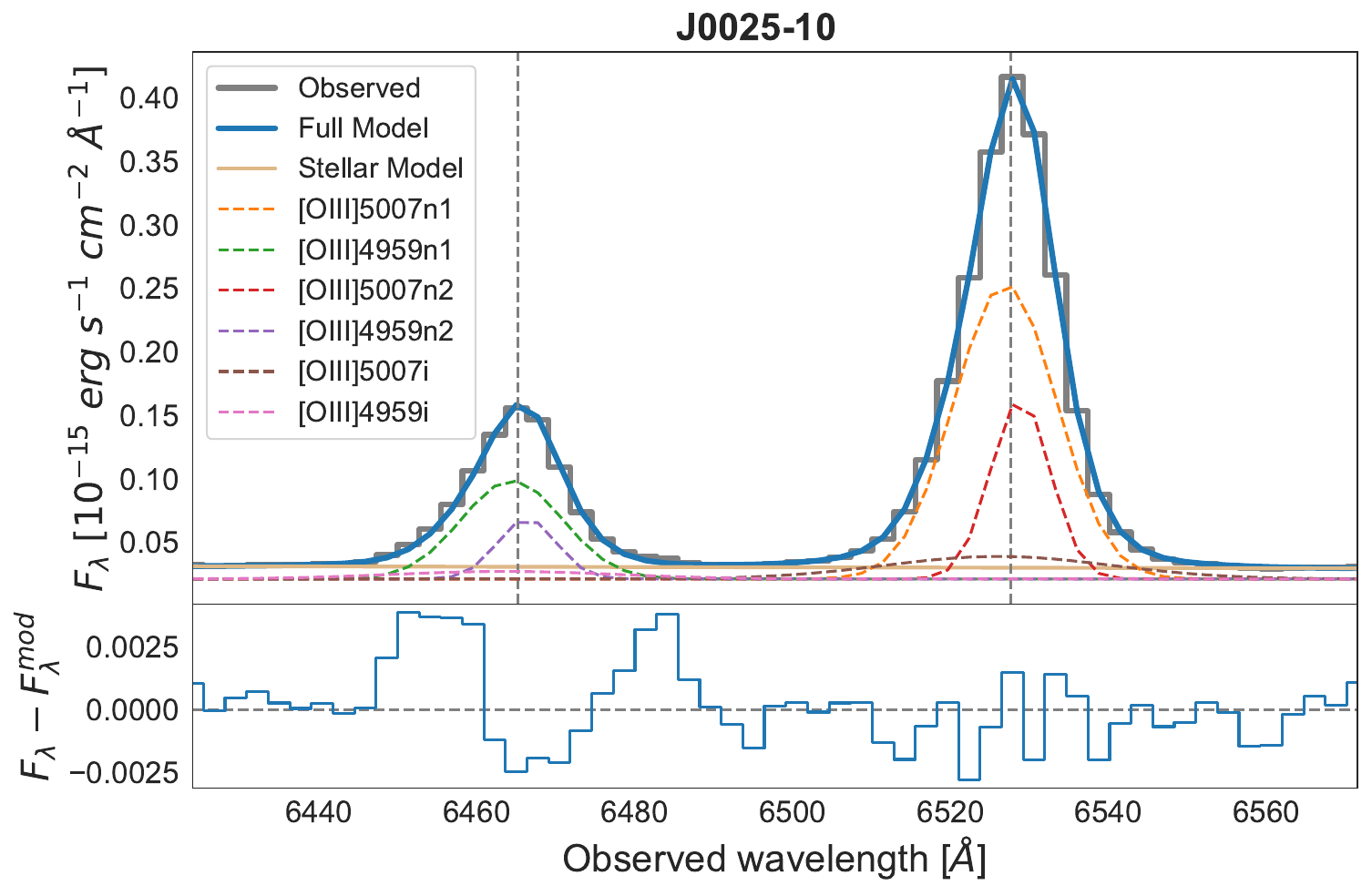}
    \includegraphics[width=0.45\textwidth]{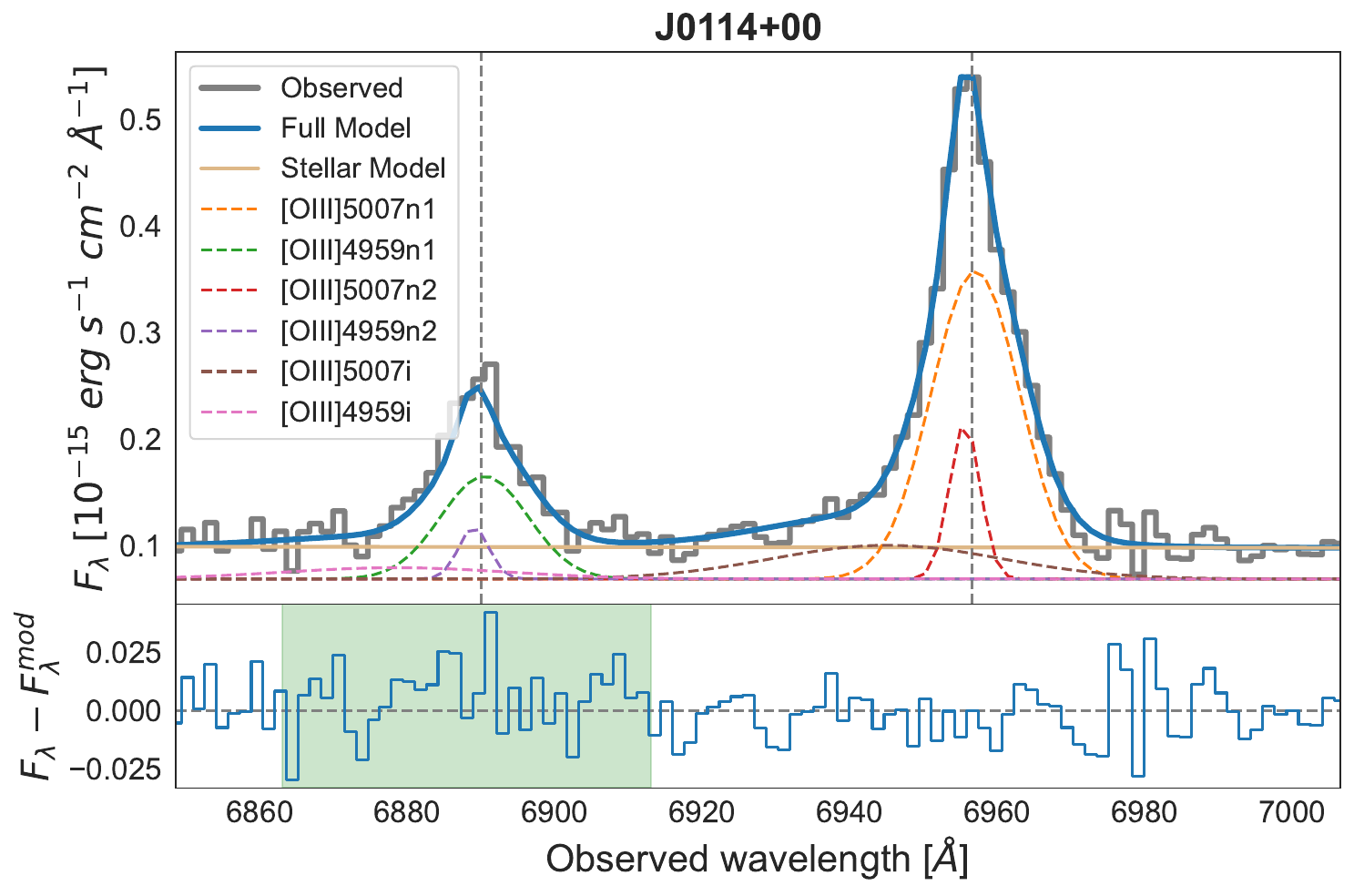}
    \includegraphics[width=0.45\textwidth]{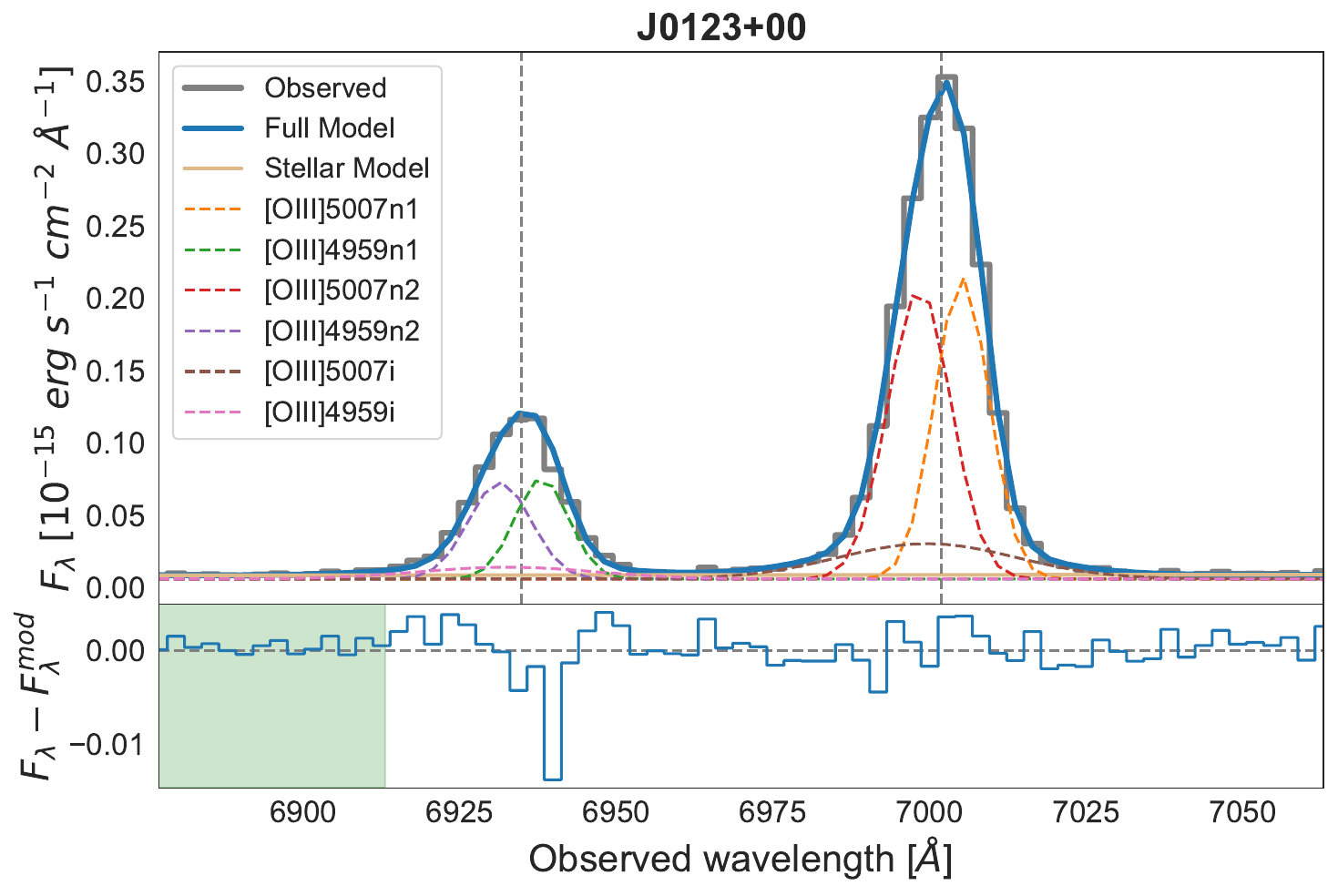}
    \includegraphics[width=0.45\textwidth]{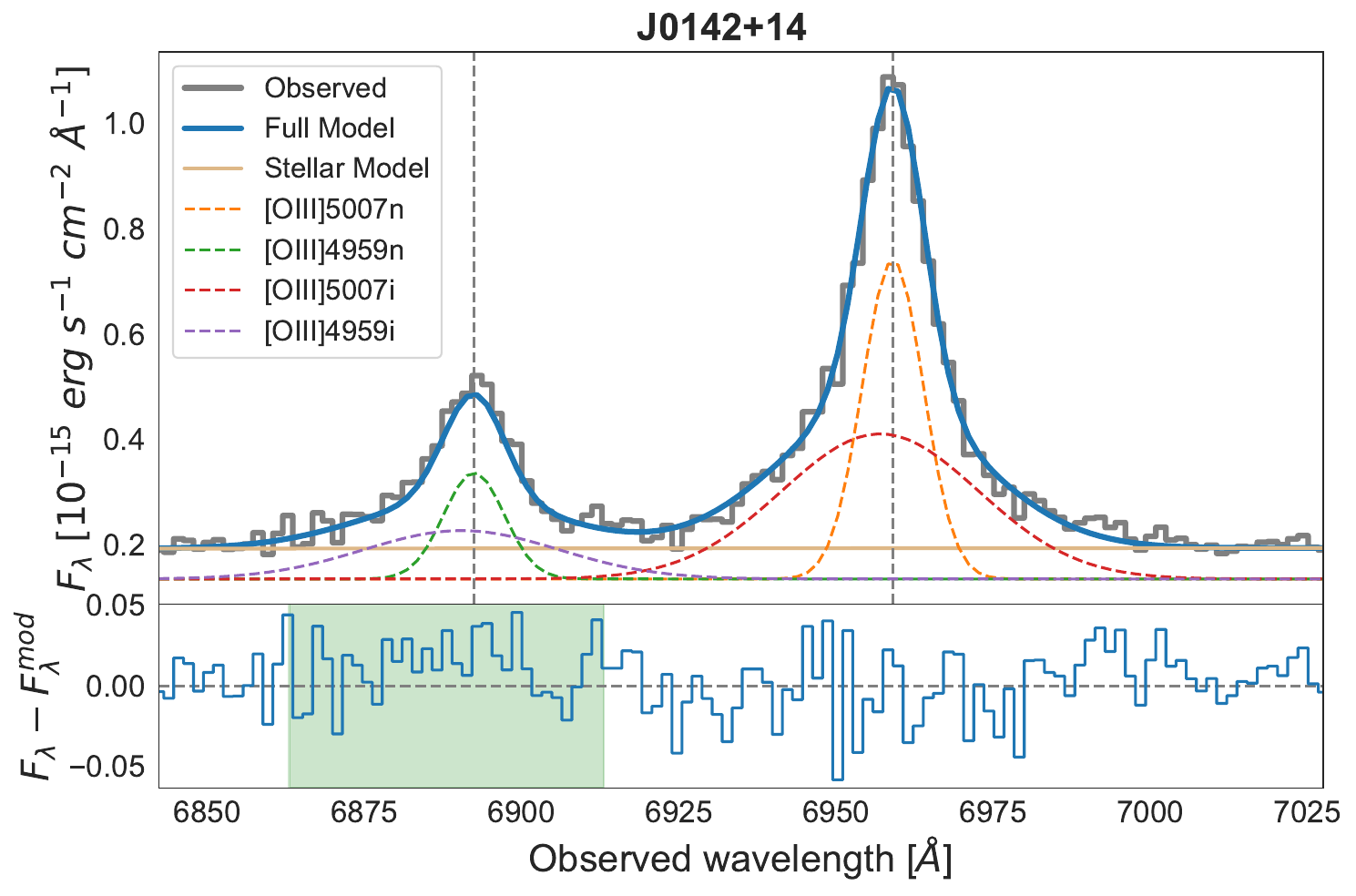}
    \includegraphics[width=0.45\textwidth]{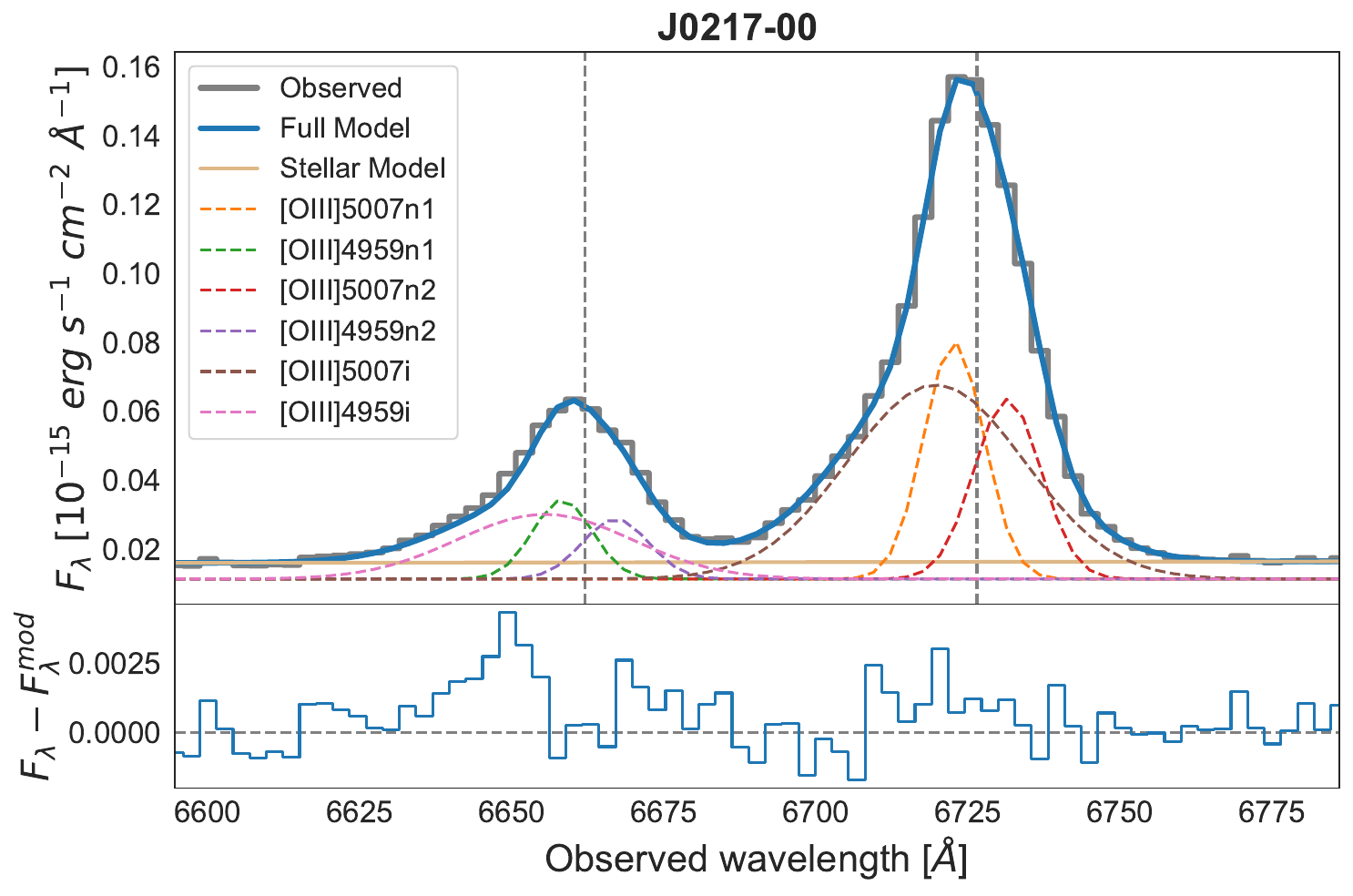}
    \includegraphics[width=0.45\textwidth]{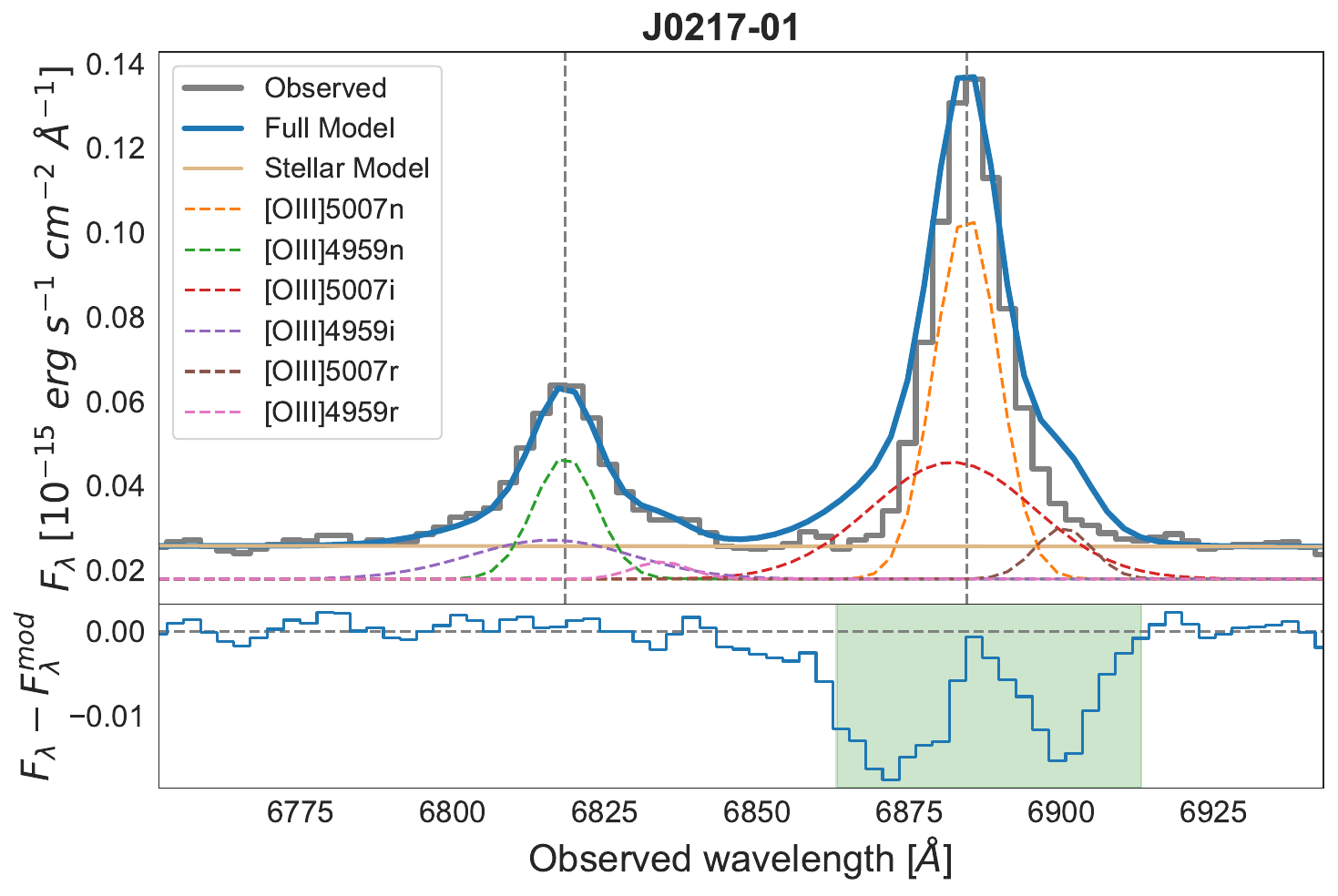}
    \includegraphics[width=0.45\textwidth]{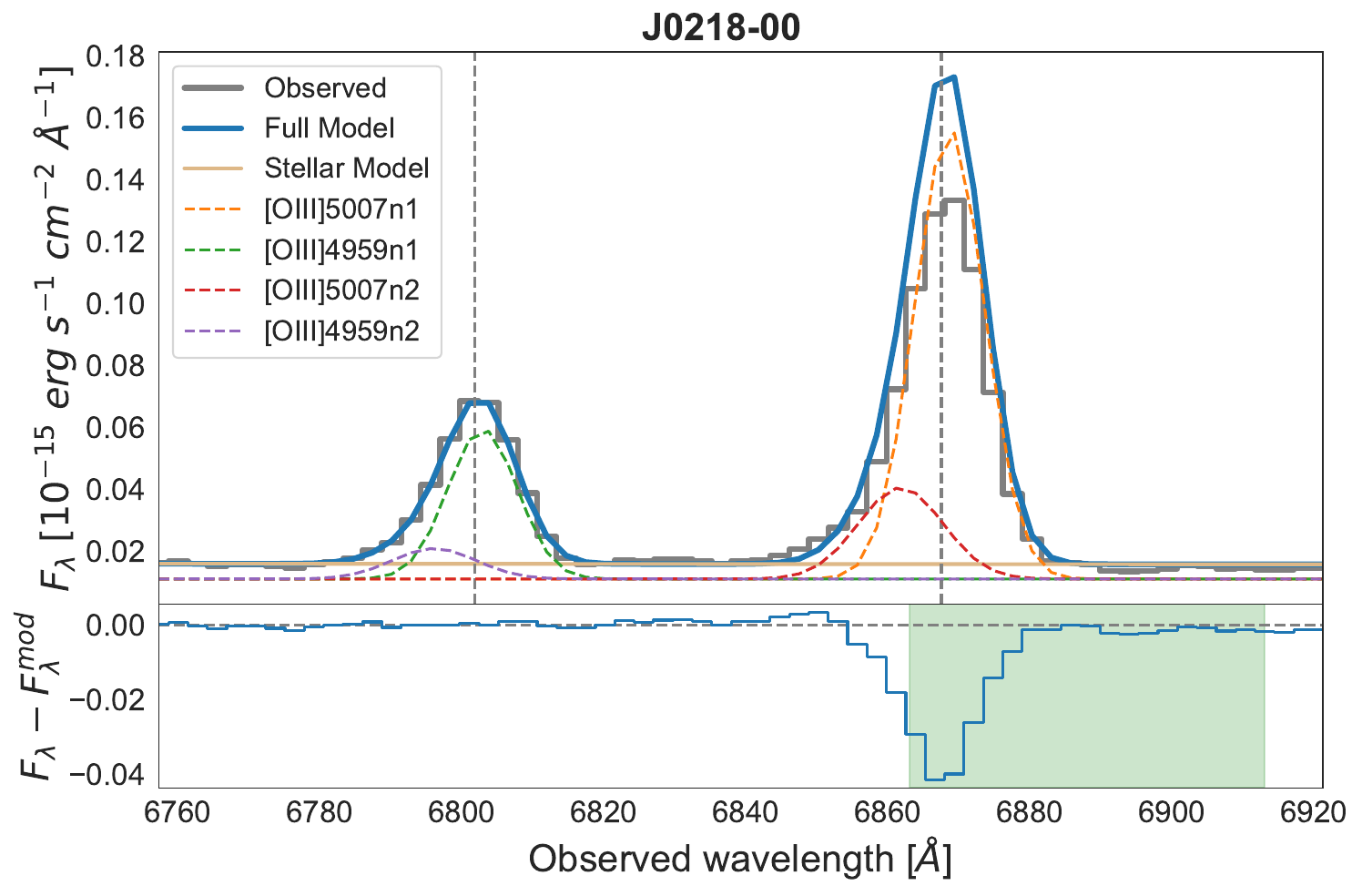}
    \includegraphics[width=0.45\textwidth]{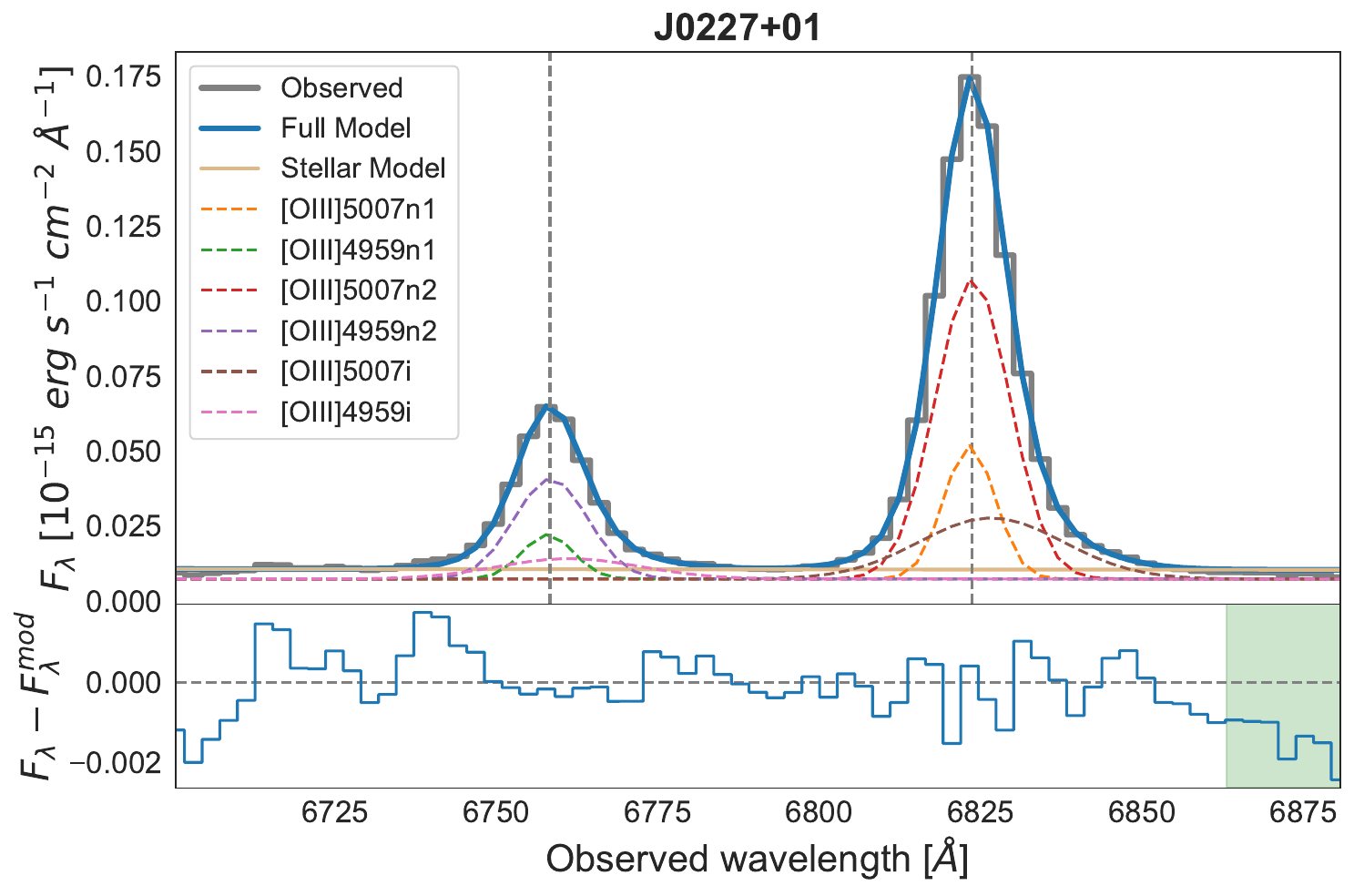} 
    \par}
\caption{Parametric fits of the [OIII] doublet for the whole QSO2 sample. Figures are identic al to the example case in Figure \ref{Fig1}.}
\label{Fig_parametric1}
\end{figure*}

\setcounter{figure}{0}
\begin{figure*}
\centering
    {\par
    \includegraphics[width=0.45\textwidth]{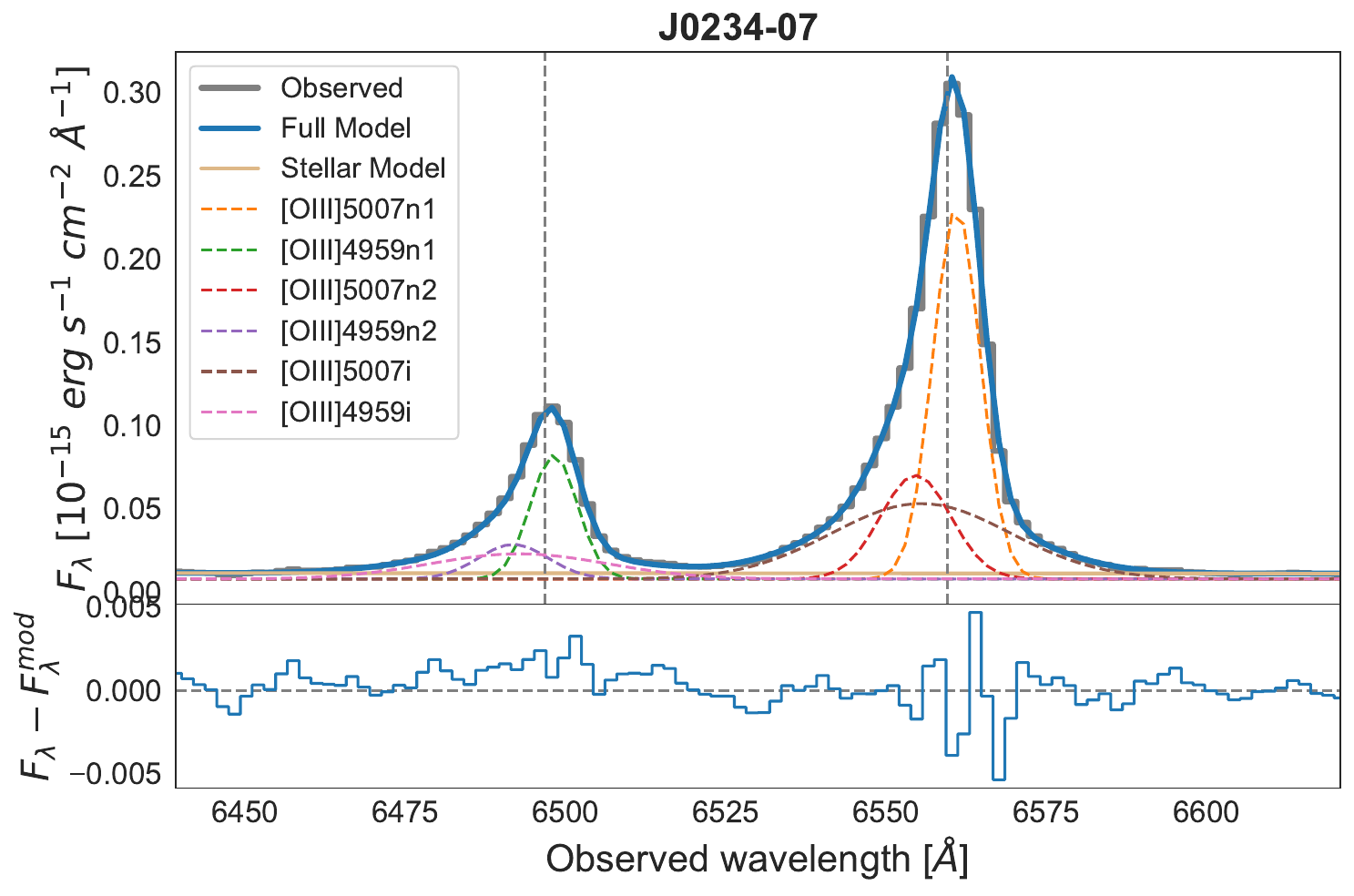}
    \includegraphics[width=0.45\textwidth]{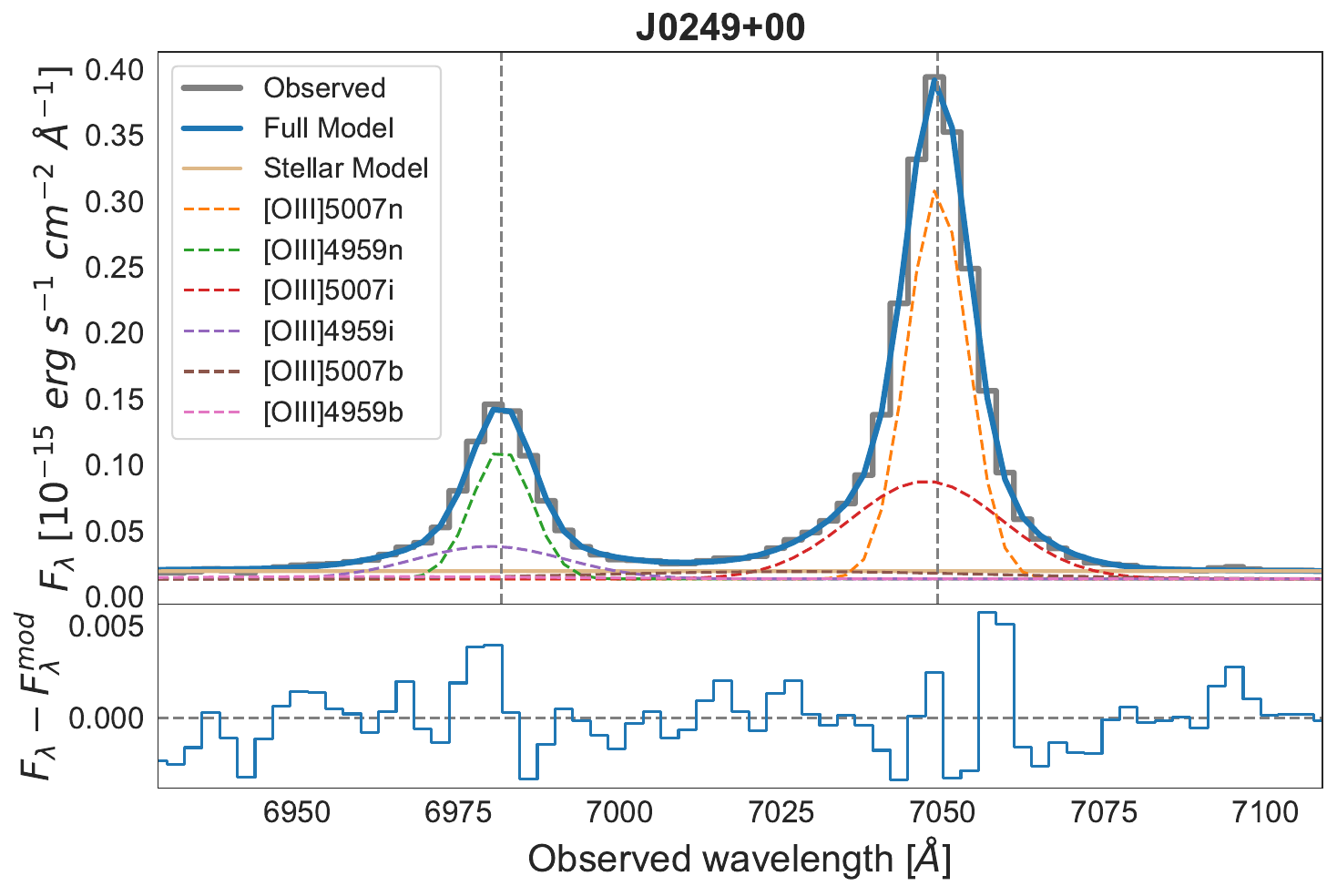}
    \includegraphics[width=0.45\textwidth]{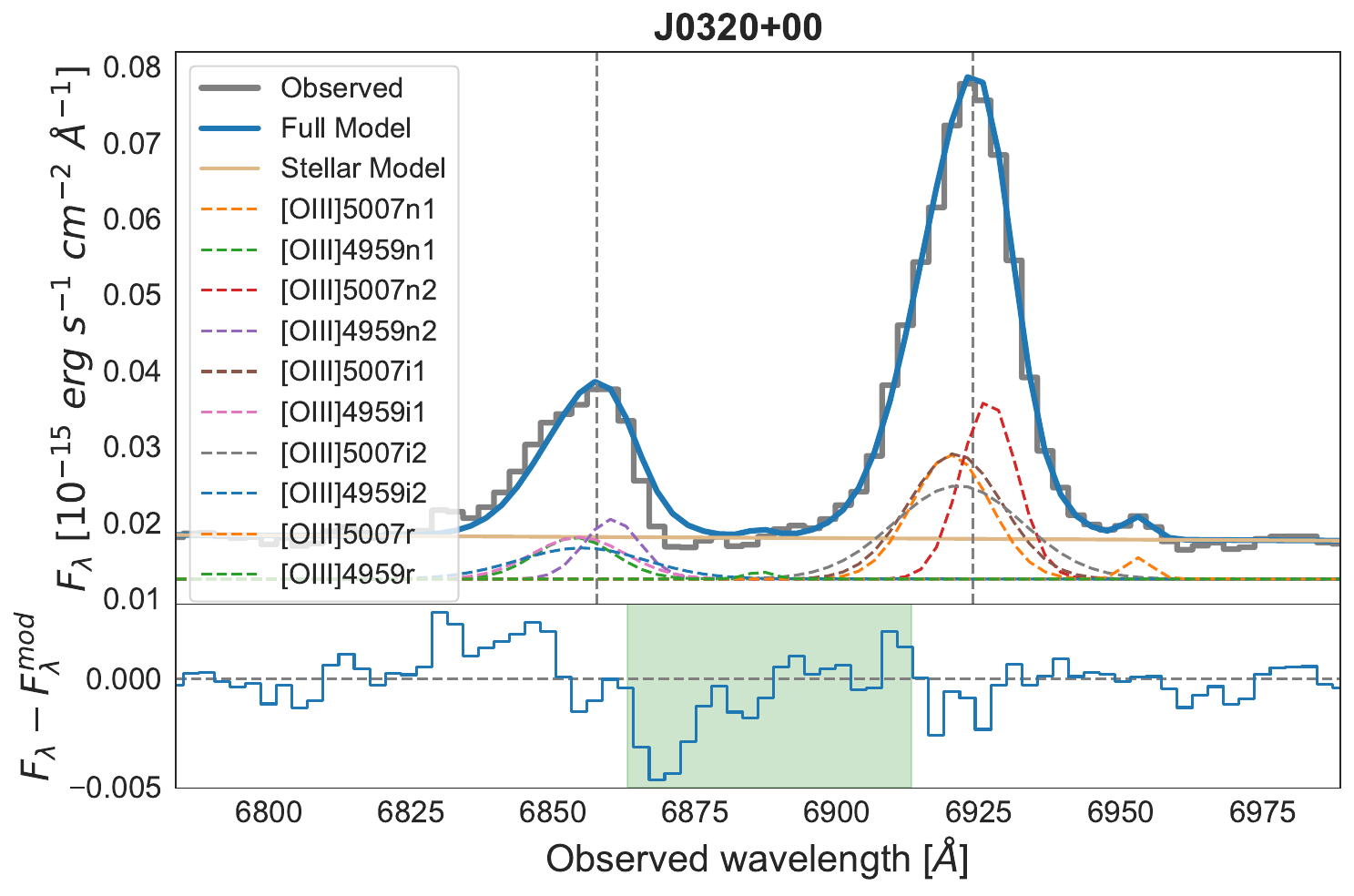}
    \includegraphics[width=0.45\textwidth]{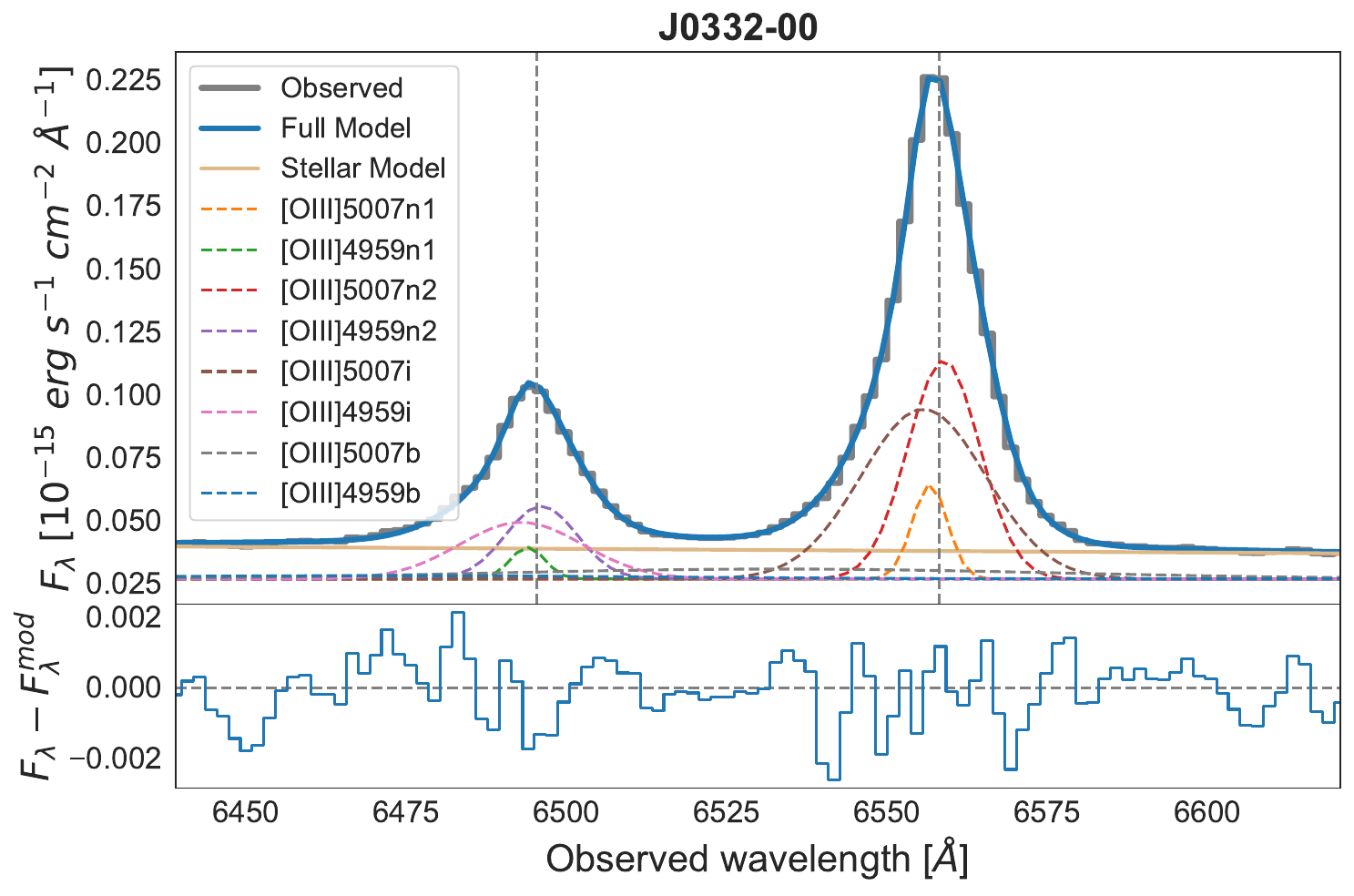}
    \includegraphics[width=0.45\textwidth]{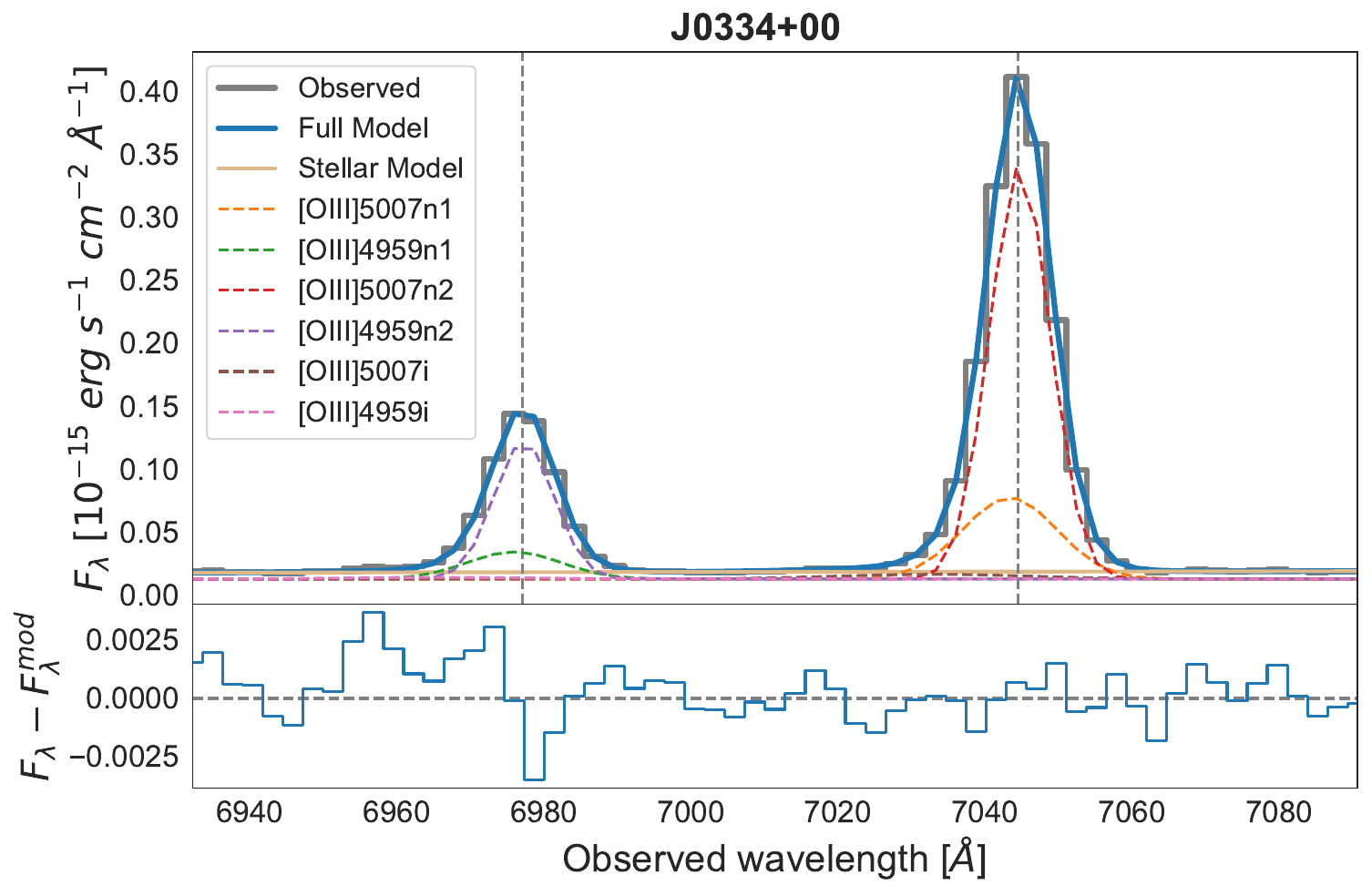}
    \includegraphics[width=0.45\textwidth]{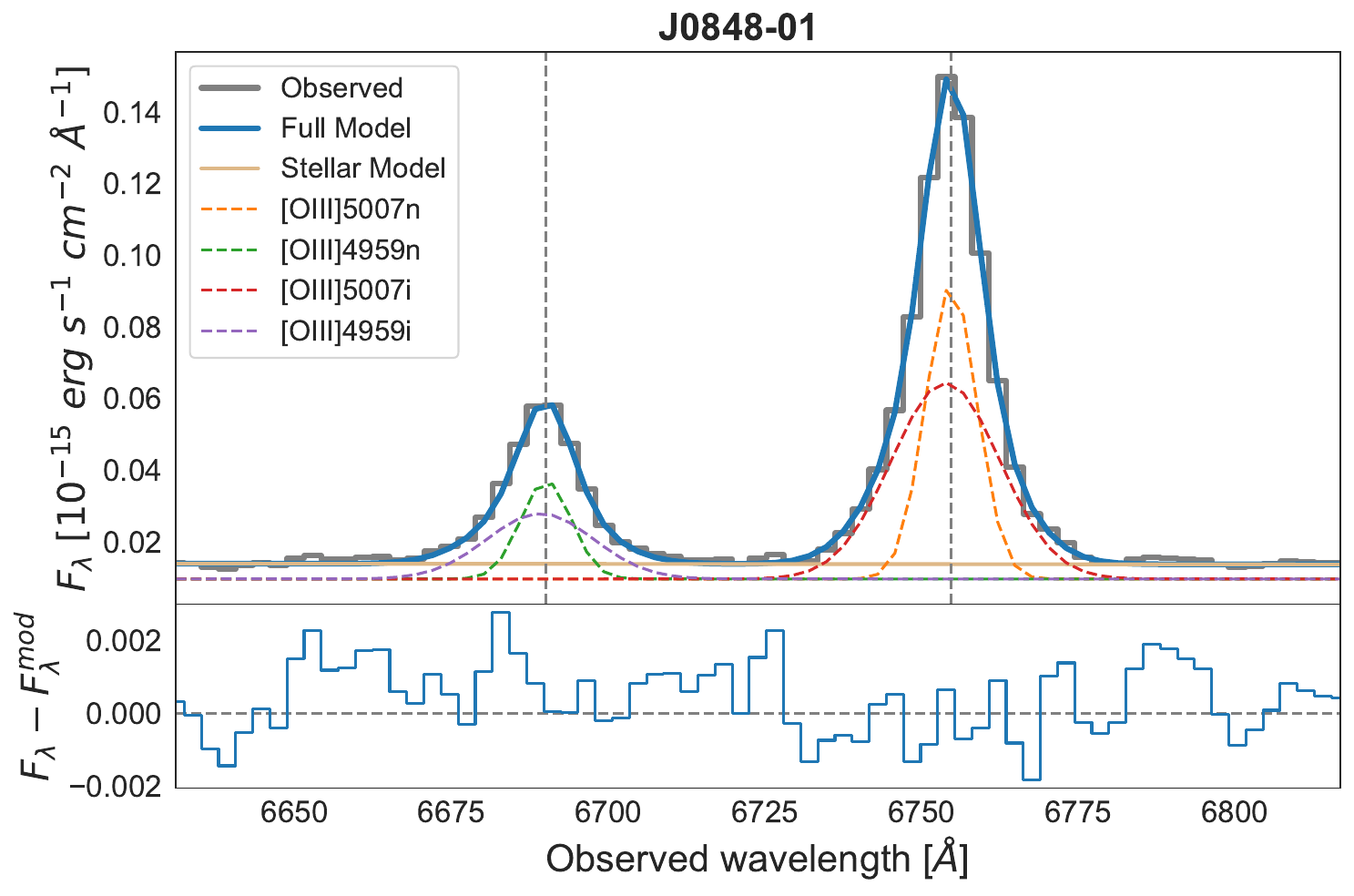}
    \includegraphics[width=0.45\textwidth]{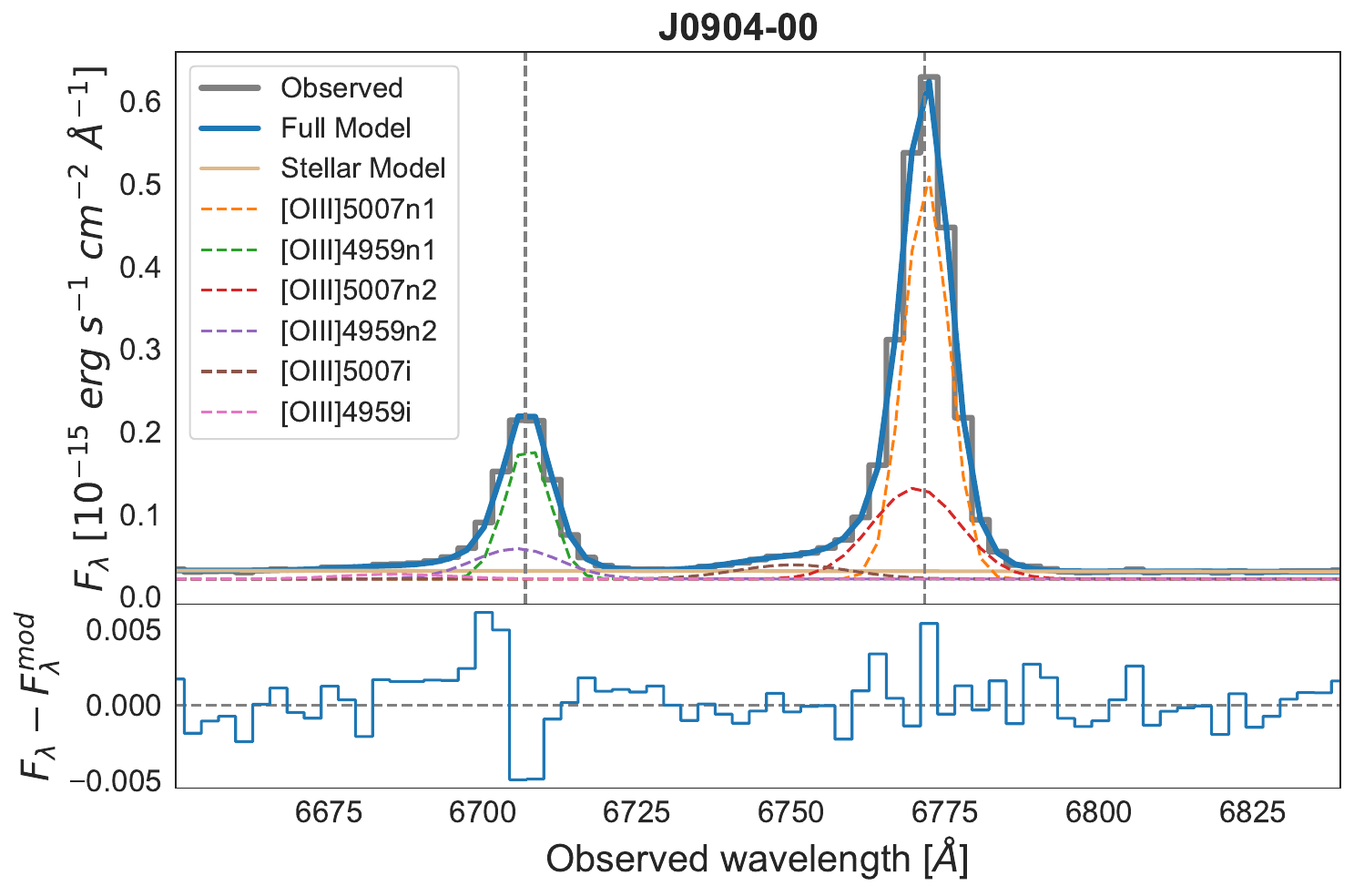}
    \includegraphics[width=0.45\textwidth]{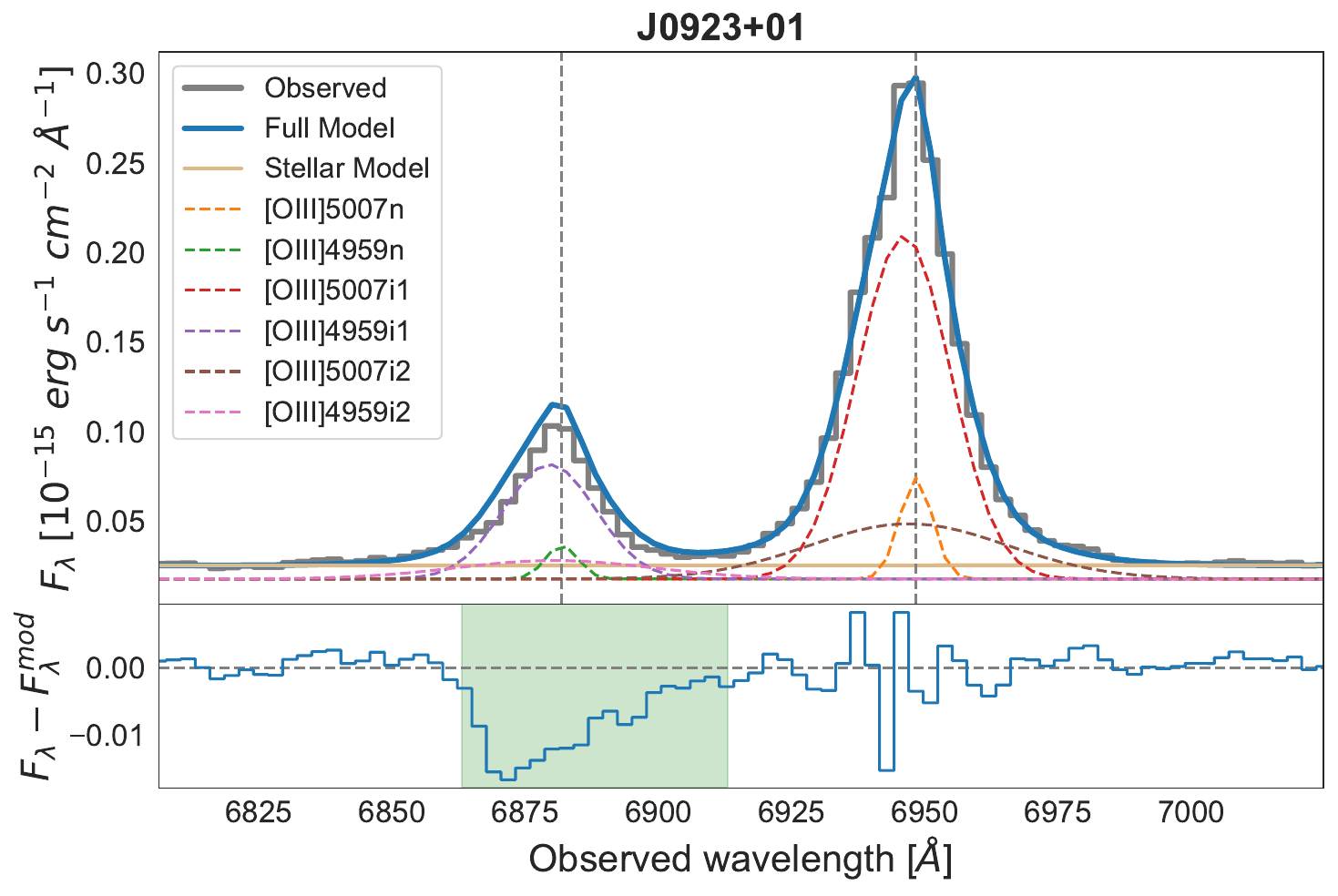}\par}
\caption{continued}
\end{figure*}

\setcounter{figure}{0}
\begin{figure*}
\centering
    {\par\includegraphics[width=0.45\textwidth]{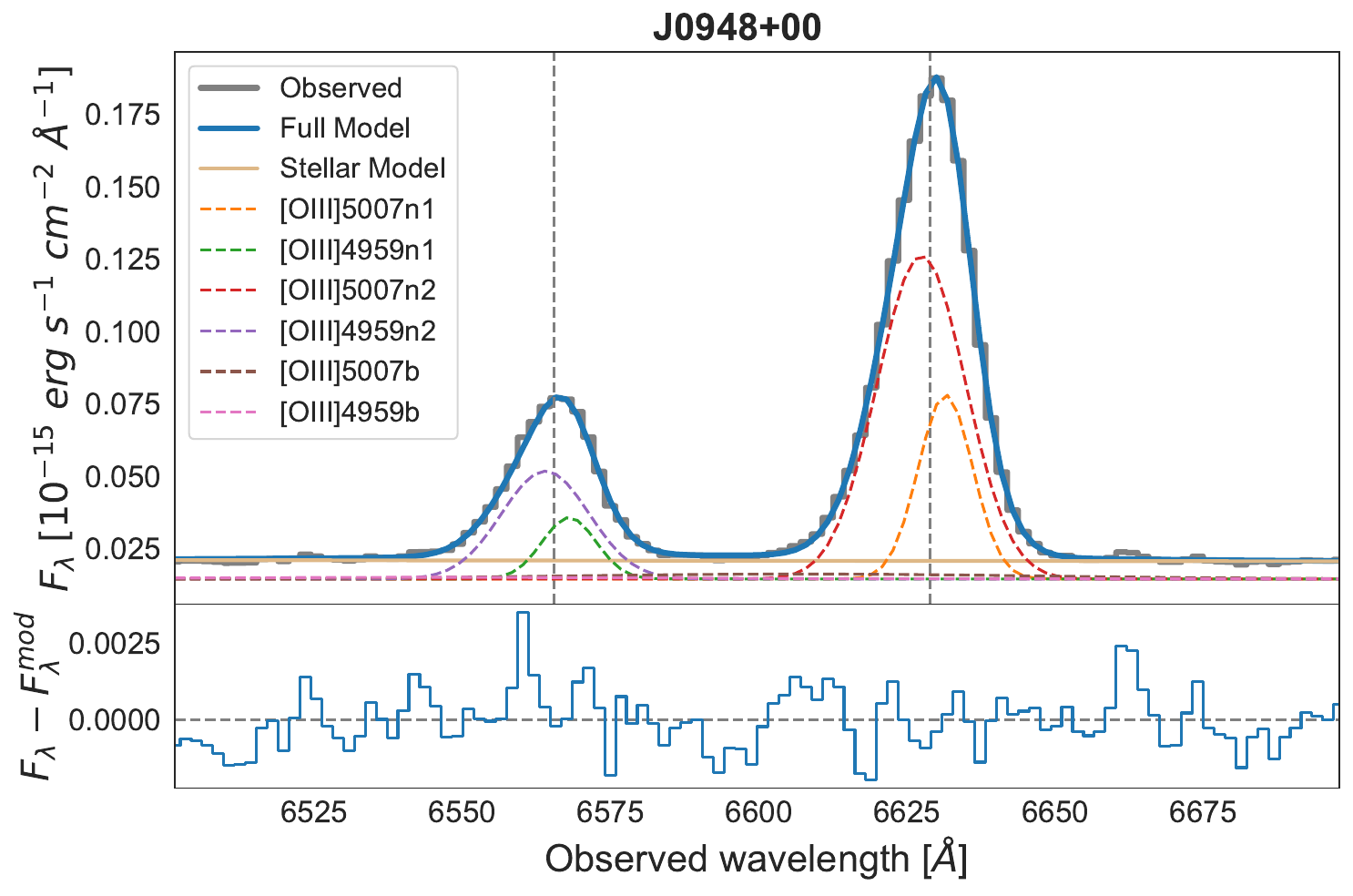}
    \includegraphics[width=0.45\textwidth]{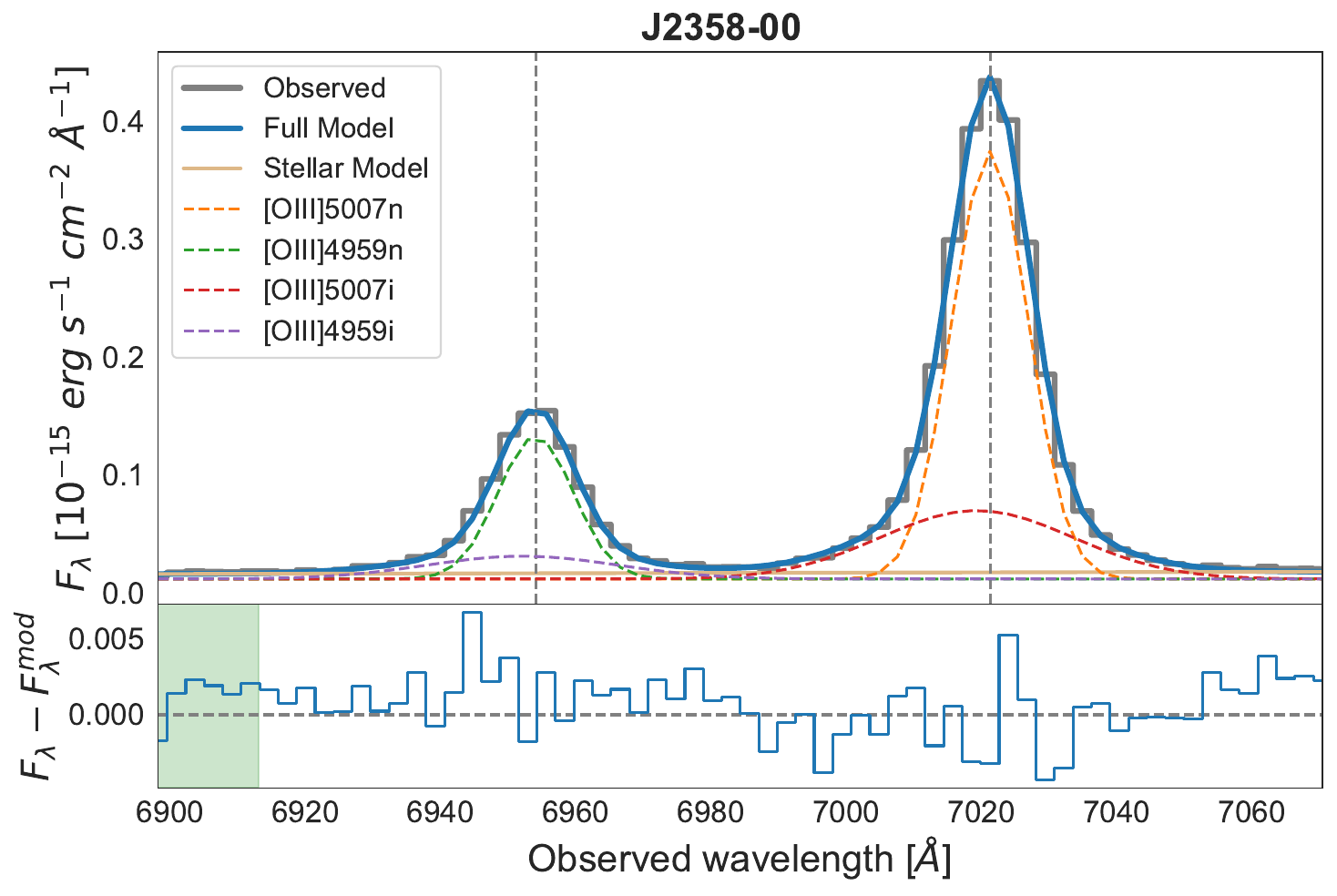}\par}
\caption{continued}
\end{figure*}


\begin{table*}[htb]
\caption{Results from the parametric fits of the $[OIII]\lambda$5007 line.}
\centering
\resizebox{!}{0.39\paperheight}{

\begin{tabular}{cccccc} \hline
Component & F$_{\lambda}$ & FWHM & v$_s$ & L$_{\rm [OIII]}$ & Luminosity\\
 & (erg cm$^{-2}$ s$^{-1}$) & (km s$^{-1}$) & (km s$^{-1}$) & (erg s$^{-1}$) & fraction \\ \hline
 
\multicolumn{6}{l}{\textbf{J0025-10}} \\ \hline
n1 & (3.76$\pm$0.12)$\cdot$10$^{-15}$ & 700$\pm$11 (280) & -33.8$\pm$3.0 & (1.109$\pm$0.035)$\cdot$10$^{42}$ & 0.655$\pm$0.020 \\
n2 & (1.348$\pm$0.065)$\cdot$10$^{-15}$ & 408.0$\pm$7.3 (280) & 54.9$\pm$2.5 & (3.98$\pm$0.19)$\cdot$10$^{41}$ & 0.235$\pm$0.011 \\
i & (6.3$\pm$1.4)$\cdot$10$^{-16}$ & 1540$\pm$110 (280) & -73$\pm$25 & (1.87$\pm$0.42)$\cdot$10$^{41}$ & 0.110$\pm$0.025 \\\hline
 
\multicolumn{6}{l}{\textbf{J0114+00}} \\ \hline
n1 & (4.30$\pm$0.25)$\cdot$10$^{-15}$ & 606$\pm$22 (120) & 21.5$\pm$9.9 & (2.27$\pm$0.13)$\cdot$10$^{42}$ & 0.680$\pm$0.040 \\
n2 & (7.65$\pm$0.88)$\cdot$10$^{-16}$ & 210$\pm$15 (120) & -41.8$\pm$6.5 & (4.05$\pm$0.47)$\cdot$10$^{41}$ & 0.121$\pm$0.014 \\
i & (1.25$\pm$0.39)$\cdot$10$^{-15}$ & 1560$\pm$280 (120) & -477$\pm$180 & (6.6$\pm$2.0)$\cdot$10$^{41}$ & 0.198$\pm$0.061 \\\hline
 
\multicolumn{6}{l}{\textbf{J0123+00}} \\ \hline
n1 & (2.29$\pm$0.19)$\cdot$10$^{-15}$ & 437.0$\pm$7.7 (240) & 142.0$\pm$9.2 & (1.28$\pm$0.11)$\cdot$10$^{42}$ & 0.404$\pm$0.033 \\
n2 & (2.50$\pm$0.19)$\cdot$10$^{-15}$ & 505$\pm$17 (240) & -150$\pm$16 & (1.40$\pm$0.11)$\cdot$10$^{42}$ & 0.442$\pm$0.034 \\
i & (8.74$\pm$1.3)$\cdot$10$^{-16}$ & 1440$\pm$83 (240) & -94$\pm$19 & (4.90$\pm$0.75)$\cdot$10$^{41}$ & 0.154$\pm$0.024 \\\hline
 
\multicolumn{6}{l}{\textbf{J0142+14}} \\ \hline
n & (7.36$\pm$0.30)$\cdot$10$^{-15}$ & 493$\pm$15 (120) & 0.0$\pm$3.5 & (3.89$\pm$0.16)$\cdot$10$^{42}$ & 0.405$\pm$0.016 \\
i & (1.081$\pm$0.077)$\cdot$10$^{-1}$4 & 1590$\pm$56 (120) & -85$\pm$15 & (5.72$\pm$0.41)$\cdot$10$^{42}$ & 0.595$\pm$0.042 \\\hline
 
\multicolumn{6}{l}{\textbf{J0217-00}} \\ \hline
n1 & (8.3$\pm$3.2)$\cdot$10$^{-16}$ & 515$\pm$91 (310) & -170$\pm$58 & (3.3$\pm$1.3)$\cdot$10$^{41}$ & 0.229$\pm$0.087 \\
n2 & (7.3$\pm$3.0)$\cdot$10$^{-16}$ & 581$\pm$110 (310) & 218$\pm$120 & (2.9$\pm$1.2)$\cdot$10$^{41}$ & 0.202$\pm$0.084 \\
i & (2.06$\pm$0.19)$\cdot$10$^{-15}$ & 1530$\pm$40 (310) & -295$\pm$32 & (8.16$\pm$0.77)$\cdot$10$^{41}$ & 0.569$\pm$0.053 \\\hline
 
\multicolumn{6}{l}{\textbf{J0217-01}} \\ \hline
n & (1.13$\pm$0.29)$\cdot$10$^{-15}$ & 538$\pm$82 (280) & 0$\pm$19 & (5.5$\pm$1.4)$\cdot$10$^{41}$ & 0.51$\pm$0.13 \\
i & (9.47$\pm$5.7)$\cdot$10$^{-16}$ & 1380$\pm$340 (280) & -100$\pm$180 & (4.6$\pm$2.7)$\cdot$10$^{41}$ & 0.43$\pm$0.26 \\
r & (1.3$\pm$2.1)$\cdot$10$^{-16}$ & 460$\pm$200 (280) & 700$\pm$200 & (7$\pm$10)$\cdot$10$^{40}$ & 0.061$\pm$0.094 \\\hline
 
\multicolumn{6}{l}{\textbf{J0218-00}} \\ \hline
n1 & (1.8$\pm$0.2)$\cdot$10$^{-15}$ & 507$\pm$21 (280) & 51$\pm$16 & (8.54$\pm$0.95)$\cdot$10$^{41}$ & 0.804$\pm$0.089 \\
n2 & (4.4$\pm$2.1)$\cdot$10$^{-16}$ & 611$\pm$58 (280) & -250$\pm$110 & (2.1$\pm$1.0)$\cdot$10$^{41}$ & 0.196$\pm$0.096 \\\hline
 
\multicolumn{6}{l}{\textbf{J0227+01}} \\ \hline
n1 & (4.5$\pm$2.4)$\cdot$10$^{-16}$ & 414$\pm$51 (220) & -14.2$\pm$8.0 & (2.0$\pm$1.1)$\cdot$10$^{41}$ & 0.184$\pm$0.097 \\
n2 & (1.43$\pm$0.32)$\cdot$10$^{-15}$ & 593$\pm$30 (220) & 6.4$\pm$8.0 & (6.4$\pm$1.4)$\cdot$10$^{41}$ & 0.59$\pm$0.13 \\
i & (5.7$\pm$1.0)$\cdot$10$^{-16}$ & 1160$\pm$55 (220) & 133$\pm$24 & (2.55$\pm$0.45)$\cdot$10$^{41}$ & 0.231$\pm$0.041 \\\hline
 
\multicolumn{6}{l}{\textbf{J0234-07}} \\ \hline
n1 & (2.059$\pm$0.031)$\cdot$10$^{-15}$ & 396.0$\pm$2.0 (320) & 63.68$\pm$0.87 & (6.4$\pm$0.095)$\cdot$10$^{41}$ & 0.4570$\pm$0.0070 \\
n2 & (8.55$\pm$0.33)$\cdot$10$^{-16}$ & 590$\pm$13 (320) & -228$\pm$11 & (2.66$\pm$0.10)$\cdot$10$^{41}$ & 0.1900$\pm$0.0070 \\
i & (1.587$\pm$0.048)$\cdot$10$^{-15}$ & 1500$\pm$18 (320) & -189.0$\pm$4.1 & (4.94$\pm$0.15)$\cdot$10$^{41}$ & 0.353$\pm$0.011 \\\hline
 
\multicolumn{6}{l}{\textbf{J0249+00}} \\ \hline
n & (3.453$\pm$0.049)$\cdot$10$^{-15}$ & 467.1$\pm$3.7 (240) & 0.00$\pm$0.59 & (2.043$\pm$0.029)$\cdot$10$^{42}$ & 0.5700$\pm$0.0080 \\
i & (2.18$\pm$0.11)$\cdot$10$^{-15}$ & 1178$\pm$38 (240) & -77.2$\pm$9.2 & (1.289$\pm$0.065)$\cdot$10$^{42}$ & 0.360$\pm$0.018 \\
b & (4.23$\pm$0.96)$\cdot$10$^{-16}$ & 3100$\pm$380 (240) & -950$\pm$180 & (2.50$\pm$0.57)$\cdot$10$^{41}$ & 0.070$\pm$0.016 \\ \hline
 
\multicolumn{6}{l}{\textbf{J0320+00}} \\ \hline
n1 & (2.8$\pm$2.4)$\cdot$10$^{-16}$ & 710$\pm$110 (250) & -141$\pm$96 & (1.4$\pm$1.2)$\cdot$10$^{41}$ & 0.22$\pm$0.19 \\
n2 & (3.0$\pm$2.3)$\cdot$10$^{-16}$ & 500$\pm$120 (250) & -116$\pm$54 & (1.53$\pm$1.2)$\cdot$10$^{41}$ & 0.23$\pm$0.18 \\
i1 & (3.4$\pm$2.4)$\cdot$10$^{-16}$ & 840$\pm$110 (250) & -141$\pm$96 & (1.8$\pm$1.2)$\cdot$10$^{41}$ & 0.27$\pm$0.19 \\
i2 & (3.4$\pm$2.3)$\cdot$10$^{-16}$ & 1150$\pm$120 (250) & -116$\pm$54 & (1.8$\pm$1.2)$\cdot$10$^{41}$ & 0.27$\pm$0.18 \\
r & (1.7$\pm$1.1)$\cdot$10$^{-17}$ & 246.0$\pm$0.0 (250) & 1260$\pm$24 & (8.7$\pm$5.4)$\cdot$10$^{39}$ & 0.0130$\pm$0.0080 \\\hline
 
\multicolumn{6}{l}{\textbf{J0332-00}} \\ \hline
n1 & (2.56$\pm$0.50)$\cdot$10$^{-16}$ & 293$\pm$30 (170) & -70.2$\pm$6.8 & (8.0$\pm$1.5)$\cdot$10$^{40}$ & 0.072$\pm$0.014 \\
n2 & (1.22$\pm$0.22)$\cdot$10$^{-15}$ & 604$\pm$26 (170) & 30$\pm$20 & (3.79$\pm$0.67)$\cdot$10$^{41}$ & 0.343$\pm$0.061 \\
i & (1.65$\pm$0.27)$\cdot$10$^{-15}$ & 1050$\pm$46 (170) & -116$\pm$14 & (5.13$\pm$0.83)$\cdot$10$^{41}$ & 0.464$\pm$0.075 \\
b & (4.33$\pm$0.51)$\cdot$10$^{-16}$ & 4750$\pm$300 (170) & -1140$\pm$150 & (1.35$\pm$0.16)$\cdot$10$^{41}$ & 0.122$\pm$0.014 \\\hline
 
\multicolumn{6}{l}{\textbf{J0334+00}} \\ \hline
n1 & (1.06$\pm$0.17)$\cdot$10$^{-15}$ & 645$\pm$28 (210) & -47$\pm$11 & (6.3$\pm$1.0)$\cdot$10$^{41}$ & 0.234$\pm$0.038 \\
n2 & (3.36$\pm$0.11)$\cdot$10$^{-15}$ & 411.0$\pm$3.7 (210) & 9.47$\pm$0.94 & (1.974$\pm$0.064)$\cdot$10$^{42}$ & 0.739$\pm$0.024 \\
i & (1.22$\pm$0.44)$\cdot$10$^{-16}$ & 1310$\pm$230 (210) & -490$\pm$210 & (7.16$\pm$2.6)$\cdot$10$^{40}$ & 0.027$\pm$0.0 \\\hline
 
\multicolumn{6}{l}{\textbf{J0848-01}} \\ \hline
n & (8.37$\pm$0.83)$\cdot$10$^{-16}$ & 426$\pm$21 (210) & 0.0$\pm$3.7 & (3.45$\pm$0.34)$\cdot$10$^{41}$ & 0.414$\pm$0.041 \\
i & (1.19$\pm$0.17)$\cdot$10$^{-15}$ & 899$\pm$45 (210) & -41.0$\pm$8.6 & (4.89$\pm$0.70)$\cdot$10$^{41}$ & 0.586$\pm$0.084 \\ \hline
 
\multicolumn{6}{l}{\textbf{J0904-00}} \\ \hline
n1 & (4.43$\pm$0.13)$\cdot$10$^{-15}$ & 376.0$\pm$5.3 (170) & 14.2$\pm$1.0 & (1.86$\pm$0.054)$\cdot$10$^{42}$ & 0.652$\pm$0.019 \\
n2 & (1.98$\pm$0.26)$\cdot$10$^{-15}$ & 744$\pm$29 (170) & -63$\pm$13 & (8.3$\pm$1.1)$\cdot$10$^{41}$ & 0.291$\pm$0.038 \\
i & (3.86$\pm$0.81)$\cdot$10$^{-16}$ & 940$\pm$180 (170) & -949$\pm$91 & (1.62$\pm$0.34)$\cdot$10$^{41}$ & 0.057$\pm$0.012 \\\hline
 
\multicolumn{6}{l}{\textbf{J0923+01}} \\ \hline
n & (4.3$\pm$1.6)$\cdot$10$^{-16}$ & 311$\pm$32 (220) & 0$\pm$11 & (2.26$\pm$0.84)$\cdot$10$^{41}$ & 0.071$\pm$0.027 \\
i1 & (4.20$\pm$0.97)$\cdot$10$^{-15}$ & 890$\pm$27 (220) & -98.6$\pm$4.5 & (2.18$\pm$0.38)$\cdot$10$^{42}$ & 0.69$\pm$0.12 \\
i2 & (1.4$\pm$1.5)$\cdot$10$^{-15}$ & 1890$\pm$250 (220) & -61$\pm$24 & (7.5$\pm$7.8)$\cdot$10$^{41}$ & 0.24$\pm$0.25 \\
\hline
 
\multicolumn{6}{l}{\textbf{J0924+01}} \\ \hline
n & (1.128$\pm$0.014)$\cdot$10$^{-15}$ & 696.0$\pm$6.1 (290) & 0.0$\pm$2.1 & (5.64$\pm$0.07)$\cdot$10$^{41}$ & 0.556$\pm$0.007 \\
i & (8.99$\pm$0.29)$\cdot$10$^{-16}$ & 2010$\pm$36 (290) & -593$\pm$26 & (4.50$\pm$0.14)$\cdot$10$^{41}$ & 0.444$\pm$0.014 \\\hline
 
\multicolumn{6}{l}{\textbf{J0948+00}} \\ \hline
n1 & (7.32$\pm$0.97)$\cdot$10$^{-16}$ & 489$\pm$26 (270) & 117.0$\pm$7.6 & (2.52$\pm$0.33)$\cdot$10$^{41}$ & 0.245$\pm$0.032 \\
n2 & (2.09$\pm$0.15)$\cdot$10$^{-15}$ & 794$\pm$10 (270) & -67$\pm$11 & (7.20$\pm$0.50)$\cdot$10$^{41}$ & 0.698$\pm$0.049 \\
b & (1.70$\pm$0.29)$\cdot$10$^{-16}$ & 4320$\pm$420 (270) & -980$\pm$270 & (5.86$\pm$0.98)$\cdot$10$^{40}$ & 0.057$\pm$0.010 \\
\hline
 
\multicolumn{6}{l}{\textbf{J2358-00}} \\ \hline
n & (5.127$\pm$0.039)$\cdot$10$^{-15}$ & 567.0$\pm$2.7 (350) & 0.00$\pm$0.64 & (2.93$\pm$0.022)$\cdot$10$^{42}$ & 0.7120$\pm$0.0050 \\ 
i & (2.075$\pm$0.099)$\cdot$10$^{-15}$ & 1440$\pm$26 (350) & -85.2$\pm$6.3 & (1.18$\pm$0.051)$\cdot$10$^{42}$ & 0.288$\pm$0.012 \\
\hline
\end{tabular}}
\tablefoot{Column 1 indicates whether a component is narrow (n), intermediate (i), broad (b), or red (r) according to the definition indicated in Section \ref{parametric}. Columns 2, 3, 4, and 5 are the integrated flux, FWHM (with the instrumental width indicated between parenthesis), velocity shift relative to systemic, and integrated luminosity of each component. Column 6 indicates the fraction of each component to the total luminosity of the line.}
\label{Tab_parametric}
\end{table*}

\begin{table*}
\caption{Outflow physical properties resulting from the parametric method.}
\centering
\begin{tabular}{cc|ccc|ccc} \hline
& & & v$_{s}$ & & & v$_{max}$ \\ \hline
\textbf{Name} & \textbf{Log(M$_{\rm OF}$)} & \textbf{$\dot{\rm\bold M}_{\rm OF}$} & \textbf{Log($\dot{\rm\bold E}_{\rm kin}$)} & \textbf{$\dot{\rm\bold E}_{\rm kin}$ / $\rm \bold L_{\rm BOL}$} & \textbf{$\dot{\rm\bold M}_{\rm OF}$} & \textbf{Log($\dot{\rm\bold E}_{\rm kin}$)} & \textbf{$\dot{\rm\bold E}_{\rm kin}$ / $\rm \bold L_{\rm BOL}$}\\ 
& (M$_{\odot}$)  & (M$_{\odot}$ yr $^{-1}$) & (erg s$^{-1}$) & (\%) & (M$_{\odot}$ yr $^{-1}$) & (erg s$^{-1}$) & (\%) \\ \hline

\textbf{J0025-10} & 6.05$\pm$0.12 & 0.246$\pm$0.099 & 38.62$\pm$0.46 & (1.4$\pm$1.5)$\cdot$ 10$^{-5}$ & 4.8$\pm$1.2 & 42.5$\pm$0.15 & (9.4$\pm$3.3)$\cdot$ 10$^{-2}$ \\
\textbf{J0114+00} & 6.59$\pm$0.16 & 5.7$\pm$2.7 & 41.66$\pm$0.56 & (3.3$\pm$4.2)$\cdot$ 10$^{-2}$ & 6.9$\pm$4.3 & 41.9$\pm$0.88 & (6$\pm$12)$\cdot$ 10$^{-2}$ \\
\textbf{J0123+00} & 6.459$\pm$0.071 & 0.86$\pm$0.21 & 39.39$\pm$0.26 & (4.1$\pm$2.4)$\cdot$ 10$^{-5}$ & 11.8$\pm$2.0 & 42.8$\pm$0.1 & (1.07$\pm$0.26)$\cdot$ 10$^{-1}$ \\
\textbf{J0142+14} & 7.535$\pm$0.032 & 9.2$\pm$1.8 & 40.36$\pm$0.24 & (5.2$\pm$2.8)$\cdot$ 10$^{-4}$ & 151.0$\pm$12.0 & 44.0$\pm$0.058 & (2.22$\pm$0.29) \\
\textbf{J0217-00} & 6.688$\pm$0.039 & 4.49$\pm$0.64 & 41.11$\pm$0.15 & (2.42$\pm$0.84)$\cdot$ 10$^{-3}$ & 23.9$\pm$2.3 & 43.3$\pm$0.056 & (3.69$\pm$0.46)$\cdot$ 10$^{-1}$ \\
\textbf{J0217-01} & 6.46$\pm$0.33 & 0.9$\pm$1.4 & 39.7$\pm$1.4 & (4$\pm$13)$\cdot$ 10$^{-4}$ & 9.8$\pm$7.6 & 42.67$\pm$0.54 & (3.4$\pm$4.2)$\cdot$ 10$^{-1}$ \\
\textbf{J0227+01} & 6.187$\pm$0.082 & 0.62$\pm$0.16 & 39.53$\pm$0.26 & (1.43$\pm$0.85)$\cdot$ 10$^{-4}$ & 5.19$\pm$0.98 & 42.3$\pm$0.1 & (8.5$\pm$2.0)$\cdot$ 10$^{-2}$ \\
\textbf{J0234-07} & 6.472$\pm$0.013 & 1.719$\pm$0.063 & 40.286$\pm$0.030 & (1.70$\pm$0.12)$\cdot$ 10$^{-3}$ & 13.32$\pm$0.42 & 42.956$\pm$0.019 & (7.92$\pm$0.34)$\cdot$ 10$^{-1}$ \\
\textbf{J0249+00} & 6.965$\pm$0.024 & 6.1$\pm$1.3 & 42.07$\pm$0.27 & (3.8$\pm$2.3)$\cdot$ 10$^{-2}$ & 25.5$\pm$1.5 & 43.967$\pm$0.044 & (2.95$\pm$0.30)$\cdot$ 10$^{-1}$ \\
\textbf{J0320+00} & 6.33$\pm$0.26 & 0.80$\pm$0.59 & 39.80$\pm$0.62 & (6$\pm$10)$\cdot$ 10$^{-4}$ & 6.3$\pm$3.2 & 42.31$\pm$0.32 & (2.7$\pm$2.0)$\cdot$ 10$^{-1}$ \\
\textbf{J0332-00} & 6.588$\pm$0.058 & 3.90$\pm$0.53 & 42.06$\pm$0.17 & (1.34$\pm$0.51)$\cdot$ 10$^{-1}$ & 22.4$\pm$2.5 & 44.045$\pm$0.095 & 12.8$\pm$2.8 \\
\textbf{J0334+00} & 5.63$\pm$0.22 & 0.58$\pm$0.36 & 40.61$\pm$0.77 & (1.3$\pm$2.2)10$^{-3}$ & 2.10$\pm$0.84 & 42.22$\pm$0.30 & (5.2$\pm$3.6)10$^{-2}$ \\
\textbf{J0848-01} & 6.464$\pm$0.062 & 0.36$\pm$0.09 & 38.27$\pm$0.29 & (1.34$\pm$0.90)$\cdot$ 10$^{-5}$ & 7.24$\pm$1.1 & 42.169$\pm$0.086 & (1.05$\pm$0.21)$\cdot$ 10$^{-1}$ \\
\textbf{J0904-00} & 5.99$\pm$0.10 & 2.84$\pm$0.70 & 41.91$\pm$0.16 & (2.47$\pm$0.93)$\cdot$ 10$^{-2}$ & 5.2$\pm$1.3 & 42.69$\pm$0.18 & (1.49$\pm$0.63)$\cdot$ 10$^{-1}$ \\
\textbf{J0923+01} & 7.24$\pm$0.14 & 4.7$\pm$1.2 & 40.12$\pm$0.11 & (4.7$\pm$1.2)$\cdot$ 10$^{-4}$ & 57$\pm$24 & 43.46$\pm$0.39 & (10.0$\pm$9.0)$\cdot$ 10$^{-4}$ \\
\textbf{J0924+01} & 6.432$\pm$0.014 & 4.89$\pm$0.29 & 41.733$\pm$0.065 & (1.14$\pm$0.17)$\cdot$ 10$^{-3}$ & 19.04$\pm$0.71 & 43.502$\pm$0.027 & (6.68$\pm$0.42)$\cdot$ 10$^{-2}$ \\
\textbf{J0948+00} & 5.544$\pm$0.075 & 1.03$\pm$0.34 & 41.50$\pm$0.44 & (1.9$\pm$2.0)$\cdot$ 10$^{-2}$ & 4.94$\pm$0.99 & 43.524$\pm$0.15 & 2.08$\pm$0.72 \\
\textbf{J2358-00} & 6.851$\pm$0.020 & 1.84$\pm$0.16 & 39.617$\pm$0.099 & (4.6$\pm$1.0)$\cdot$ 10$^{-5}$ & 28.5$\pm$1.3 & 43.184$\pm$0.029 & (1.68$\pm$0.11)$\cdot$ 10$^{-1}$ \\ \hline
\end{tabular}
\label{Tab_parametric_energetics}
\end{table*}


\section{Flux-weighted non-parametric analysis}
\label{Appendix nonpar}

In Figure \ref{Fig_non_parametric1} we include all the fits performed using the flux-weighted non-parametric method (except the one shown in Figure \ref{Fig2}), which is described in full in Section \ref{flux-non-parametric}. The corresponding parameters from the fits of the [OIII] line are listed in Table \ref{Tab_non_parametric}, and the outflow physical properties in Table \ref{Tab_nonparametric_energetics}.

\begin{figure*}
\centering
{\par\includegraphics[width=0.43\textwidth]{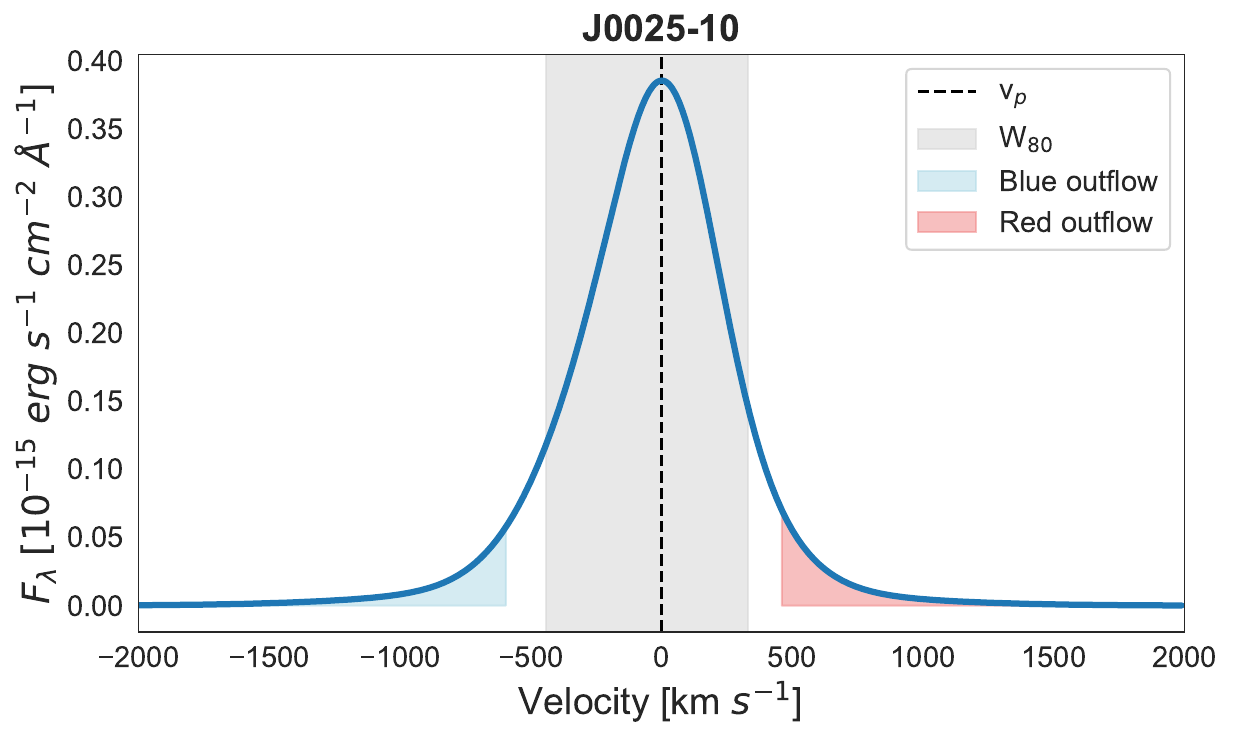}
\includegraphics[width=0.43\textwidth]{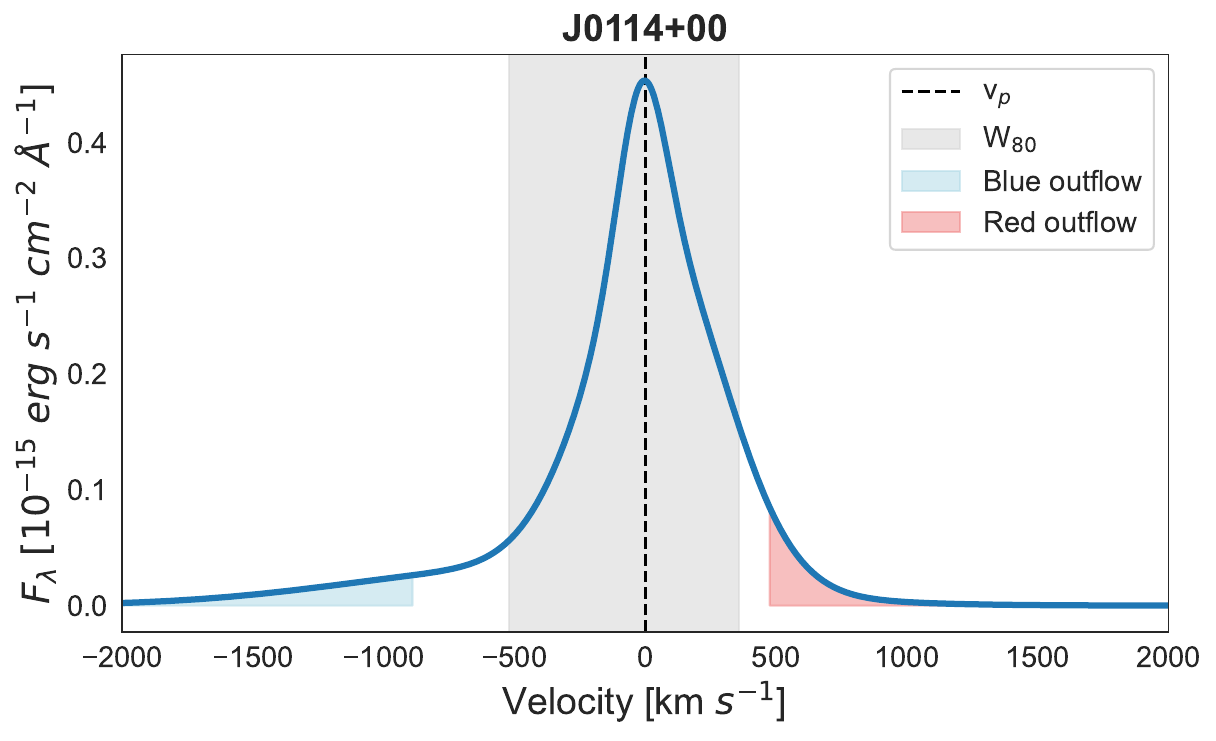}
\includegraphics[width=0.43\textwidth]{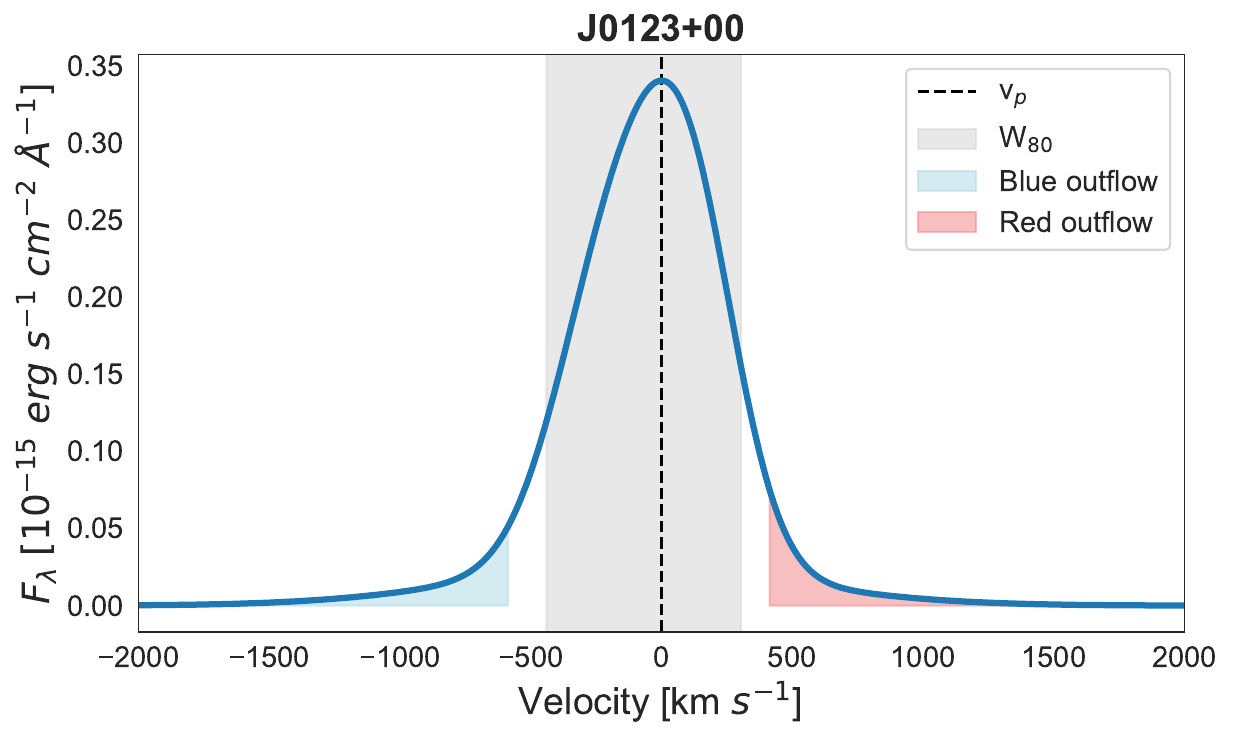}
\includegraphics[width=0.43\textwidth]{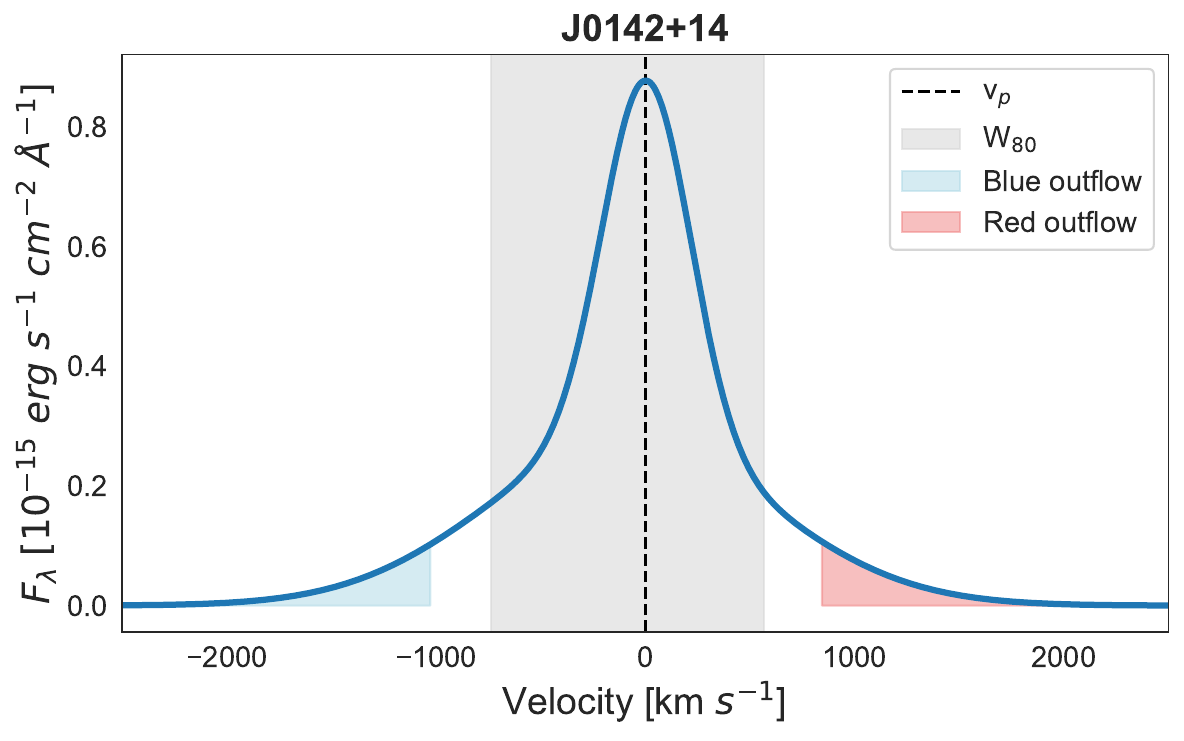}
\includegraphics[width=0.43\textwidth]{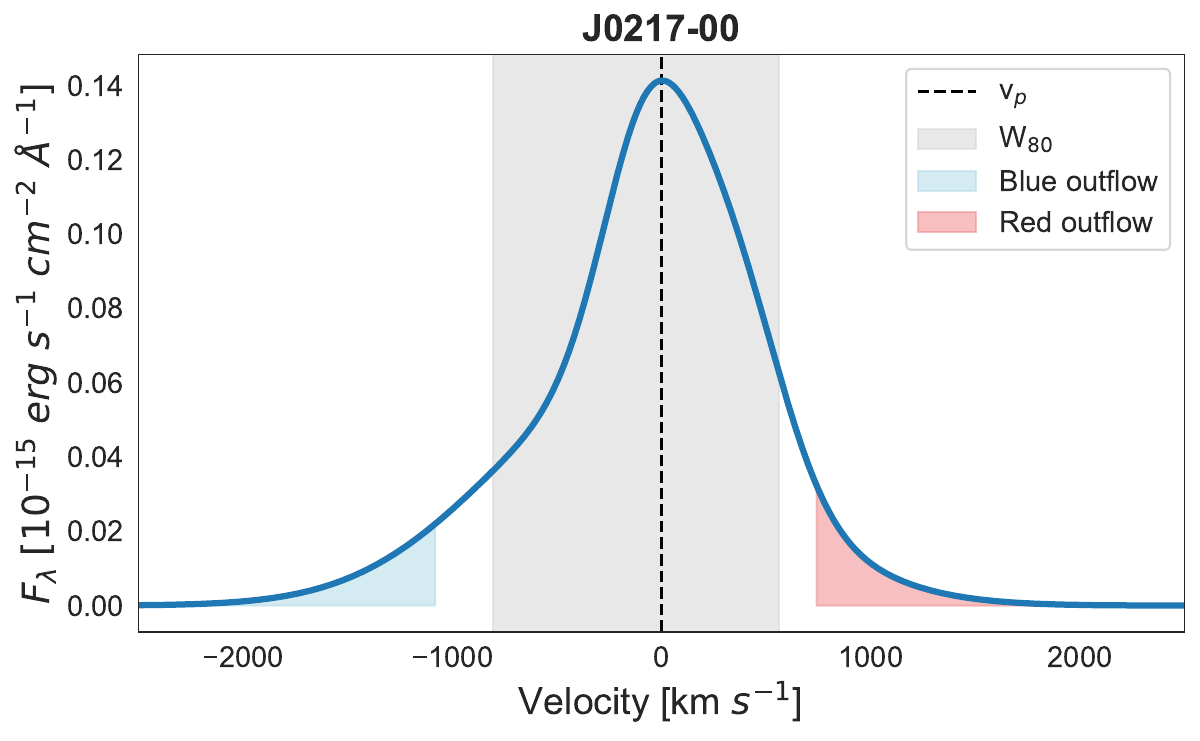}
\includegraphics[width=0.43\textwidth]{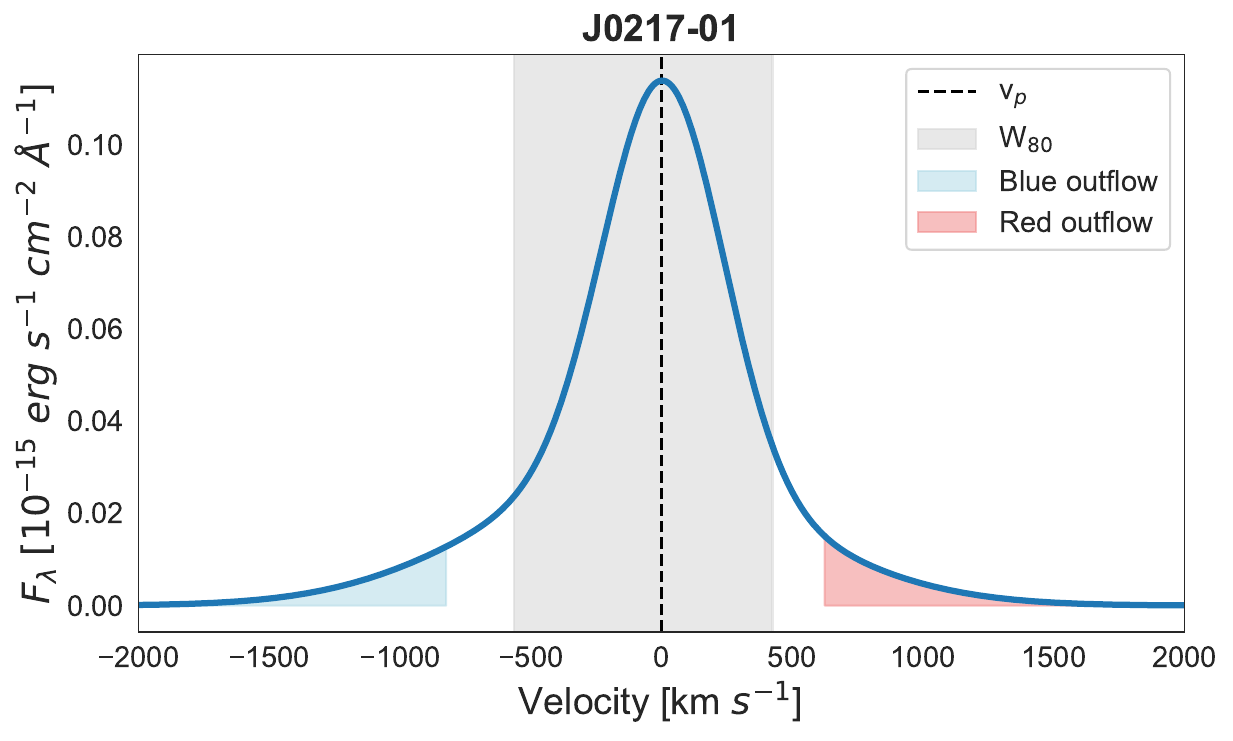}
\includegraphics[width=0.43\textwidth]{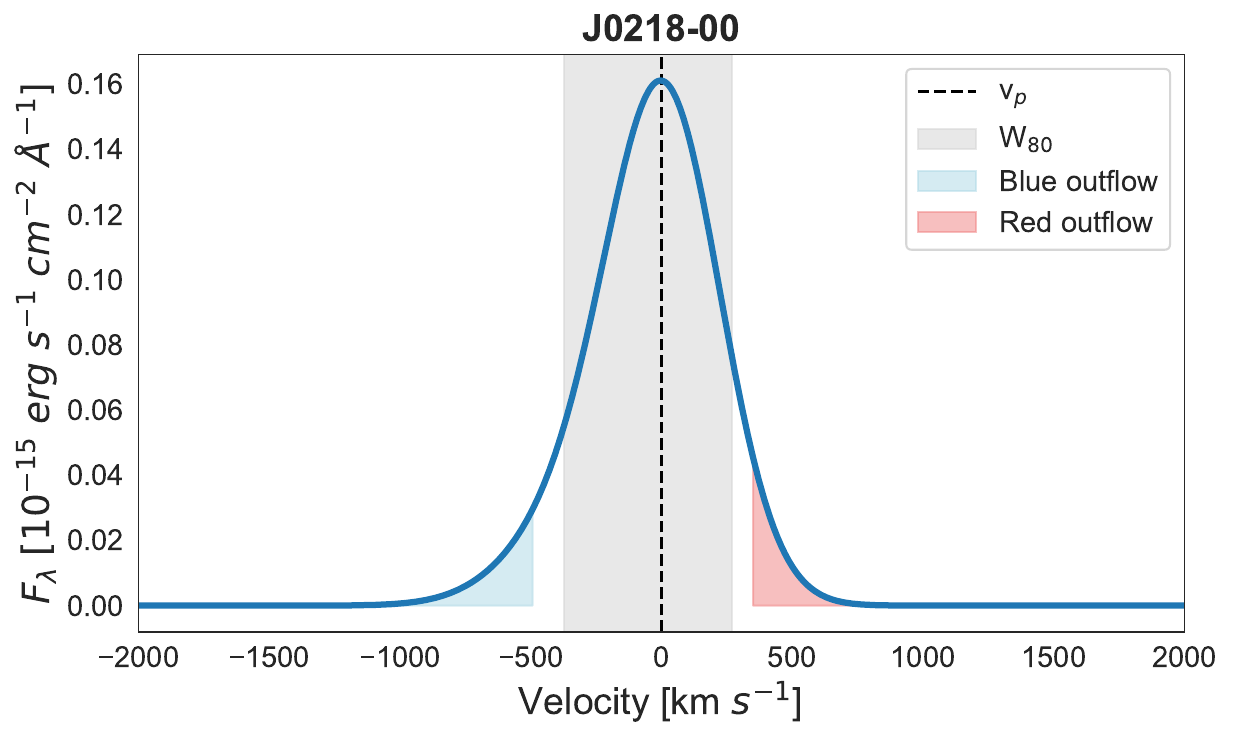}
\includegraphics[width=0.43\textwidth]{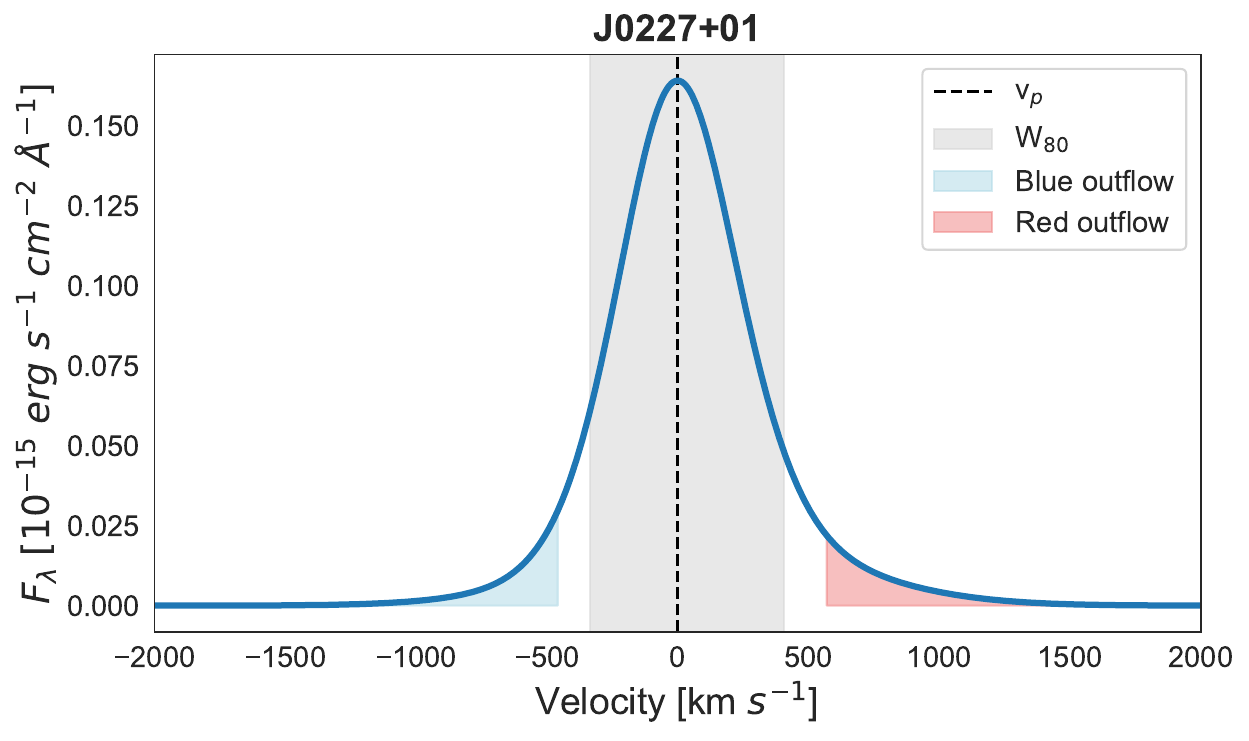}
\includegraphics[width=0.43\textwidth]{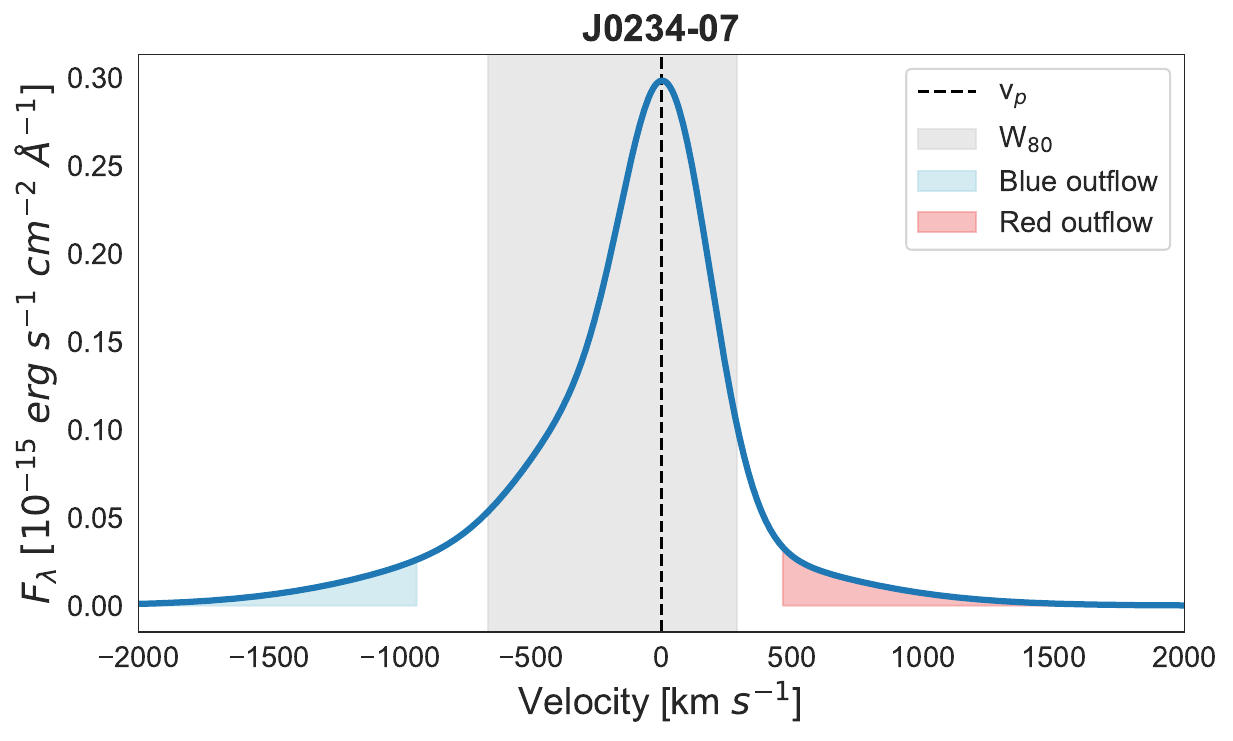}
\includegraphics[width=0.43\textwidth]{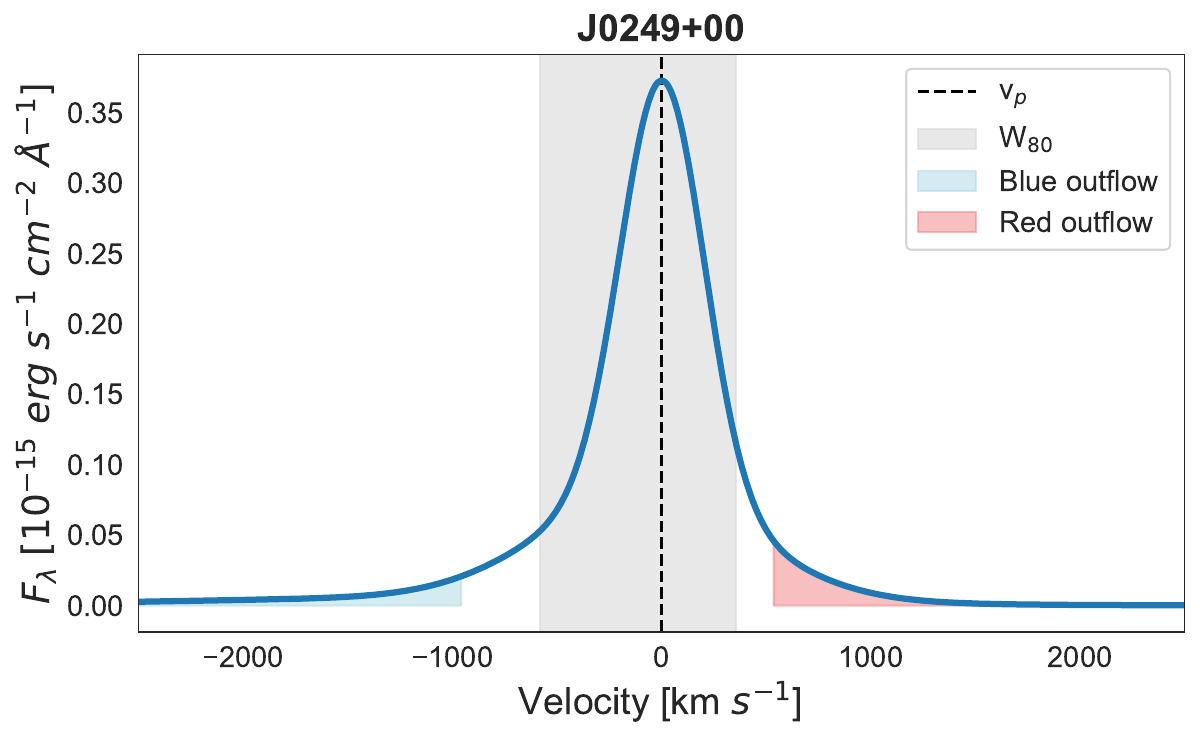}\par}
\caption{Flux-weighted non-parametric fits of the [OIII]$\lambda$5007 \AA~emission line shown for the whole sample. Figures are identical to the example in Figure \ref{Fig2}.}
\label{Fig_non_parametric1}
\end{figure*}

\setcounter{figure}{0}
\begin{figure*}
\centering
{\par\includegraphics[width=0.43\textwidth]{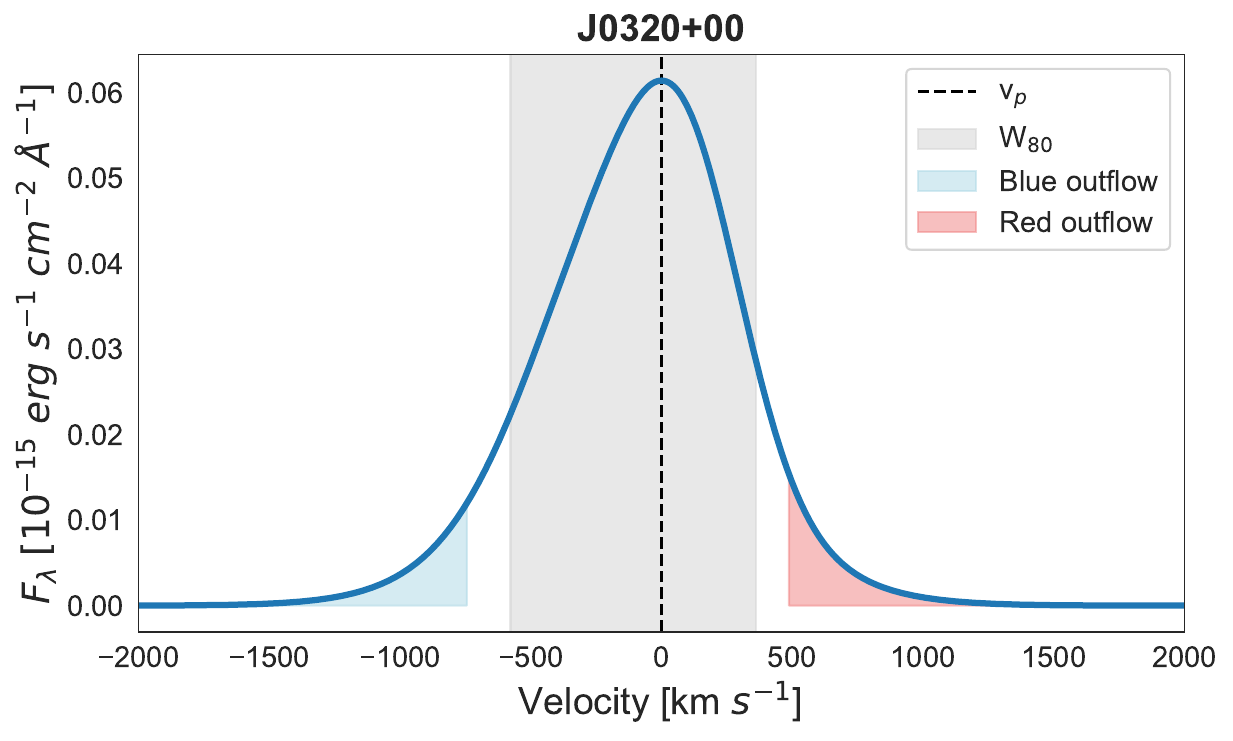}
\includegraphics[width=0.43\textwidth]{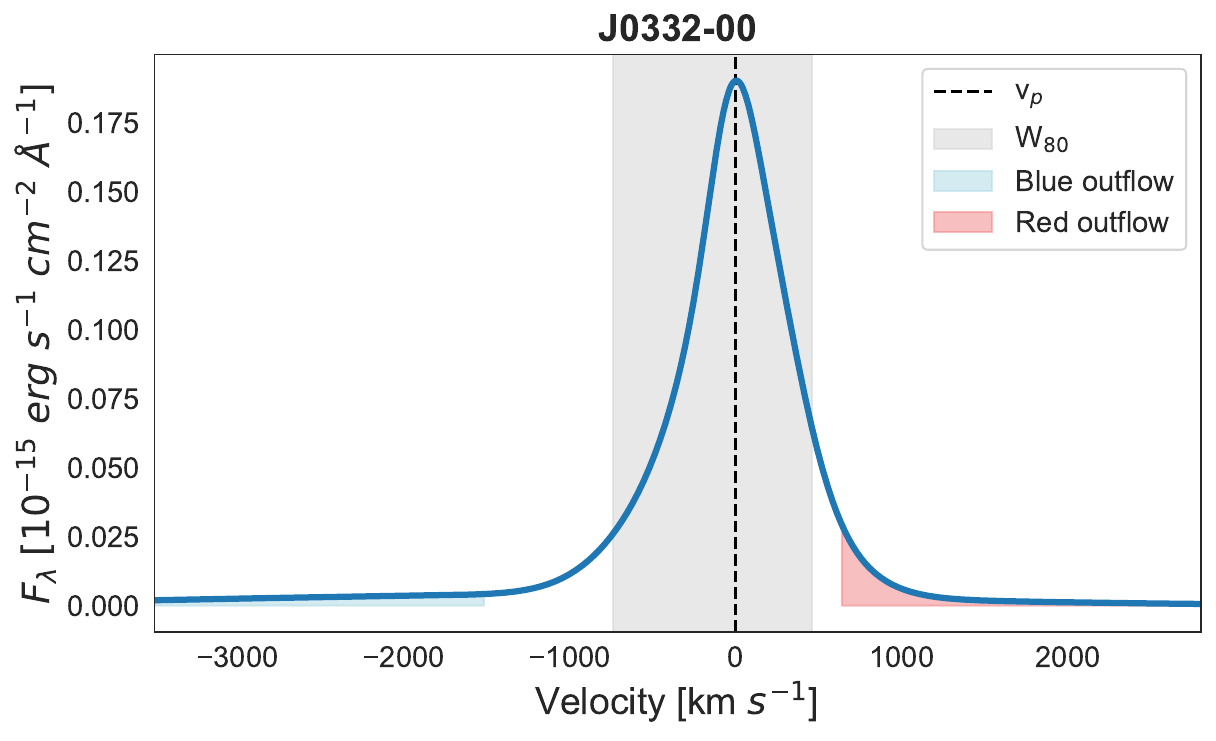}
\includegraphics[width=0.43\textwidth]{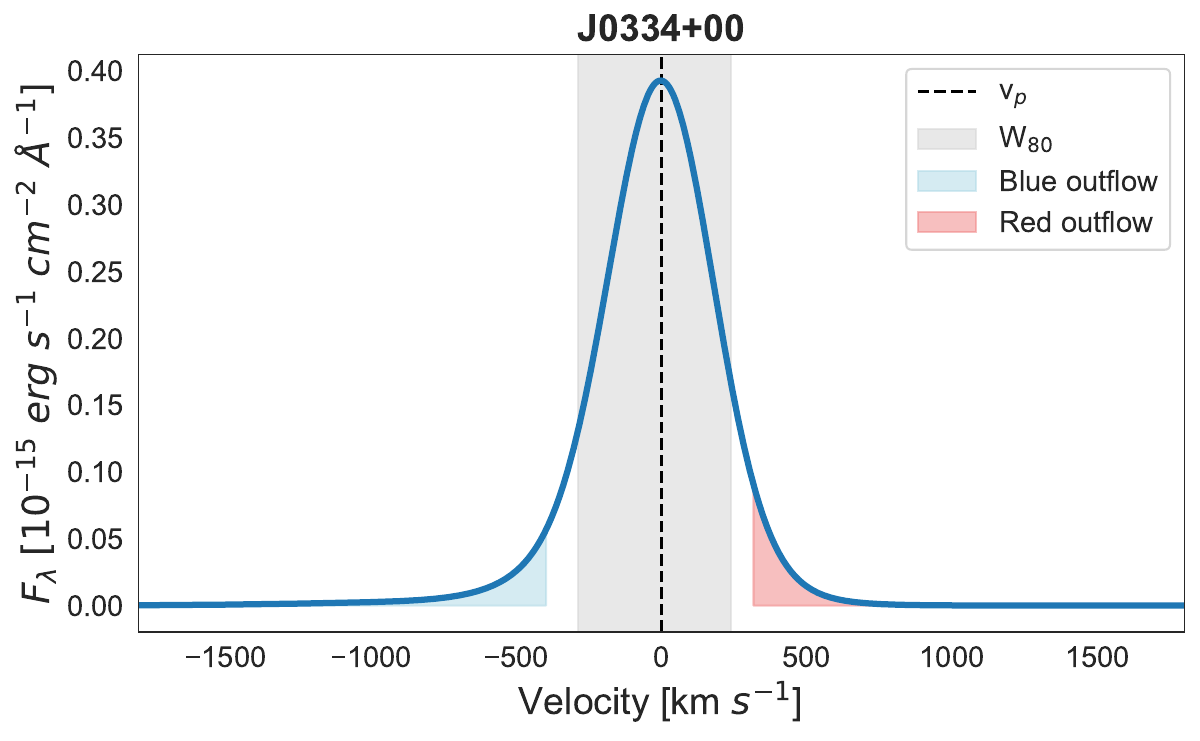}
\includegraphics[width=0.43\textwidth]{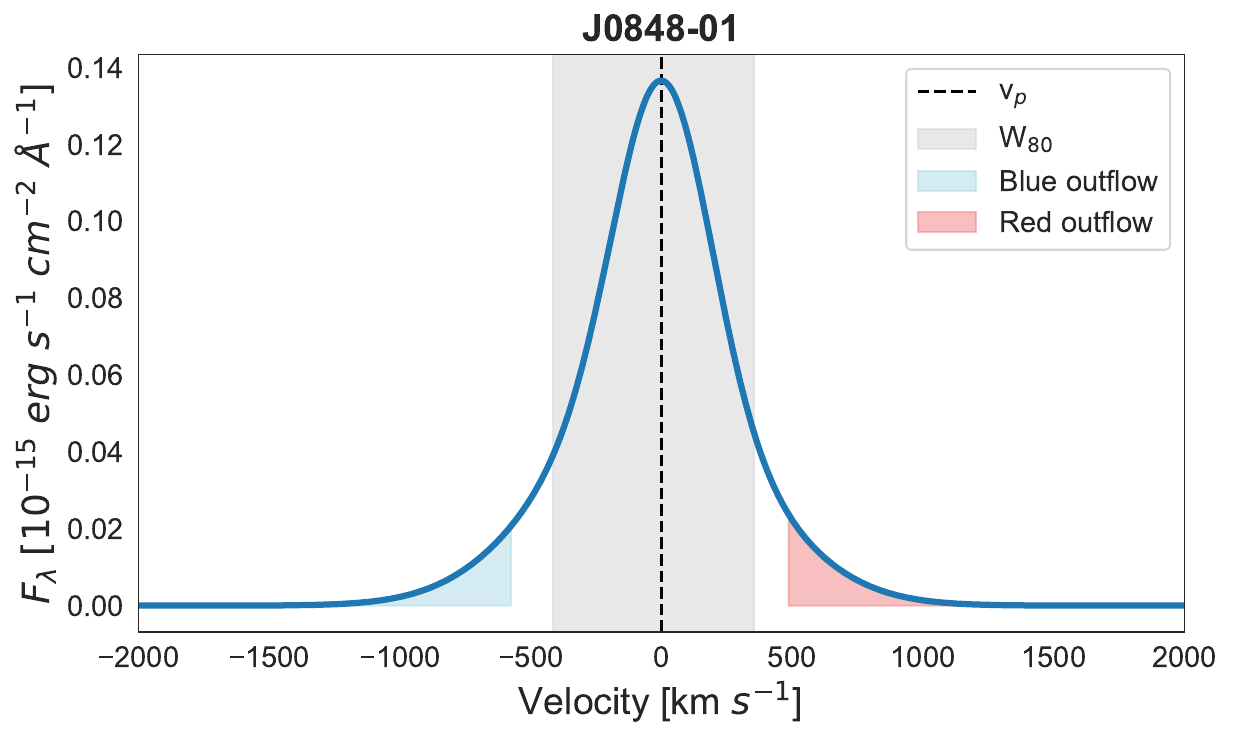}
\includegraphics[width=0.43\textwidth]{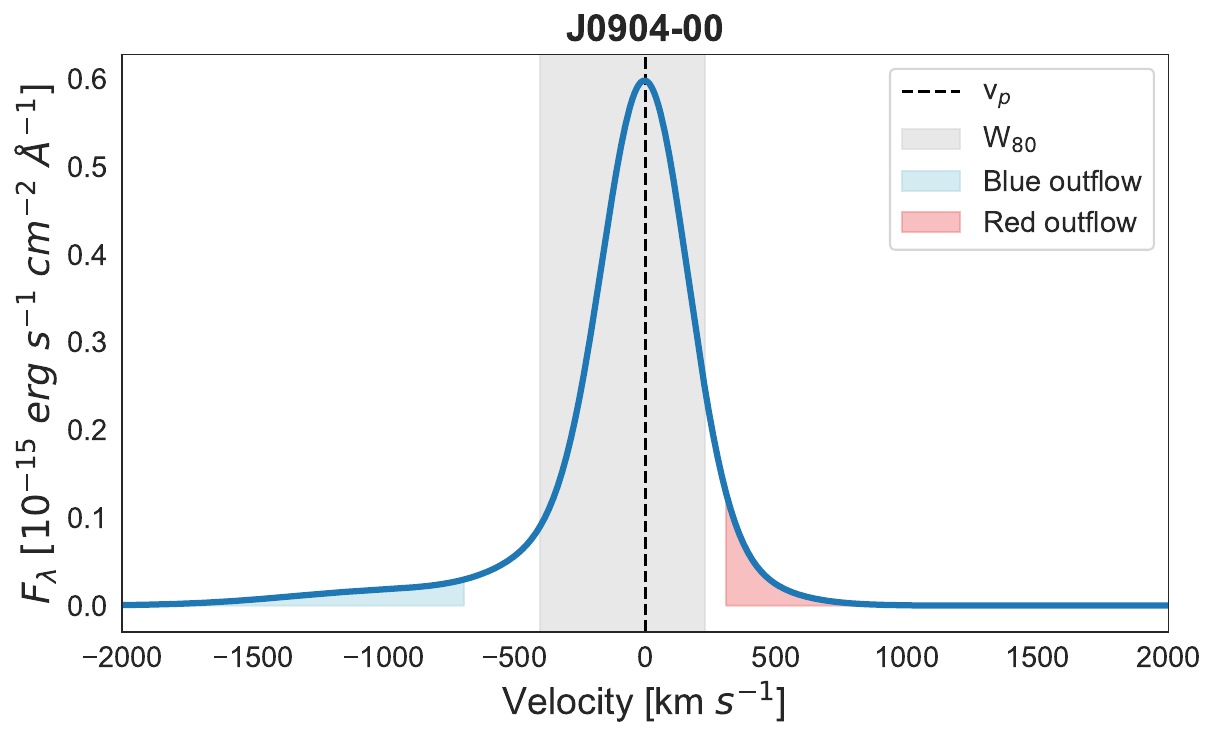}
\includegraphics[width=0.43\textwidth]{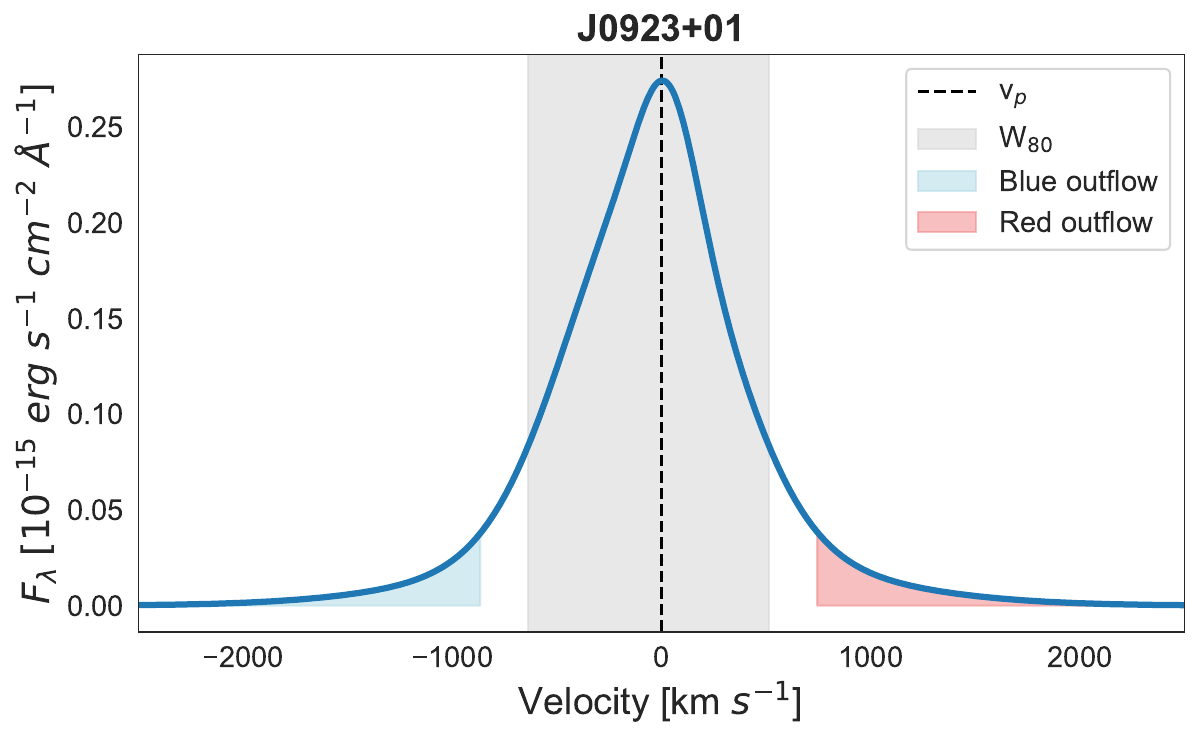}
\includegraphics[width=0.43\textwidth]{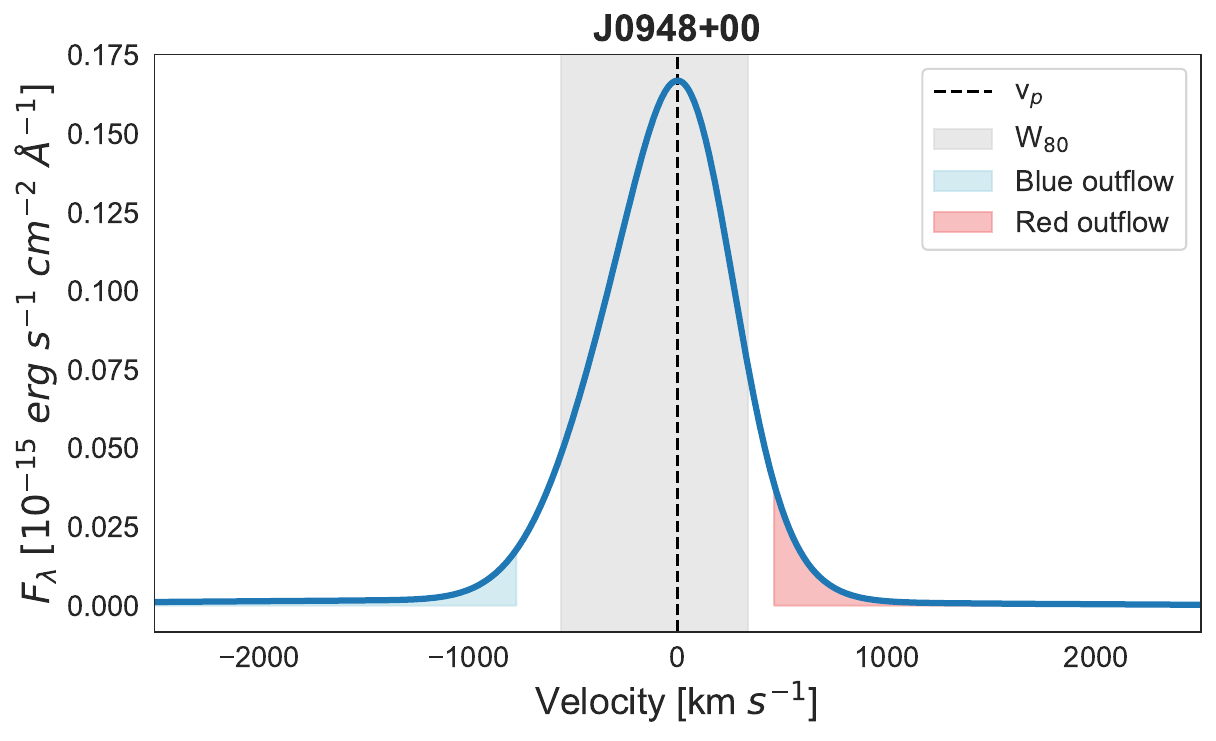}
\includegraphics[width=0.43\textwidth]{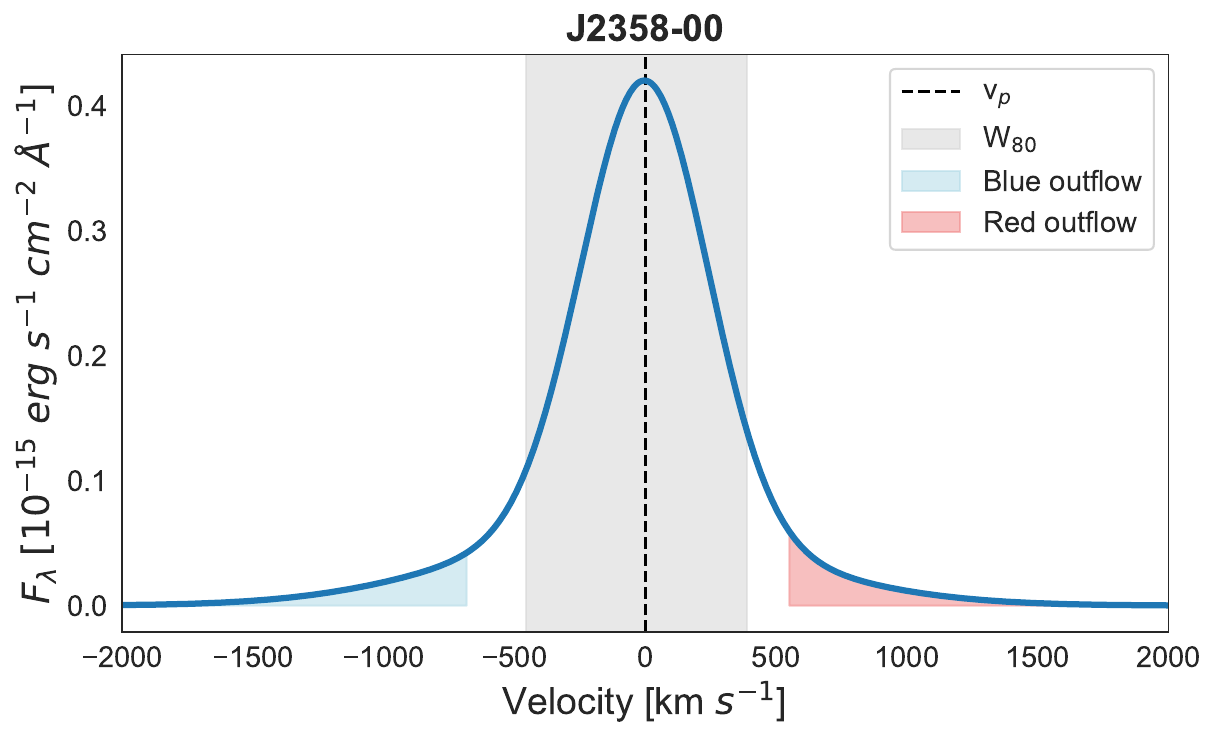}\par}
\caption{continued}
\end{figure*}

\begin{table*}[]
\caption{Results from the flux-weighted non-parametric fits of the $[OIII]\lambda$5007 line.}
\centering
\begin{tabular}{ccccccccccccccc} \hline
\textbf{Name} & \textbf{$\rm\bold \lambda _c$} & \textbf{W$_{80}$} & \textbf{$\rm\bold\Delta$V} & \textbf{V$_{05}$} & \textbf{V$_{95}$} & \textbf{V$_{med}$} & \textbf{a} \\ 
 & (\AA) & (km s$^{-1}$) & (km s$^{-1}$) & (km s$^{-1}$) & (km s$^{-1}$) & (km s$^{-1}$) & (km s$^{-1}$) \\ \hline
\textbf{J0025-10} & 6528.139$\pm$0.014 & 770.4$\pm$2.8 & -66.0$\pm$3.3 & -589.3$\pm$4.9 & 457.2$\pm$3.0 & -30.43$\pm$0.83 & -51.0$\pm$2.2 \\
\textbf{J0114+00} & 6955.832$\pm$0.099 & 875$\pm$51 & -194$\pm$65 & -870$\pm$130 & 479$\pm$23 & -2.3$\pm$6.2 & -146$\pm$47 \\
\textbf{J0123+00} & 7002.574$\pm$0.043 & 742.5$\pm$3.4 & -85.5$\pm$4.2 & -581.3$\pm$6.5 & 410.7$\pm$4.2 & -45.2$\pm$1.8 & -48.7$\pm$2.4 \\
\textbf{J0142+14} & 6958.925$\pm$0.074 & 1303$\pm$28 & -85$\pm$16 & -1015$\pm$30 & 845$\pm$29 & -27.0$\pm$4.5 & -111$\pm$22 \\
\textbf{J0217-00} & 6724.11$\pm$0.27 & 1363$\pm$15 & -162$\pm$22 & -1069$\pm$29 & 746$\pm$22 & -21$\pm$12 & -188$\pm$15 \\
\textbf{J0217-01} & 6884.24$\pm$0.32 & 990$\pm$120 & -97$\pm$76 & -820$\pm$160 & 623$\pm$97 & -24$\pm$17 & -93$\pm$90 \\
\textbf{J0218-00} & 6868.18$\pm$0.11 & 642$\pm$11 & -67$\pm$11 & -486$\pm$18 & 352.1$\pm$9.7 & -26.5$\pm$4.1 & -47.5$\pm$8.8 \\
\textbf{J0227+01} & 6823.574$\pm$0.039 & 742.5$\pm$5.5 & 60.0$\pm$6.0 & -452.0$\pm$6.7 & 572$\pm$10 & 12.4$\pm$1.7 & 47.5$\pm$4.2 \\
\textbf{J0234-07} & 6560.469$\pm$0.010 & 950.9$\pm$2.8 & -229.0$\pm$3.4 & -925.5$\pm$5.1 & 466.6$\pm$4.1 & -69.96$\pm$0.57 & -234.9$\pm$2.9 \\
\textbf{J0249+00} & 7049.048$\pm$0.011 & 941$\pm$11 & -206$\pm$15 & -949$\pm$29 & 535.3$\pm$8.7 & -28.9$\pm$1.2 & -167$\pm$10 \\
\textbf{J0320+00} & 6924.11$\pm$0.16 & 961$\pm$12 & -97$\pm$13 & -734$\pm$18 & 540$\pm$20 & -64.9$\pm$6.3 & -65.6$\pm$8.9 \\
\textbf{J0332-00} & 6557.284$\pm$0.048 & 1197$\pm$16 & -433$\pm$54 & -1500$\pm$110 & 641$\pm$11 & -25.0$\pm$2.9 & -235$\pm$17 \\
\textbf{J0334+00} & 7044.5970$\pm$0.0090 & 526.4$\pm$2.2 & -39.3$\pm$2.5 & -392.8$\pm$4.6 & 314.2$\pm$2.0 & -12.3$\pm$0.5 & -24.3$\pm$1.5 \\
\textbf{J0848-01} & 6754.591$\pm$0.057 & 768$\pm$11 & -37.4$\pm$8.7 & -564$\pm$15 & 490$\pm$13 & -15.3$\pm$2.8 & -31.5$\pm$8.1 \\
\textbf{J0904-00} & 6771.839$\pm$0.014 & 627.7$\pm$7.5 & -188$\pm$11 & -685$\pm$22 & 308.9$\pm$3.3 & -27.96$\pm$0.88 & -117.8$\pm$7.1 \\
\textbf{J0923+01} & 6947.64$\pm$0.14 & 1153$\pm$16 & -56$\pm$12 & -854$\pm$13 & 742$\pm$28 & -50.4$\pm$5.4 & -25.2$\pm$8.4 \\
\textbf{J0924+01} & 6906.315$\pm$0.042 & 1645$\pm$25 & -509$\pm$18 & -1616$\pm$33 & 598$\pm$12 & -116.0$\pm$3.8 & -575$\pm$24 \\
\textbf{J0948+00} & 6629.820$\pm$0.071 & 894.9$\pm$9.0 & -154$\pm$15 & -767$\pm$26 & 458.2$\pm$8.3 & -59.0$\pm$3.3 & -103.2$\pm$7.9 \\
\textbf{J2358-00} & 7021.214$\pm$0.013 & 843.7$\pm$3.5 & -61.5$\pm$4.3 & -674.8$\pm$7.6 & 552.4$\pm$5.4 & -13.36$\pm$0.71 & -39.1$\pm$3.0 \\ \hline
\end{tabular}
\label{Tab_non_parametric}
\end{table*}

\begin{table*}
\caption{Outflow physical properties resulting from the flux-weighted non-parametric method.}
\centering
\begin{tabular}{ccccccccccc} \hline
\textbf{Name} & \textbf{v$_{\rm OF}$} & \textbf{Log(M$_{\rm OF}$)} & \textbf{$\dot{\rm\bold M}_{\rm OF}$} & \textbf{Log($\dot{\rm\bold E}_{\rm kin}$)} & \textbf{$\dot{\rm\bold E}_{\rm kin}$ / $\rm \bold L_{\rm BOL}$}\\ 
& (km s$^{-1}$) & (M$_{\odot}$)  & (M$_{\odot}$ yr $^{-1}$) & (erg s$^{-1}$) & (\%)\\ \hline
\textbf{J0025-10} &  -751$\pm$13 / 589.6$\pm$7.9 & 5.9965$\pm$0.0040 & 2.030$\pm$0.028 & 41.472$\pm$0.016 & (9.89$\pm$0.37)$\cdot$10$^{-3}$ \\
\textbf{J0114+00} & -1179$\pm$190 / 585$\pm$44 & 6.2994$\pm$0.0072 & 5.39$\pm$0.60 & 42.25$\pm$0.19 & (1.27$\pm$0.56)$\cdot$10$^{-1}$ \\
\textbf{J0123+00} & -770$\pm$18 / 530$\pm$10 & 6.2709$\pm$0.0056 & 3.687$\pm$0.074 & 41.729$\pm$0.023 & (8.95$\pm$0.48)$\cdot$10$^{-3}$ \\
\textbf{J0142+14} & -1256$\pm$38 / 1072$\pm$38 & 6.7538$\pm$0.0045 & 20.19$\pm$0.55 & 42.942$\pm$0.032 & (1.97$\pm$0.15)$\cdot$10$^{-1}$ \\
\textbf{J0217-00} & -1305$\pm$35 / 917$\pm$33 & 5.9334$\pm$0.0044 & 2.913$\pm$0.065 & 42.088$\pm$0.028 & (2.29$\pm$0.15)$\cdot$10$^{-2}$ \\
\textbf{J0217-01} & -1040$\pm$210 / 820$\pm$120 & 5.7720$\pm$0.0048 & 1.73$\pm$0.22 & 41.72$\pm$0.18 & (3.7$\pm$1.6)$\cdot$10$^{-2}$ \\
\textbf{J0218-00} & -595$\pm$26 / 414$\pm$11 & 5.8001$\pm$0.0016 & 0.962$\pm$0.027 & 40.915$\pm$0.044 & (3.97$\pm$0.40)$\cdot$10$^{-3}$ \\
\textbf{J0227+01} & -578$\pm$14 / 740$\pm$18 & 5.8061$\pm$0.0049 & 1.297$\pm$0.026 & 41.272$\pm$0.025 & (7.85$\pm$0.45)$\cdot$10$^{-3}$ \\
\textbf{J0234-07} & -1179.0$\pm$7.1 / 692.5$\pm$7.1 &5.9177$\pm$0.0026 & 2.356$\pm$0.018 & 41.8882$\pm$0.0072 & (6.78$\pm$0.11)$\cdot$10$^{-2}$ \\
\textbf{J0249+00} & -1499$\pm$110 / 730$\pm$19 & 6.3289$\pm$0.0029 & 7.26$\pm$0.34 & 42.59$\pm$0.08 & (1.23$\pm$0.23)$\cdot$10$^{-1}$ \\
\textbf{J0320+00} & -883$\pm$28 / 736$\pm$48 & 5.5796$\pm$0.0037 & 0.937$\pm$0.033 & 41.294$\pm$0.041 & (2.61$\pm$0.24)$\cdot$10$^{-2}$ \\
\textbf{J0332-00} & -2660$\pm$130 / 863$\pm$35 & 5.8196$\pm$0.0029 & 3.54$\pm$0.13 & 42.79$\pm$0.06 & (7.06$\pm$0.98)$\cdot$10$^{-1}$ \\
\textbf{J0334+00} & -525$\pm$13 / 379.5$\pm$2.4 & 6.1932$\pm$0.0083 & 2.152$\pm$0.054 & 41.169$\pm$0.026 & (4.59$\pm$0.27)$\cdot$10$^{-5}$ \\
\textbf{J0848-01} & -704$\pm$22 / 610$\pm$18 & 5.6881$\pm$0.0067 & 0.979$\pm$0.025 & 41.129$\pm$0.029 & (9.59$\pm$0.63)$\cdot$10$^{-3}$ \\
\textbf{J0904-00} & -1029$\pm$34 / 382.9$\pm$2.2 & 6.2328$\pm$0.0066 & 3.654$\pm$0.087 & 41.965$\pm$0.039 & (2.80$\pm$0.25)$\cdot$10$^{-2}$ \\
\textbf{J0923+01} & -1102$\pm$23 / 987$\pm$62 & 6.2644$\pm$0.0031 & 6.5.86$\pm$0.19 & 42.304$\pm$0.038 & (7.02$\pm$0.62)$\cdot$10$^{-2}$ \\
\textbf{J0924+01} & -1940$\pm$37 / 791$\pm$22 & 5.7792$\pm$0.0045 & 2.49$\pm$0.042 & 42.346$\pm$0.022 & (4.66$\pm$0.23)$\cdot$10$^{-3}$ \\
\textbf{J0948+00} & -1360$\pm$260 / 576$\pm$18 & 5.7859$\pm$0.0054 & 1.76$\pm$0.22 & 41.87$\pm$0.22 & (4.6$\pm$2.3)$\cdot$10$^{-2}$ \\
\textbf{J2358-00} & -923$\pm$13 / 740.6$\pm$9.7 & 6.3870$\pm$0.0050 & 6.199$\pm$0.086 & 42.142$\pm$0.014 & (1.530$\pm$0.049)$\cdot$10$^{-2}$ \\ \hline
\end{tabular}
\label{Tab_nonparametric_energetics}
\end{table*}


\section{Peak-weighted non-parametric analysis}
\label{Appendix speranza}

In Figure \ref{Fig_non_parametric3} we include all the fits performed using the peak-weighted non-parametric method from \citealt{speranza2021murales} (except the one shown in Figure \ref{Fig3}), which is described in full in Section \ref{peak-non-parametric}. The corresponding outflow velocities derived from the fits of the [OIII] line and the outflow physical properties are listed in Table \ref{Tab_speranza_energetics}.

\begin{figure*}[htbp]
\centering
{\par\includegraphics[width=0.42\textwidth]{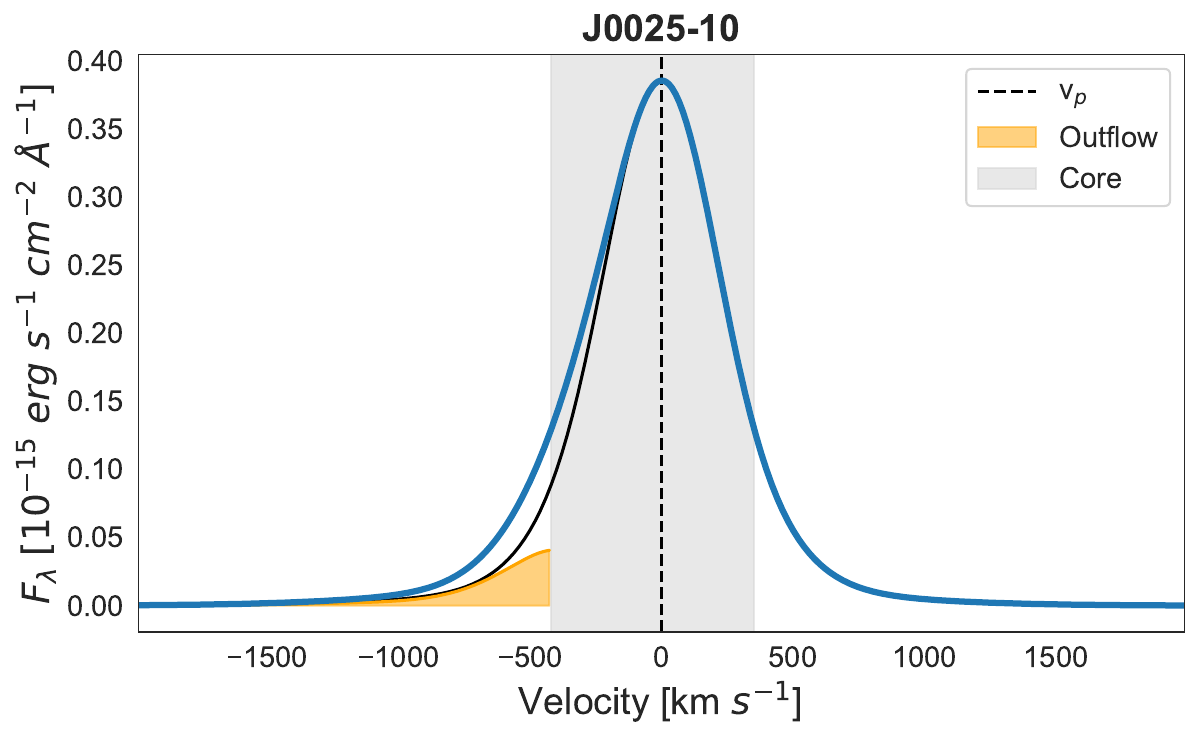}
\includegraphics[width=0.42\textwidth]{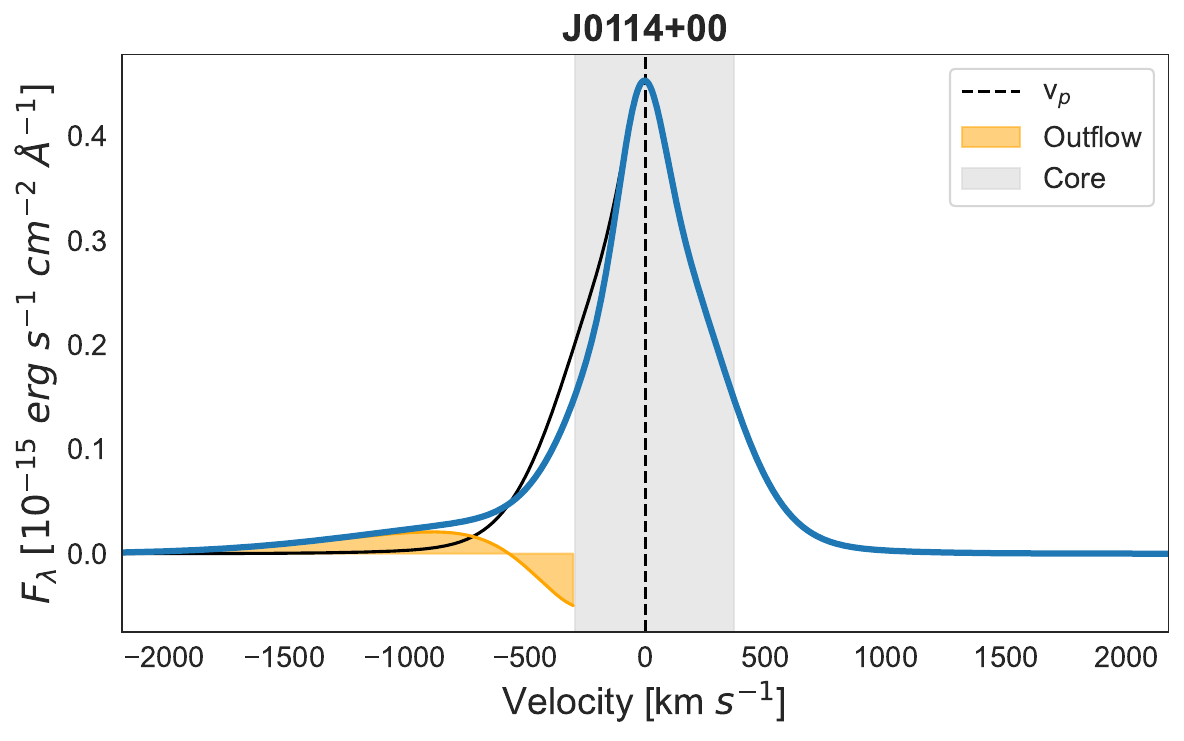}
\includegraphics[width=0.42\textwidth]{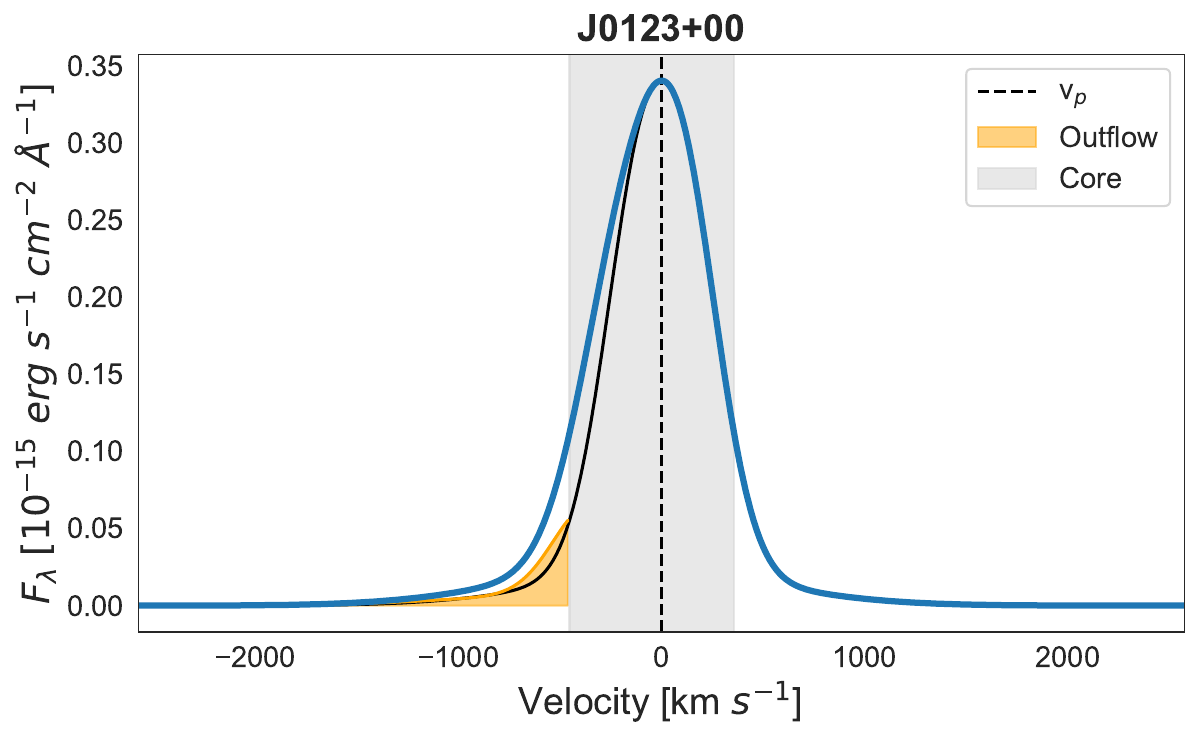}
\includegraphics[width=0.42\textwidth]{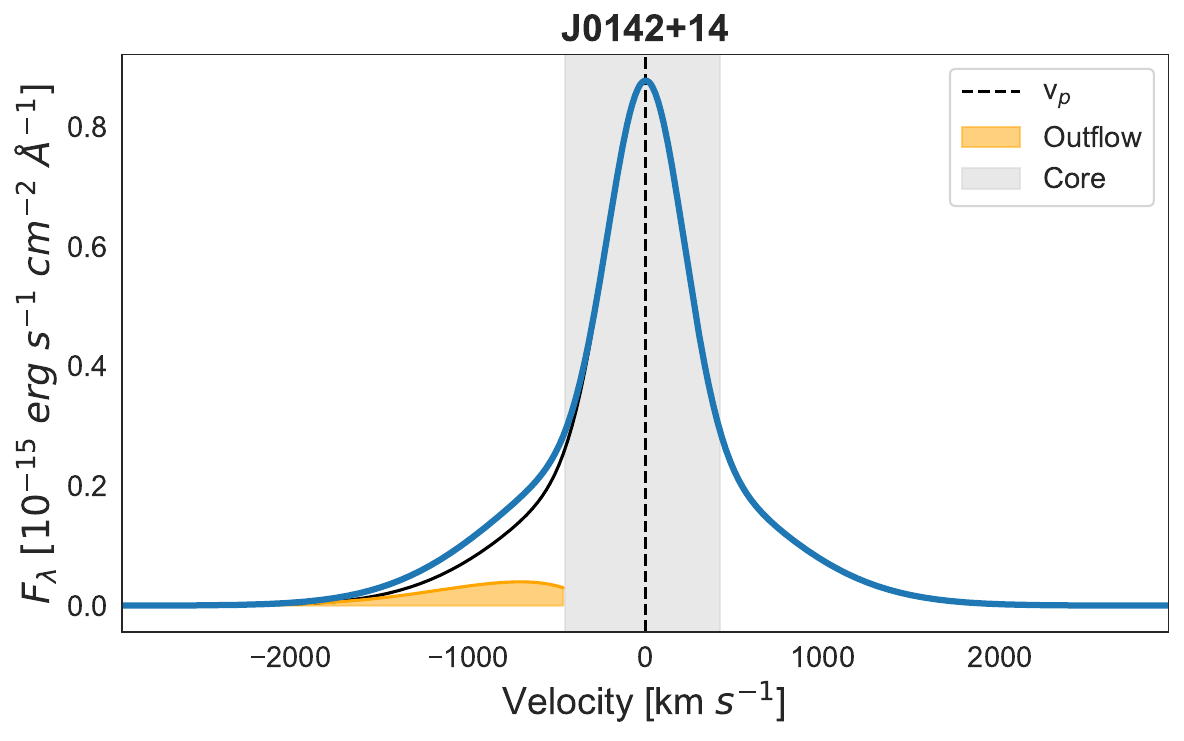}
\includegraphics[width=0.42\textwidth]{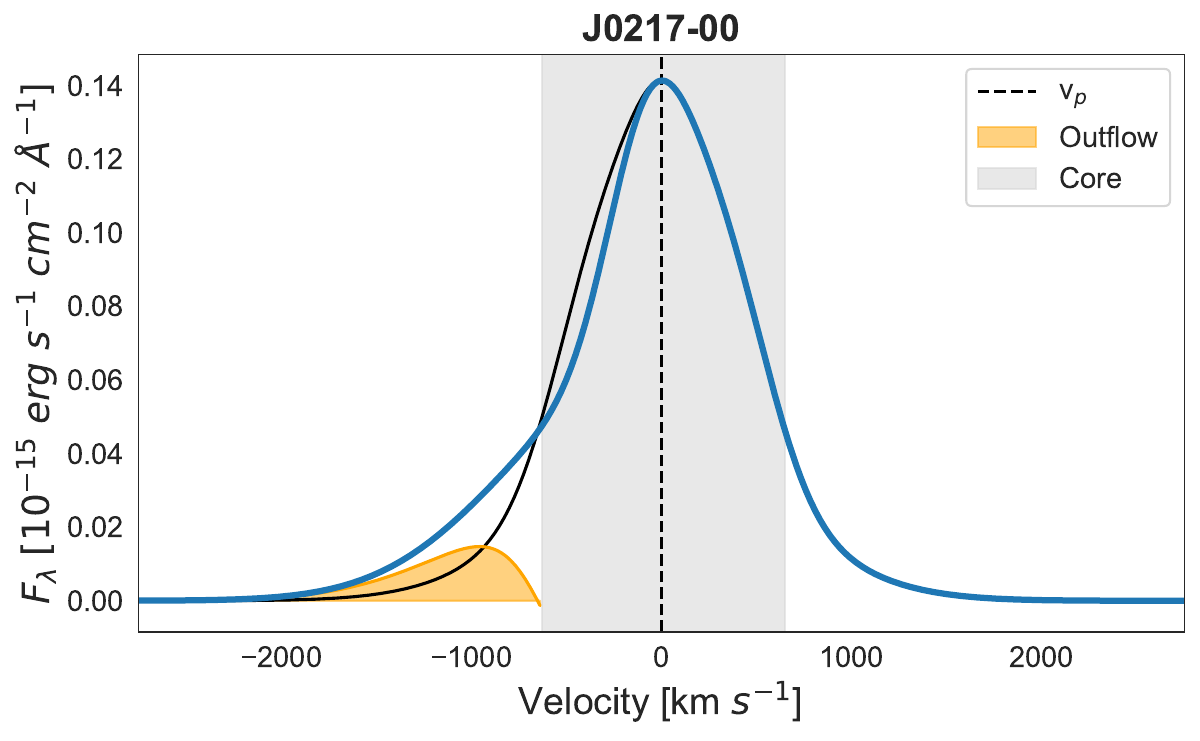}
\includegraphics[width=0.42\textwidth]{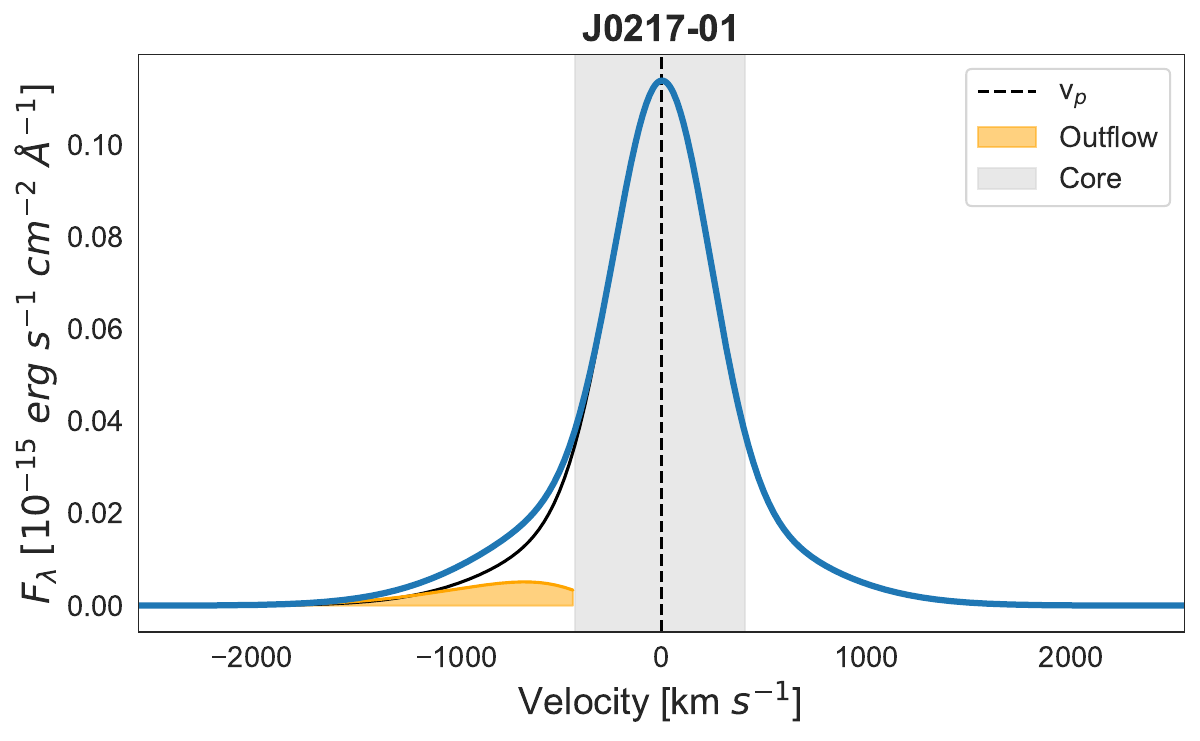}
\includegraphics[width=0.42\textwidth]{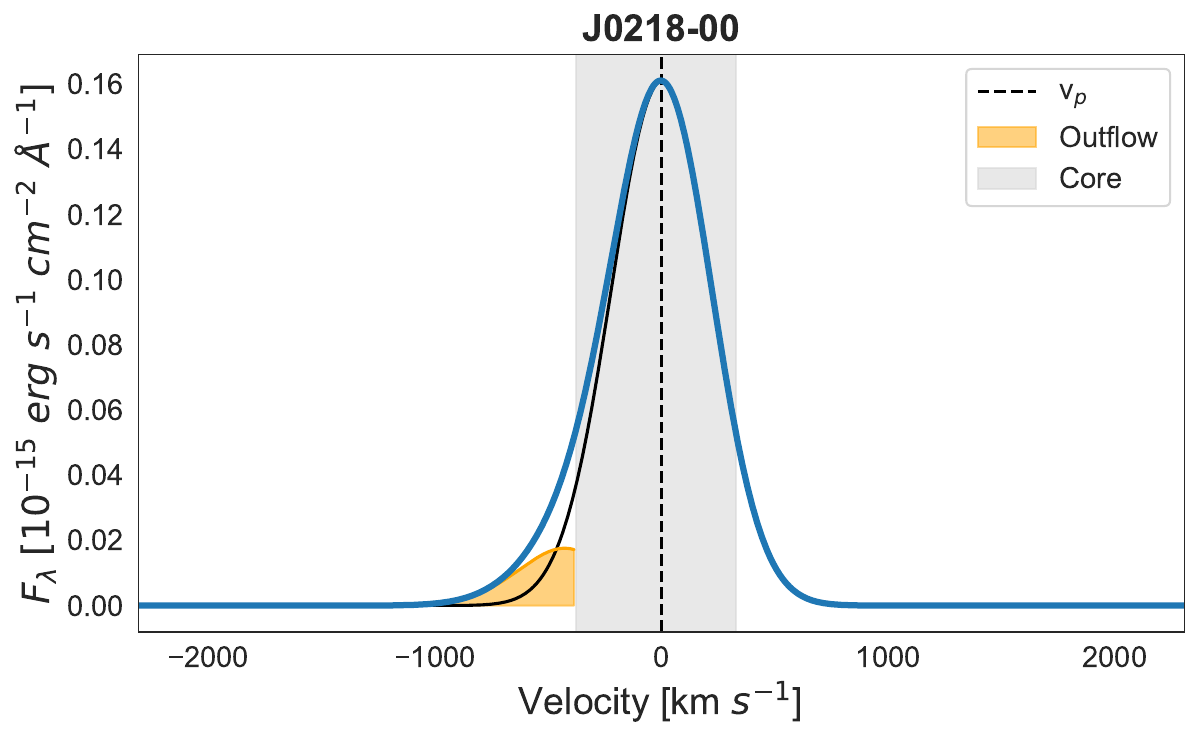}
\includegraphics[width=0.42\textwidth]{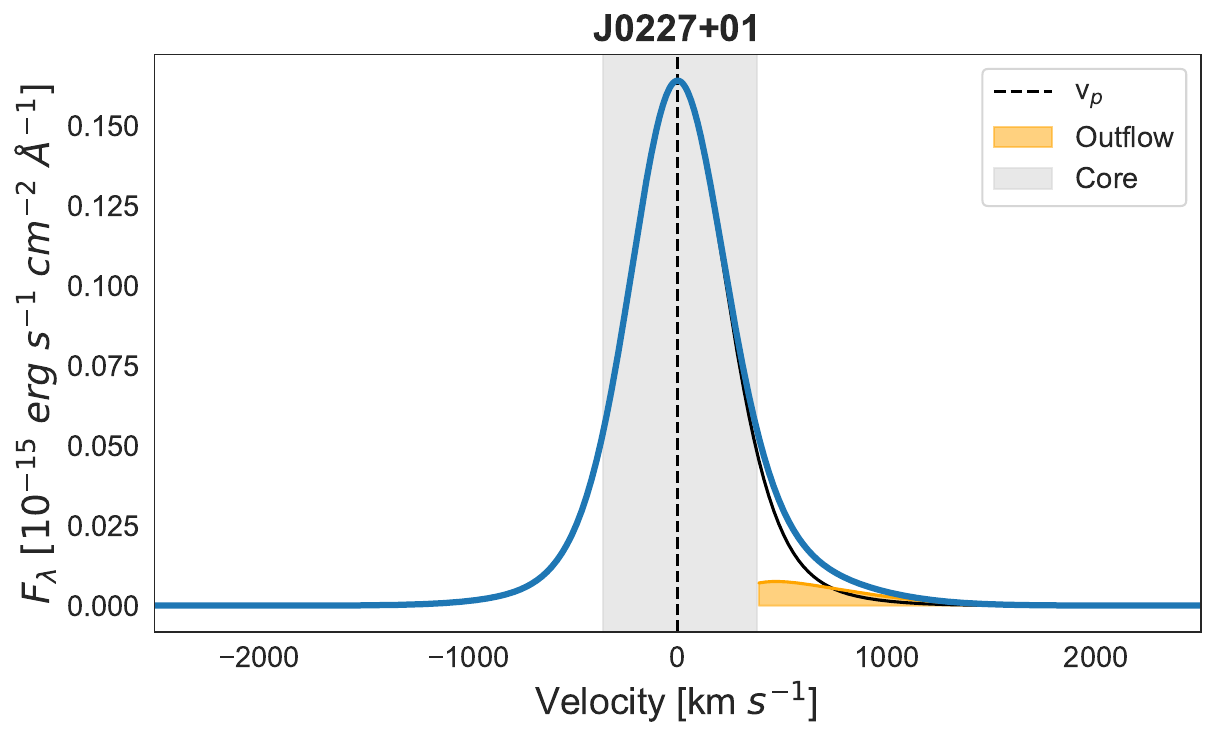}
\includegraphics[width=0.42\textwidth]{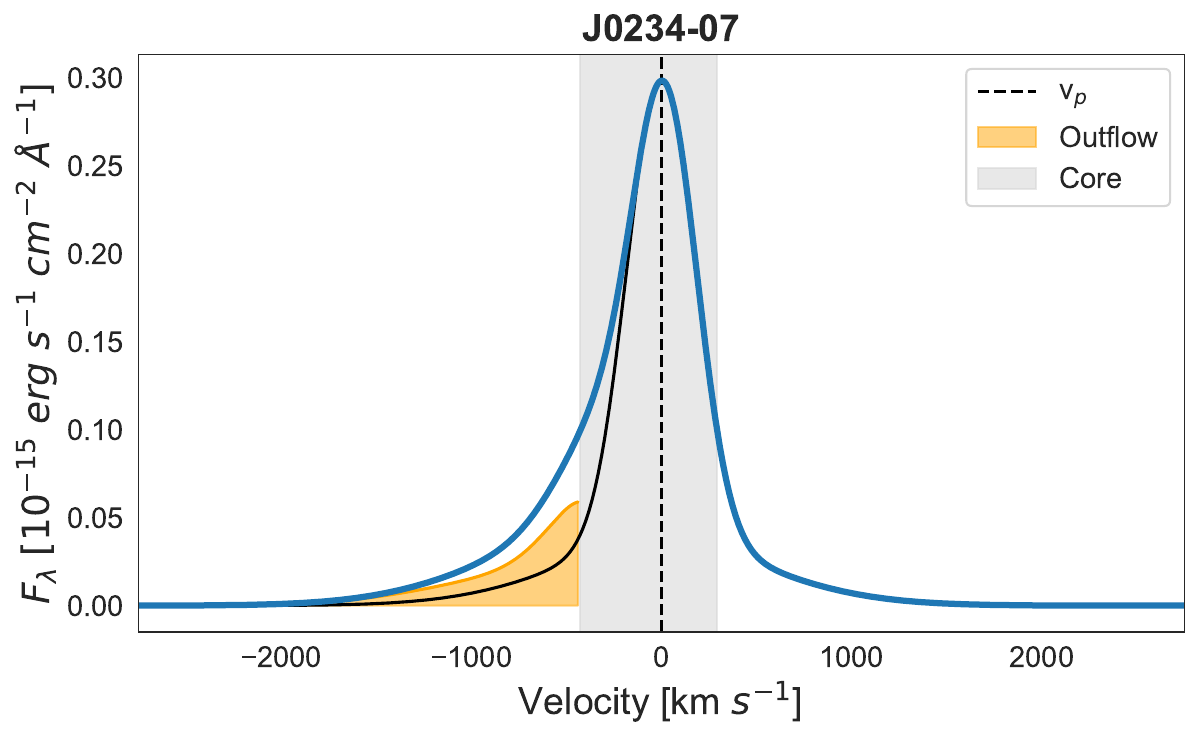}
\includegraphics[width=0.42\textwidth]{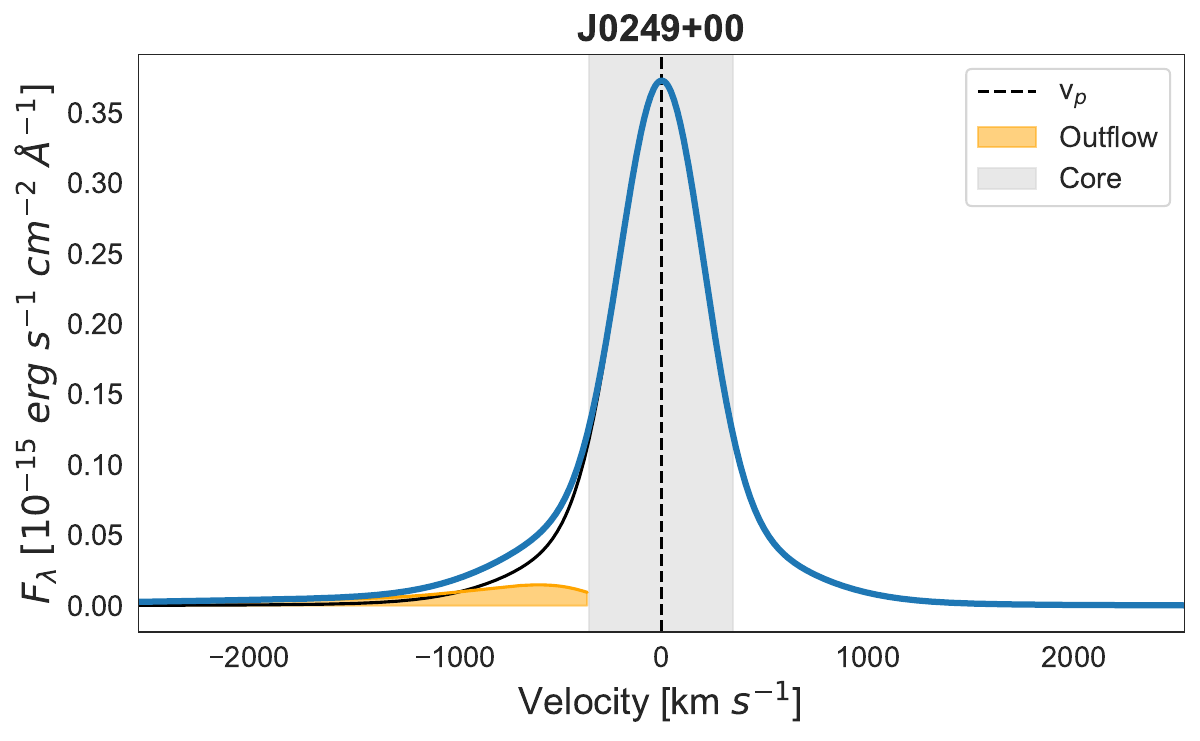}\par}
\caption{Peak-weighted non-parametric fits of the [OIII]$\lambda$5007 \AA~emission line shown for the whole sample. Figures are identical to the example in Figure \ref{Fig3}.}
\label{Fig_non_parametric3}
\end{figure*}

\setcounter{figure}{0}
\begin{figure*}[htbp]
\centering
{\par\includegraphics[width=0.42\textwidth]{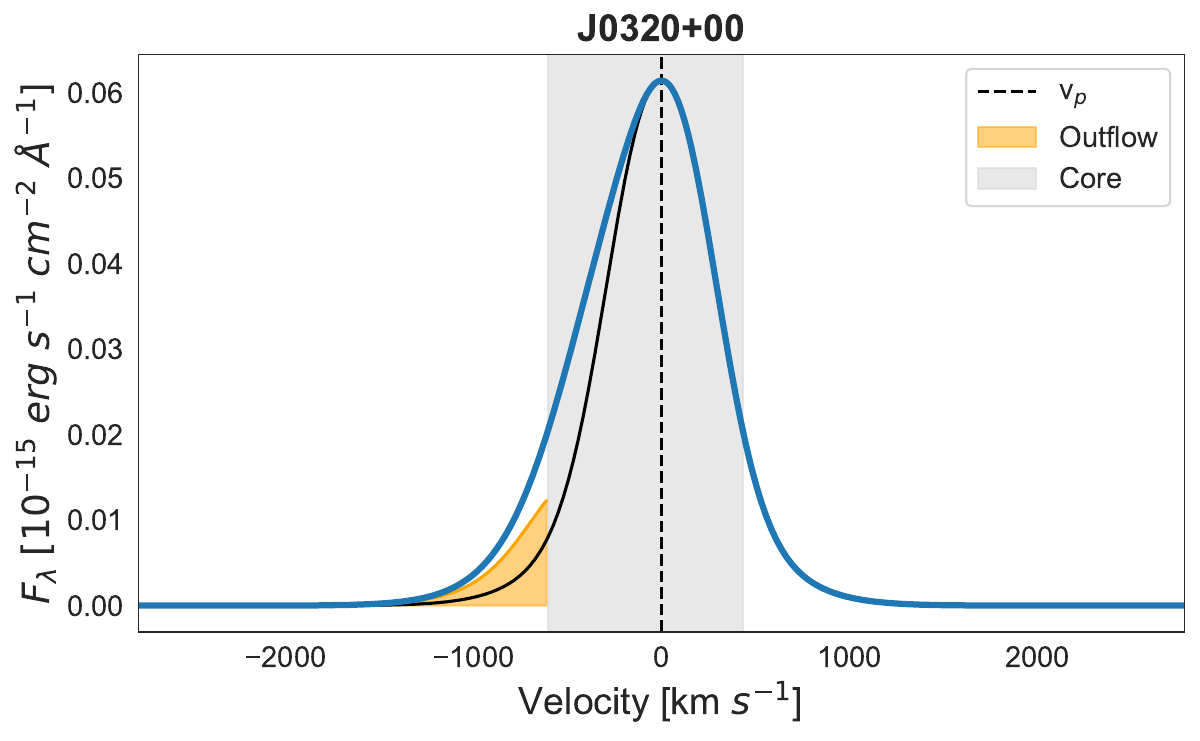}
\includegraphics[width=0.42\textwidth]{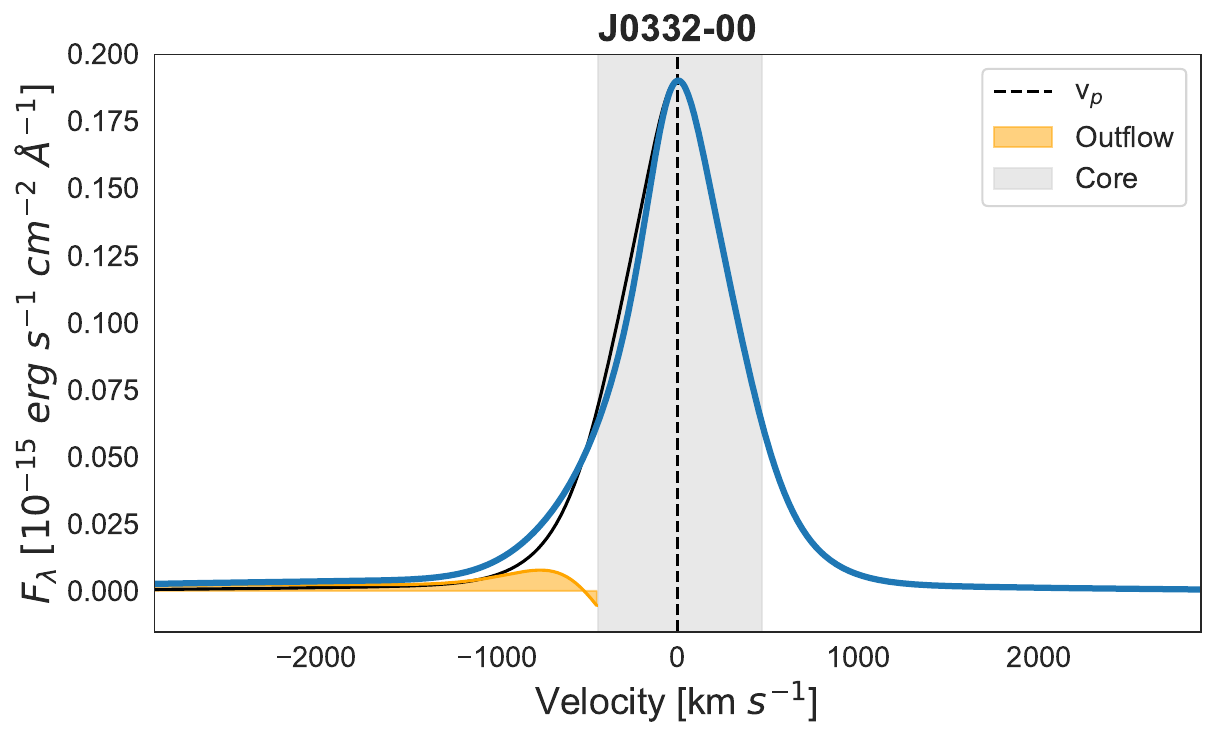}
\includegraphics[width=0.42\textwidth]{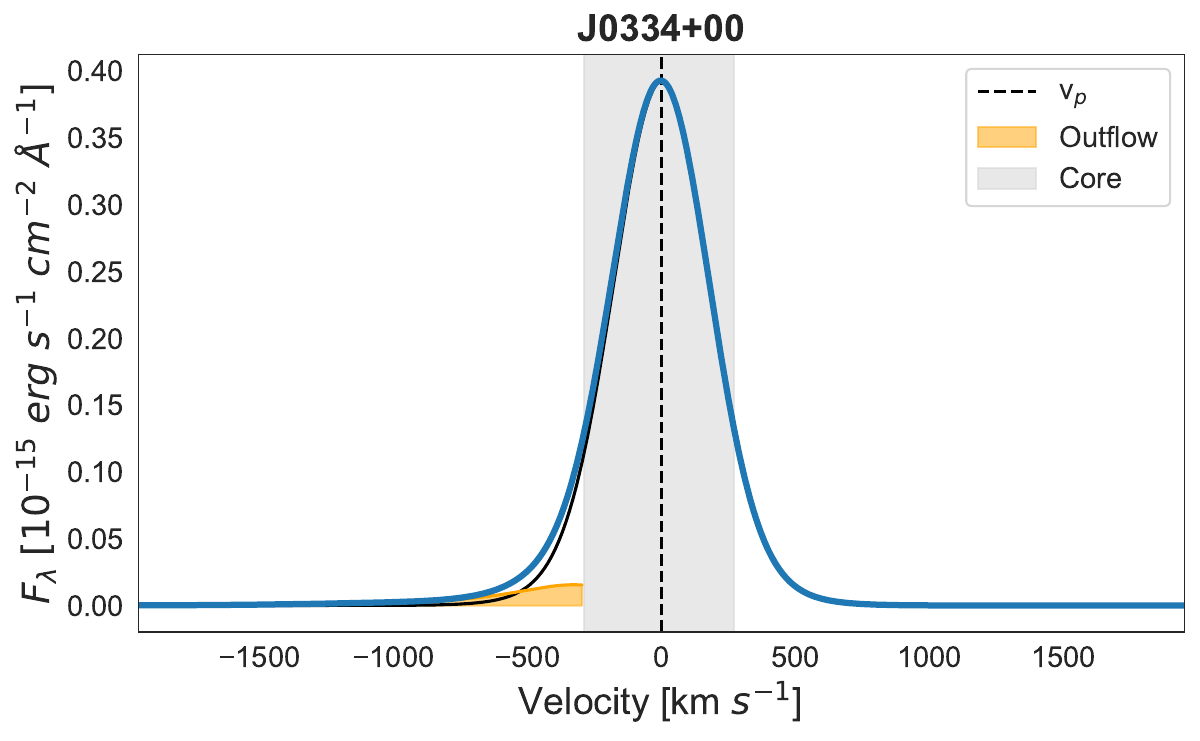}
\includegraphics[width=0.42\textwidth]{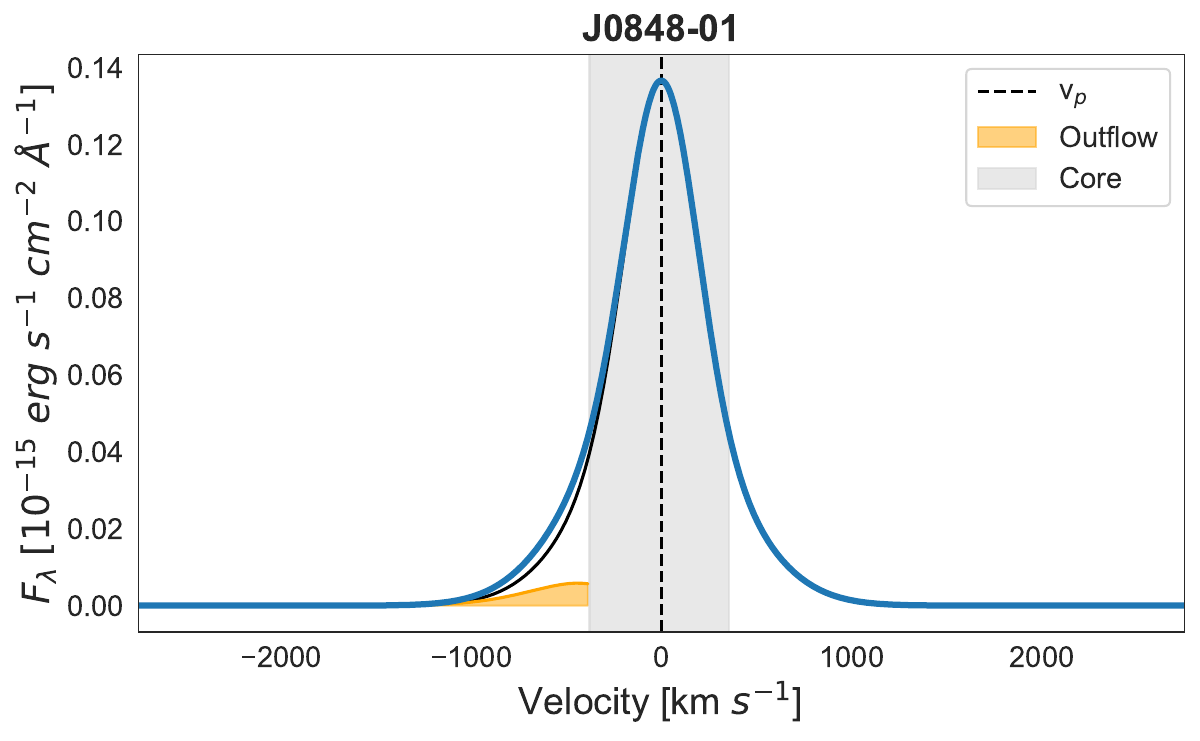}
\includegraphics[width=0.42\textwidth]{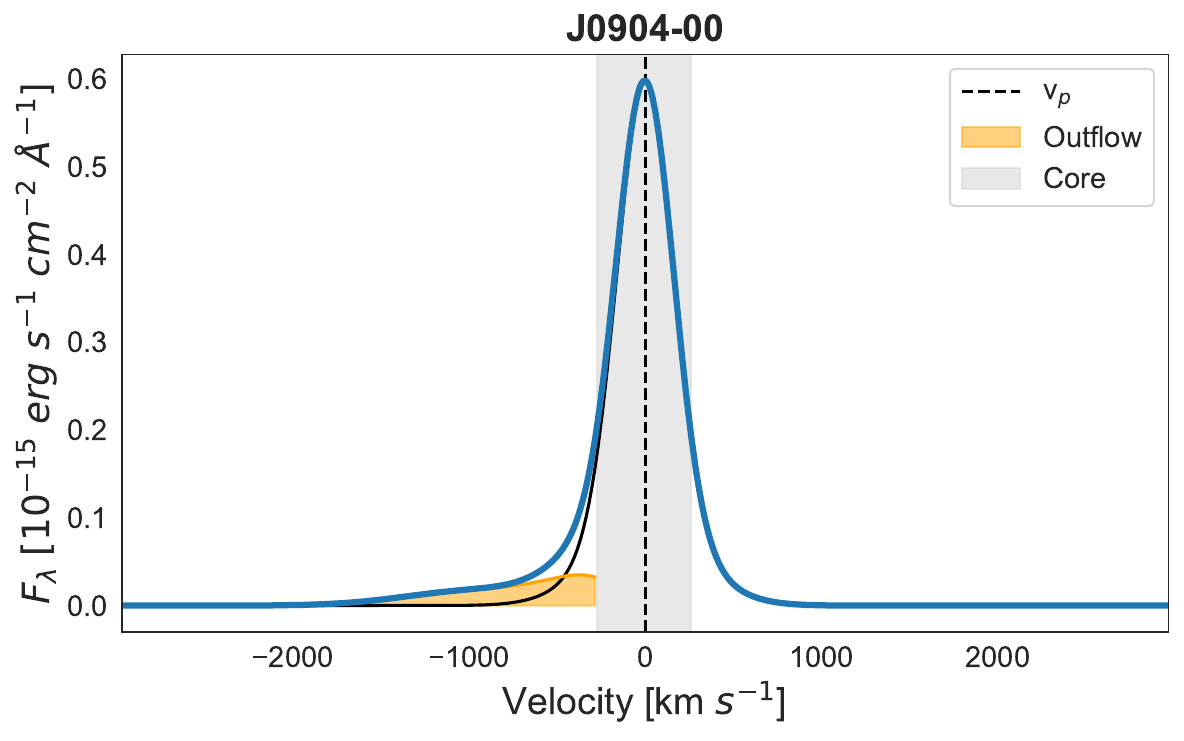}
\includegraphics[width=0.42\textwidth]{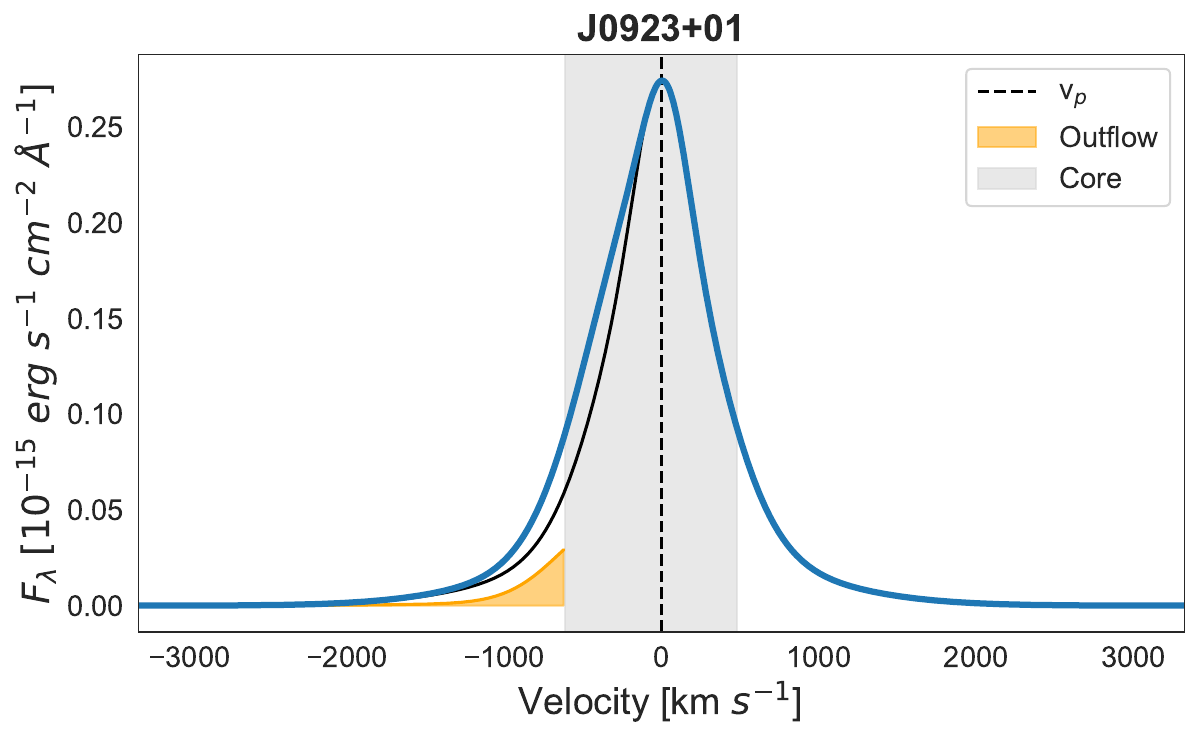}
\includegraphics[width=0.42\textwidth]{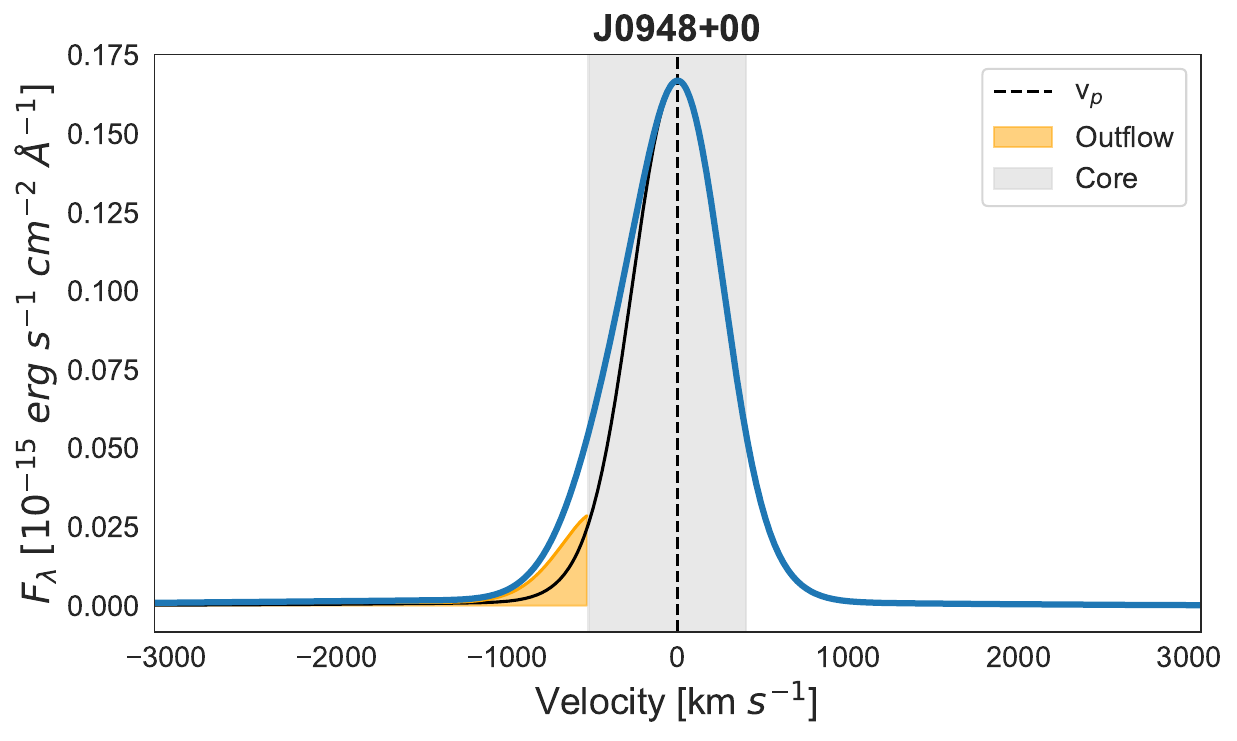}
\includegraphics[width=0.42\textwidth]{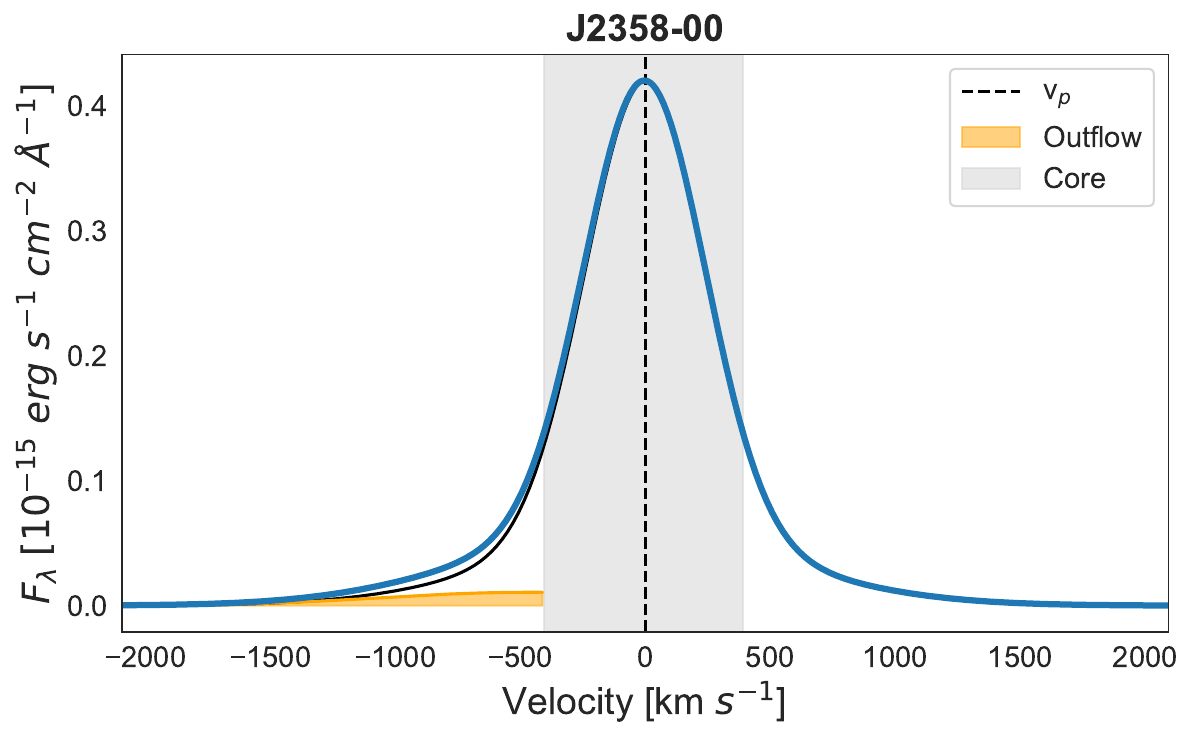}\par}
\caption{continued}
\end{figure*}

\begin{table*}
\caption{Outflow properties derived from the peak-weighted non-parametric fits of the $[OIII]\lambda$5007 line.}
\centering
\begin{tabular}{ccccccccccc} \hline
\textbf{Name} & \textbf{v$_{\rm OF}$} & \textbf{Log(M$_{\rm OF}$)} & \textbf{$\dot{\rm\bold M}_{\rm OF}$} & \textbf{Log($\dot{\rm\bold E}_{\rm kin}$)} & \textbf{$\dot{\rm\bold E}_{\rm kin}$ / $\rm \bold L_{\rm BOL}$}\\ 
& (km s$^{-1}$) & (M$_{\odot}$)  & (M$_{\odot}$ yr $^{-1}$) & (erg s$^{-1}$) & (\%) \\ \hline
\textbf{J0025-10} & -594.4$\pm$7.6 & 5.652$\pm$0.015 & 0.815$\pm$0.031 & 40.959$\pm$0.023 & (3.03$\pm$0.16)$\cdot$10$^{-3}$ \\
\textbf{J0114+00} & -1360$\pm$220 & 5.76$\pm$0.34 & 2.4$\pm$1.4 & 42.1$\pm$0.4 & (9.4$\pm$8.6)$\cdot$10$^{-2}$ \\
\textbf{J0123+00} & -603.1$\pm$7.7 & 5.974$\pm$0.020 & 1.748$\pm$0.082 & 41.302$\pm$0.025 & (3.35$\pm$0.19)$\cdot$10$^{-3}$ \\
\textbf{J0142+14} & -925$\pm$23 & 6.408$\pm$0.100 & 7.2$\pm$1.5 & 42.289$\pm$0.11 & (4.4$\pm$1.1)$\cdot$10$^{-2}$ \\
\textbf{J0217-00} & -1110$\pm$29 & 5.734$\pm$0.063 & 1.88$\pm$0.27 & 41.864$\pm$0.072 & (1.37$\pm$0.23)$\cdot$10$^{-2}$ \\
\textbf{J0217-01} & -840$\pm$860 & 5.410$\pm$0.082 & 0.63$\pm$0.58 & 41.2$\pm$1.6 & (1.1$\pm$4.0)$\cdot$10$^{-2}$ \\
\textbf{J0218-00} & -551$\pm$21 & 5.508$\pm$0.025 & 0.541$\pm$0.036 & 40.716$\pm$0.054 & (2.50$\pm$0.31)$\cdot$10$^{-3}$ \\
\textbf{J0227+01} & 676$\pm$21 & 5.412$\pm$0.030 & 0.534$\pm$0.040 & 40.887$\pm$0.050 & (3.24$\pm$0.37)$\cdot$10$^{-3}$ \\
\textbf{J0234-07} & -709$\pm$4.1 & 6.0148$\pm$0.0046 & 2.254$\pm$0.026 & 41.5532$\pm$0.0086 & (3.130$\pm$0.062)$\cdot$10$^{-2}$ \\
\textbf{J0249+00} & -951$\pm$32 & 6.103$\pm$0.019 & 3.7$\pm$0.2 & 42.018$\pm$0.050 & (3.32$\pm$0.38)$\cdot$10$^{-2}$ \\
\textbf{J0320+00} & -730$\pm$15 & 5.258$\pm$0.048 & 0.401$\pm$0.046 & 40.829$\pm$0.056 & (9.0$\pm$1.1)$\cdot$10$^{-3}$ \\
\textbf{J0332-00} & -1230$\pm$100 & 5.511$\pm$0.044 & 1.21$\pm$0.15 & 41.76$\pm$0.11 & (6.7$\pm$1.7)$\cdot$10$^{-2}$ \\
\textbf{J0334+00} & -542$\pm$19 & 5.732$\pm$0.029 & 0.893$\pm$0.068 & 40.919$\pm$0.056 & (2.58$\pm$0.33)$\cdot$10$^{-3}$ \\
\textbf{J0848-01} & -602$\pm$17 & 5.103$\pm$0.099 & 0.235$\pm$0.054 & 40.43$\pm$0.11 & (1.93$\pm$0.48)$\cdot$10$^{-3}$ \\
\textbf{J0904-00} & -718$\pm$19 & 6.187$\pm$0.017 & 3.38$\pm$0.16 & 41.737$\pm$0.039 & (1.66$\pm$0.15)$\cdot$10$^{-2}$ \\
\textbf{J0923+01} & -791$\pm$14 & 5.744$\pm$0.046 & 1.34$\pm$0.15 & 41.421$\pm$0.053 & (9.2$\pm$1.1)$\cdot$10$^{-3}$ \\
\textbf{J0924+01} &  -1240$\pm$16 &6.032$\pm$0.014 & 4.11$\pm$0.14 & 42.302$\pm$0.021 & (4.22$\pm$0.21)$\cdot$10$^{-3}$ \\
\textbf{J0948+00} & -732$\pm$17 & 5.622$\pm$0.026 & 0.943$\pm$0.059 & 41.197$\pm$0.039 & (9.77$\pm$0.88)$\cdot$10$^{-3}$ \\
\textbf{J2358-00} & -827$\pm$33 & 5.793$\pm$0.088 & 1.58$\pm$0.28 & 41.534$\pm$0.097 & (3.77$\pm$0.85)$\cdot$10$^{-3}$ \\ \hline
\end{tabular}
\label{Tab_speranza_energetics}
\end{table*}

\end{appendix}

\end{document}